\documentclass[useamsfonts]{pasj01}
\usepackage{natbib}
\usepackage{aas_macros}
\usepackage{url}
\usepackage{bm}
\usepackage{xcolor}

\newcommand{\simgt}{\lower.5ex\hbox{$\; \buildrel > \over \sim \;$}}
\newcommand{\simlt}{\lower.5ex\hbox{$\; \buildrel < \over \sim \;$}}

\begin{document}
\SetRunningHead{T. Hamana et al.}{Cosmological constraints from
  HSC survey first year data}
\Received{2019/6/14}
\Accepted{2019/11/21}
\Published{\today}

\title{Cosmological constraints from cosmic shear two-point correlation
  functions with HSC survey first-year data\thanks{This paper was
    published in PAPJ 72 (1), 16 (1-32) (2020), and its arXiv
    version is arXiv:1906.06041v2. After the publication, we discovered a
    couple of bugs in the software used for numerical computations in
    the original paper, and we have redone all the computations affected
    by the bugs with the corrected software. The bugs and their effects
    on results are described in the erratum (to be published in
    PASJ). In this arXiv version (v3), we present revised results
    obtained from corrected computations in the same structure of the
    original paper. Note that no methodology was changed.}}

%
%
\author{Takashi \textsc{Hamana}\altaffilmark{1,2}}%
\author{Masato \textsc{Shirasaki}\altaffilmark{1}}
\author{Satoshi \textsc{Miyazaki}\altaffilmark{1,2}}
\author{Chiaki \textsc{Hikage}\altaffilmark{3}}%
\author{Masamune \textsc{Oguri}\altaffilmark{4,5,3}}%
\author{Surhud \textsc{More}\altaffilmark{6,3}}
\author{Robert \textsc{Armstrong}\altaffilmark{7}}
\author{Alexie \textsc{Leauthaud}\altaffilmark{8}}
\author{Rachel \textsc{Mandelbaum}\altaffilmark{9}}
\author{Hironao \textsc{Miyatake}\altaffilmark{10,11,3,12}}
\author{Atsushi J. \textsc{Nishizawa}\altaffilmark{10,11}}
\author{Melanie \textsc{Simet}\altaffilmark{13,12}}
\author{Masahiro \textsc{Takada}\altaffilmark{3}}
\author{Hiroaki \textsc{Aihara}\altaffilmark{5}}
\author{James \textsc{Bosch}\altaffilmark{14}}
\author{Yutaka \textsc{Komiyama}\altaffilmark{1,2}}
\author{Robert \textsc{Lupton}\altaffilmark{14}}
\author{Hitoshi \textsc{Murayama}\altaffilmark{3,15,16}}
\author{Michael~A. \textsc{Strauss}\altaffilmark{14}}
\author{Masayuki \textsc{Tanaka}\altaffilmark{1}}
\altaffiltext{1}{National Astronomical Observatory of Japan, Mitaka,
  Tokyo 181-8588, Japan}
\altaffiltext{2}{The Graduate University for Advanced Studies, SOKENDAI, Mitaka, Tokyo 181-8588, Japan}
\altaffiltext{3}{Kavli Institute for the Physics and Mathematics of the Universe (Kavli IPMU, WPI), University of Tokyo, Chiba 277-8582, Japan}
\altaffiltext{4}{Research Center for the Early Universe, University of Tokyo, Tokyo 113-0033, Japan}
\altaffiltext{5}{Department of Physics, University of Tokyo, Tokyo 113-0033, Japan}
\altaffiltext{6}{The Inter-University Center for Astronomy and Astrophysics, Post bag 4, Ganeshkhind, Pune, 411007, India}
\altaffiltext{7}{Lawrence Livermore National Laboratory, Livermore, CA
  94551, USA}
\altaffiltext{8}{University of California Santa Cruz, 
1156 High St., Santa Cruz, CA 95064, USA}
\altaffiltext{9}{McWilliams Center for Cosmology, Department of Physics, Carnegie Mellon University, Pittsburgh, PA 15213, USA}
\altaffiltext{10}{Institute for Advanced Research, Nagoya University,
  Nagoya 464-8602, Aichi, Japan}
\altaffiltext{11}{Division of Particle and Astrophysical Science,
  Graduate School of Science, Nagoya University, Nagoya 464-8602, Aichi, Japan}
\altaffiltext{12}{Jet Propulsion Laboratory, California Institute of Technology, Pasadena, CA 91109, USA}
\altaffiltext{13}{University of California, Riverside, 900 University Avenue, Riverside, CA 92521, USA}
\altaffiltext{14}{Department of Astrophysical Sciences, Princeton
  University, 4 Ivy Lane, Princeton, NJ 08544, USA}
\altaffiltext{15}{Department of Physics and Center for Japanese Studies,
  University of California, Berkeley, CA 94720, USA}
\altaffiltext{16}{Theoretical Physics Group, Lawrence Berkeley National
  Laboratory, MS 50A-5104, Berkeley, CA 94720, USA}
%
%

\KeyWords{cosmology: observations --- dark matter --- cosmological
  parameters --- large-scale structure of universe }

\maketitle

\begin{abstract}
We present measurements of cosmic shear two-point correlation functions
(TPCFs) from Hyper Suprime-Cam Subaru Strategic Program 
(HSC SSP) first-year data, and derived cosmological constraints 
based on a blind analysis.
The HSC first-year shape catalog is divided into four
tomographic redshift bins ranging from $z=0.3$ to 1.5 with equal
widths of $\Delta z =0.3$.
The unweighted galaxy number densities in each tomographic bin are
6.1, 6.1, 4.6, 2.7 arcmin$^{-2}$ from the lowest to highest
redshifts, respectively.
We adopt the standard TPCF estimators, $\xi_\pm$, for our
cosmological analysis, given that we find no evidence of the significant
B-mode shear. 
The TPCFs are detected at high significance for all ten combinations of
auto- and cross-tomographic bins over a wide angular range, yielding 
a total signal-to-noise ratio of 19 in the angular ranges adopted in
the cosmological analysis, $7'<\theta<56'$ for $\xi_+$
and $28'<\theta<178'$ for $\xi_-$.
We perform the standard Bayesian likelihood analysis for 
cosmological inference from the measured cosmic shear TPCFs, including 
contributions from intrinsic alignment of galaxies as well as 
systematic effects from PSF model errors, 
shear calibration uncertainty, and 
source redshift distribution errors.
We adopt a covariance matrix derived from realistic mock catalogs
constructed from full-sky gravitational lensing simulations that fully
account for survey geometry and measurement noise.
For a flat $\Lambda$ cold dark matter model, we find $S_8 \equiv
\sigma_8\sqrt{\Omega_m/0.3}=0.823_{-0.028}^{+0.032}$, and
$\Omega_m=0.332_{-0.096}^{+0.050}$.
We carefully check the robustness of the cosmological results
against astrophysical modeling uncertainties and systematic
uncertainties in measurements, and find that none of them has 
a significant impact on the cosmological constraints.
\end{abstract}

%
%
\section{Introduction}
The $\Lambda$ cold dark matter ($\Lambda$CDM) model is now
considered to be the standard theoretical framework for the expansion
history of the Universe and for cosmic structure formation.
The standard $\Lambda$CDM model is described by only a handful of
cosmological parameters.
Measuring values of the cosmological parameters, as well as checking 
their consistency between different cosmological observations, 
is one of the most important goals of modern cosmology.
Multiple probes, such as the cosmic microwave background (CMB; e.g.,
\citealt{2013ApJS..208...19H,2016A&A...594A..13P,2020A&A...641A...6P}), 
high redshift type-Ia supernovae (e.g.,
\citealt{2012ApJ...746...85S,2014A&A...568A..22B}, and
\citealt{2013PhR...530...87W} for a review)), baryon acoustic
oscillations (BAOs; e.g.,
\citealt{2014MNRAS.441...24A,2017MNRAS.470.2617A}), and weak lensing
as described in detail below, have been utilized for this purpose.
Different methods probe different cosmic epochs through a 
measurement of the growth of cosmic structure formation and/or the
distance-redshift relation of the Universe.
In addition, the methods have different parameter degeneracies and are
affected by different systematic effects.
For these reasons, it is common practice to combine multiple probes to
infer tighter and more reliable cosmological constraints.
More importantly, if a discordance between cosmological constraints from
different probes is found, it may indicate physics beyond the
$\Lambda$CDM model.
Therefore it is of fundamental importance to infer improved cosmological
constraints from each probe.
This is exactly the purpose of this study, which uses weak lensing
observations from the HSC SSP. 

Weak lensing is one of the most powerful tools for cosmology, as it
provides a unique means to study the matter distribution in the Universe.
The cosmic shear is the coherent distortion of the shapes of distant
galaxies caused by the gravitational lensing of intervening large-scale
structures,  including the dark matter component.
Statistical measures of cosmic shear, such as 
the two-point correlation function (TPCF) or the power
spectrum, depend both on the time evolution of the cosmic
structures and on the cosmic expansion history, and thus serve 
as a unique cosmological probe. They probe the large-scale, 
linear to weakly non-linear, matter power spectrum at relatively recent
epochs ($z < 1$), and thus are most sensitive to the normalization of
matter fluctuation ($\sigma_8$) and the mean matter density parameter
($\Omega_m$) \citep{1997ApJ...484..560J}.
Because of the degeneracy between these two parameters, the combination
$S_8 = \sigma_8 (\Omega_m/0.3)^{\alpha}$ with a degeneracy direction of
$\alpha\sim 0.5$ is commonly used to quantify the constraints from
cosmic shear. 

Cosmological constraints from cosmic shear are improved primarily by
increasing the survey volume as well as the number density of source galaxies, 
along with a proper control of systematic effects. 
Currently, three wide-field imaging surveys that will each eventually
cover over 1000 square degrees are underway; the Dark Energy Survey 
\citep[DES,][]{2016MNRAS.460.1270D}, the Kilo-Degree 
survey \citep[KiDS,][]{2013ExA....35...25D}, and the 
Hyper Suprime-Cam Subaru Strategic Program 
\citep[hereafter the HSC survey;][]{2018PASJ...70S...4A}.
All three projects have published initial cosmic shear analyses with
early data, yielding $4-8$ percent constraints on $S_8$
\citep{2018PhRvD..98d3528T, 2017MNRAS.465.1454H, 2017MNRAS.471.4412K, 2020A&A...633A..69H, 2019PASJ...71...43H}.
They also demonstrated that none of the systematic effects examined in
the papers seriously affected the resulting constraints.  

Among the three surveys, the unique advantage of the HSC survey is its
higher galaxy number density\footnote{The number densities given in this
  paragraph are the effective number density of galaxies used for cosmic
  shear analyses defined in \citet{2013MNRAS.434.2121C} and are taken
  from Table 1 and 2 of \citet{2019PASJ...71...43H}.} of $16.5$
arcmin${}^{-2}$ compared to that of DES ($5.14$ arcmin${}^{-2}$) and KiDS
($6.85$ arcmin${}^{-2}$), due to the combination of its depth
($5\sigma$ point-source depth of the Wide layer of $i\sim 26$ AB mag)
and excellent image quality (typical $i$-band seeing of $0\farcs58$,
\citealt{2018PASJ...70S...8A,2018PASJ...70S..25M}).
\citet{2019PASJ...71...43H} measured the tomographic cosmic shear power 
spectra using the HSC survey first-year data over 137~deg${}^2$.
They selected galaxies from the HSC first-year weak lensing shear 
catalog \citep{2018PASJ...70S..25M} with photometric redshifts
\citep{2018PASJ...70S...9T} ranging from
0.3 to 1.5, and divided them into four tomographic
redshift bins with equal widths of $\Delta z =  0.3$.
Even the highest redshift tomographic bin contains 2.0 galaxies 
per arcmin${}^{2}$.
They detected cosmic shear power spectra with high
signal-to-noise ratios ($SN$) of $SN=4.9$, $9.2$, $12.3$, and $11.5$ for
auto-power spectra of each tomographic bin (from the lowest to highest
redshift) and $SN=15.6$ for combined auto- and cross-power spectra. 

In this paper, we present the cosmic shear TPCFs measured from the HSC
survey first-year data, and derive cosmological constraints with them.
We use the same data set as that used in \citet{2019PASJ...71...43H}, but
use a completely different analysis scheme, namely the real-space TPCFs 
instead of Fourier-space power spectra, using an independent cosmological
inference pipeline.
In principle, those two estimates provide almost the same information, but 
different treatments of actual observational effects, such as 
discrete galaxy sampling and the correction of the irregular survey
geometry, which can affect the measured signal and the cosmological
inference in different ways. 
Also, the two approaches have different noise properties and different
sensitivities to systematic effects, and are sensitive to different scales.
Therefore this study provides an important cross-check of the robustness
of the Fourier-space analysis by \citet{2019PASJ...71...43H}.
Furthermore, our analysis indicates that our TPCF analysis probes
a slightly different range of multipole from that used in
\citet{2019PASJ...71...43H}, and therefore contains some 
complementary cosmological information.

The structure of this paper is as follows. 
In Section~\ref{sec:data}, we briefly summarize the HSC survey
first-year shear catalog and the photometric redshift data
used in this study. We also describe our blind analysis scheme.
In Section~\ref{sec:measurements}, we describe the method to measure the
TPCFs of the cosmic shear, and present our measurements.
We also present TPCFs of the measured shapes of stars and
residuals between those shapes and the point spread function (PSF) model,
which allow us to estimate the residual systematics
in the cosmic shear TPCFs.
In Section~\ref{sec:models}, we summarize model ingredients for the
cosmic shear TPCFs and covariance.
Our method for cosmological inference is described in
Section~\ref{sec:analyses} along with our methods to take into account
various systematics in our cosmological analysis.
Our cosmological constraints and tests for systematics are
presented in Section~\ref{sec:results}.
Finally, we summarize and discuss our results in Section~\ref{sec:summary}.
In  Appendix~\ref{appendix:residual_PSF}, we describe the impact of the 
PSF leakage and the residual PSF model error on the measurement of shear
TPCFs.
In  Appendix~\ref{appendix:B-mode}, we present E/B-mode TPCFs measured
from the HSC survey data.
In Appendix~\ref{appendix:mock}, we describe mock simulation data that
are used to derive the covariance matrix and to test our cosmological inference
pipeline.
In Appendix~\ref{appendix:powerspectrum}, the difference of the
information content in the measured cosmic shear statistics between this
study and \citet{2019PASJ...71...43H} is examined.
In Appendix~\ref{appendix:photoz_uncertainties}, we discuss a
possible impact of an error in the outlier fraction of galaxy redshift
distributions on cosmological constraints on.

Throughout this paper we quote 68\% credible intervals for 
parameter uncertainties unless otherwise stated.

%
%
\section{HSC survey data}
\label{sec:data}

In this section, we briefly summarize the HSC survey products used in this
study.
\citet{2019PASJ...71...43H} describe the dataset we use in detail; here we focus 
on those aspects that are directly relevant to this study.
We refer the readers to
\citet{2018PASJ...70S...4A} for an overview of the HSC survey and 
survey design, \citet{2018PASJ...70S...8A} for the first public 
data release,
\citet{2018PASJ...70S...1M,2018PASJ...70S...2K,2018PASJ...70...66K,2018PASJ...70S...3F}
for  the performance of 
the HSC instrument itself, \citet{2018PASJ...70S...5B} for the optical imaging data 
processing pipeline used for the first-year data, 
\citet{2018PASJ...70S..25M} for the first-year shape catalog, 
\citet{2018MNRAS.481.3170M} for the
calibration of galaxy shape measurements with image simulations, 
and \citet{2018PASJ...70S...9T} for photometric redshifts derived for
the first data. 

%
%
\subsection{HSC first-year shape catalog}
\label{sec:shape-catalog}

We use the HSC first-year shape catalog \citep{2018PASJ...70S..25M}, in
which the shapes of galaxies are estimated on the $i$-band coadded image
using the re-Gaussianization PSF correction method 
\citep{2003MNRAS.343..459H}.
Only galaxies that pass our selection criteria are contained in the catalog.
Among others, the four major criteria for galaxies to be selected are, 
\begin{enumerate}
\renewcommand{\labelenumi}{(\arabic{enumi})}
\item {\it full-color and full-depth cut}: the object should be located in regions reaching 
approximately full survey depth in all five ($grizy$) broad bands,
\item {\it magnitude cut}: $i$-band cmodel magnitude (corrected for
  extinction) should be brighter than 24.5 AB mag, 
\item {\it resolution cut}: the galaxy size normalized by the PSF size
  defined by the re-Gaussianization method should be larger than a given
  threshold of {\tt ishape\_hsm\_regauss\_resolution $\ge$ 0.3}, 
\item {\it bright object mask cut}: the object should 
not be located within the bright object masks.
\end{enumerate}
 See Table~4 of
\citet{2018PASJ...70S..25M} for the full description of the selection 
criteria.
As a result, the final weak lensing shear catalog covers 136.9 deg$^2$,
consisting of 6 disjoint regions (named  XMM, GAMA09H, WIDE12H,
GAMA15H, VVDS, and HECTOMAP) and contains $\sim$12.1M galaxies.

%
%
\subsection{Photometric redshifts}
\label{sec:photo-z}

Since spectroscopic redshifts have been obtained for only a small
fraction of galaxies in the HSC shape catalog, we utilize 
photometric redshift (hereafter photo-$z$) information to divide
galaxies into tomographic redshift bins.

Utilizing the HSC five-band photometry, photo-$z$s were estimated with
six independent codes,  described in detail in \citet{2018PASJ...70S...9T}.
Three of the six photo-$z$'s used the PSF-matched aperture photometry
(called the {\tt afterburner} photometry; see \citealt{2018PASJ...70S...8A}),
which we adopt in this study: They are (1) an empirical polynomial
fitting method ({\tt DEmP}) \citep{2014ApJ...792..102H}, (2) a neural
network code ({\tt Ephor AB}), and (3) a hybrid code combining machine
learning with template fitting ({\tt FRANKEN-Z}). 

The accuracy of HSC photo-$z$'s were examined in detail in
\citet{2018PASJ...70S...9T}, who concluded that HSC photo-$z$'s ($z_p$) 
are most accurate at $0.2\simlt z_p \simlt 1.5$.
Given the smaller lensing signals for lower redshift galaxies,
we set the redshift range of our cosmic shear analysis from 0.3 to 1.5.
We adopt the {\tt best} estimate of {\tt Ephor AB} for the point
estimator of photo-$z$'s to define tomographic bins.
Specifically, we select galaxies with the point estimator being 
within that redshift range, and divide them into four 
tomographic redshift bins with equal redshift width of $\Delta z =0.3$, 
again based on the point estimator.
After the redshift cut, the final number of galaxies used in this study 
is $\sim$9.6 million, which are split into four tomographic bins, 
containing 3.0M, 3.0M, 2.3, and 1.3M galaxies respectively from the lowest to 
highest redshift bins.

%
%
\subsubsection{Redshift distribution of galaxies in each tomographic bin}
\label{sec:pz}

%
%
\begin{figure}
\begin{center}
 \includegraphics[width=82mm]{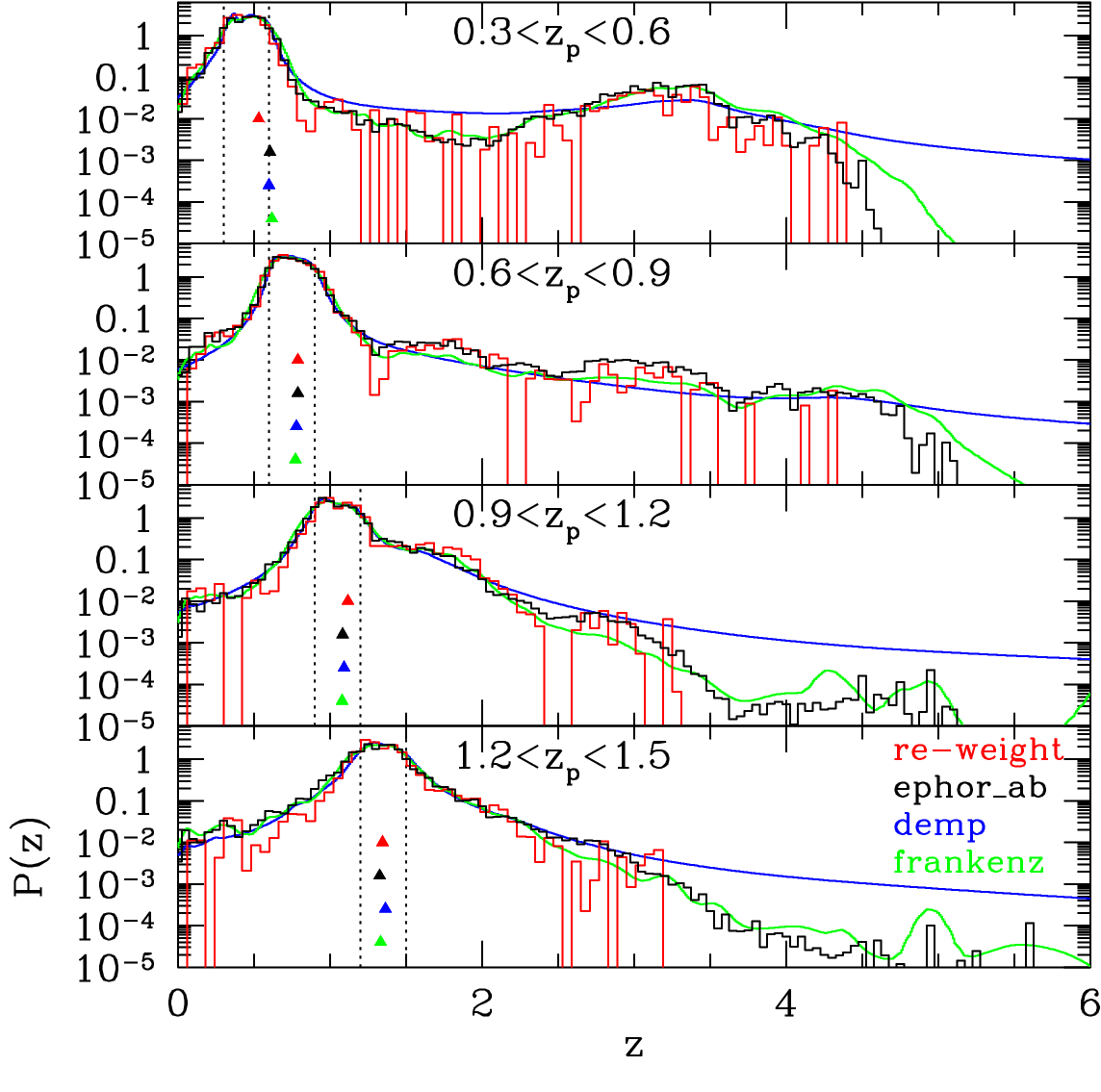}
\end{center}
\caption{Histograms show galaxy redshift distributions for the four
  tomographic redshift bins; $0.3<z<0.6$,
  $0.6<z<0.9$, $0.9<z<1.2$, and $1.2<z<1.5$, from the top to bottom
  panels, respectively.
  The triangles show the mean
  redshift of each redshift distribution.
  The vertical dotted lines show the boundaries of the four tomographic
  bins. 
  The redshift distributions are computed up to $z=6$.
  Different colors indicate different methods:
  The COSMOS-reweighted method (red) is our principal method.
  In order to test the robustness of our results against uncertainties
  in the redshift distributions, we will use stacked-$P(z)$ with three
  photo-$z$ methods, {\tt Ephor AB} (black), {\tt DEmP} (blue), and  
  {\tt FRANKEN-Z} (green).
  Different binning for different methods originates from their 
  different redshift resolutions, except for the COSMOS-reweighted
  method for which a three-times under-sampled binning is shown for
  clarity (the original resolution is $\Delta z = 0.02$).
  \label{fig:pz}}
\end{figure}

Since the photo-$z$ point estimator is a noisy estimator of the true
redshifts of galaxies, the true redshift distribution of galaxies in
individual tomographic bins must be separately and reliably estimated.
We follow the methodology described in \citet{2019PASJ...71...43H} 
to infer the true redshift distribution as well as to test 
the robustness of derived cosmological
results against uncertainty in the adopted redshift distributions.
They adopted the reweighting method based on the HSC's five-band 
photometry and COSMOS 30-band photo-$z$ catalog
\citep{2009ApJ...690.1236I,2016ApJS..224...24L}.
We refer the readers to Section~5.2 of \citet{2019PASJ...71...43H} and
references therein for a full detail of the method. Here we only present
the derived redshift distributions as the red histograms in 
Figure~\ref{fig:pz}, which are the same as those used in 
\citet{2019PASJ...71...43H}.
The distributions computed to $z=6$. 
We use them as our fiducial redshift distributions in our cosmological
analysis.
In our model description in Section~\ref{sec:models}, these redshift
distributions are denoted as $p^a(z)$, where $a=1-4$ runs over 
the four tomographic bins.
We note that for the lowest tomographic bin, the mean redshift
shown by the red triangle looks not to match up with the
histograms. This is due to outliers located at higher redshifts.
The $3\sigma$ clipped mean redshifts are summarized in Table 4 of
\citet{2019PASJ...71...43H}. For the lowest bin, the clipped mean is
$z=0.44$ which is very close to the median redshift of $z=0.43$.

We also infer the stacked photo-$z$ probability distribution functions 
(PDFs), which are obtained by
stacking the full PDFs of photo-$z$'s for individual galaxies
($P_j(z)$) with their shear weight ($w_j$),
$p^a(z)=\sum_j w_j P_j(z) / \sum_j w_j$, where the summation runs 
over all galaxies in individual tomographic bins.
The stacked photo-$z$ PDFs for the three photo-$z$ methods are 
shown in the three bottom panels of Figure~\ref{fig:pz}.
Since stacking $P_j(z)$ is not a mathematically sound
way to infer the true redshift distribution (see Section~5.2 of
\citealt{2019PASJ...71...43H}), we do not adopt the stacked $p(z)$ as a
fiducial choice, but use it merely for testing the impact of redshift
distribution uncertainties in Section \ref{sec:impact_dz}.

%
%
\subsection{Weak lensing shear estimation}
\label{sec:shear-estimation}

The HSC shape catalog described in Section~\ref{sec:shape-catalog}
contains all the basic parameters needed to estimate the weak lensing
shear with the re-Gaussianization method, including corrections for biases.
The following five sets of parameters for each galaxy are directly 
relevant to this study; (1) the two-component distortion, 
$\bm{e}=(e_1,e_2)$, which represents the shape of each galaxy image, 
(2) shape weight, $w$, (3) intrinsic shape dispersion per component,
$e_{\mbox{rms}}$, (4) multiplicative bias, $m$, and (5) additive bias,
$(c_1, c_2)$.
Following Appendix~A of \citet{2018PASJ...70S..25M}, 
an estimator for the shear is obtained for each galaxy as
\begin{equation}
  \label{eq:gamma-e}
        \hat{\gamma}_i = {1\over {1+\bar{m}}}
        \left[ {{e_i}\over{2 \cal{R}}}-c_i\right], 
\end{equation}
with the weighted-average multiplicative bias factor,
\begin{equation}
  \label{eq:hatm}
        \bar{m}={{\sum_i w_i m_i} \over {\sum_i w_i}},
\end{equation}
and the shear responsivity $\cal{R}$ representing the response 
of the distortion to a small shear \citep{1995ApJ...449..460K,
  2002AJ....123..583B} given by
\begin{equation}
  \label{eq:Resp}
        {\cal{R}}=1-{{\sum_i w_i e_{\mbox{rms}}^2} \over {\sum_i w_i}}.
\end{equation}
In the above expressions, the subscript $i$ denotes each galaxy, and
the summation is taken over all galaxies in each tomographic redshift bin.

%
%
\subsubsection{Selection bias}
\label{sec:selection-bias}

In addition to the shear calibration mentioned above, which is based on
the full galaxy sample in the shape catalog, we take account of the additional
multiplicative biases arising from the tomographic redshift galaxy
selection.
To do so, we follow \citet{2019PASJ...71...43H} and we refer the readers
to the paper and references therein for details.
In short, there are two sources of biases: One is the selection bias
that is due to the difference in galaxy size distributions for different
tomographic samples. The other is the correction to the shear
responsivity due to the dependence of the intrinsic ellipticity
variation on redshift (see subsection 5.3 of \citet{2018MNRAS.481.3170M}
for details). Both biases vary with the tomographic bins.
The former is denoted by $m_{\mbox{sel}}^a$ and the latter is denoted by
$m_R^a$, where the superscript $a$ labels the tomographic bin.
As we use exactly the same data set as that used in
\citet{2019PASJ...71...43H} with the same tomographic binning, we
adopt the same values of those biases given in Table~3 of
\citet{2019PASJ...71...43H}.
We apply these corrections to the {\it theoretical prediction} of cosmic
shear TPCFs (see Section~\ref{sec:model_correction}) as
$\xi_{\pm}^{ab}(\theta) \rightarrow
(1+m_{\mbox{sel}}^a+m_R^a)(1+m_{\mbox{sel}}^b+m_R^b)
\xi_{\pm}^{ab}(\theta)$. 

%
%
\subsection{Blinding}
\label{sec:blinding}

In order to avoid confirmation bias, we perform our cosmological
analysis in a blind fashion.
The HSC weak lensing team defined blinding and unblinding procedures, 
and agreed that they must be followed in cosmological analysis of the
weak lensing data (see Section~3.2 of \citealt{2019PASJ...71...43H} for
the overall description).
Here we give a brief overview of the blinding scheme 
we adopt for our analysis.

We use a two-level blinding scheme similar to
\citet{2019PASJ...71...43H}. 
The first is a catalog-level blinding, while the
second is the analysis-level blinding which is adopted during the
cosmological analysis.
At the catalog-level, we  blind the real shear values by modifying 
the multiplicative bias as
\begin{equation}
  \label{eq:blind}
        m_{\mbox{cat}}^i=m_{\mbox{true}} +dm_1^i+dm_2^i,
\end{equation}
where $m_{\mbox{true}}$ denotes the array of true multiplicative bias values
in the HSC shape catalog for each galaxy, and the index $i$ runs from 0
to 2 and denotes the three different shear catalog versions.
There are multiple cosmological analyses that are being conducted 
by the HSC team, each with different analysis leads. Each analysis
lead receives a separate set of three catalogs.
The analysis team carried out the same analysis for all the three catalogs.
The values of $dm_1^i$ are different for each of the three catalogs as well as for 
each analysis team, and are encrypted.
Only the PI of each analysis team can decrypt them, and this term is removed 
before performing the analysis. 
This prevents an accidental comparison of blinded catalogs by 
another analysis team. 
The values of $dm_2^i$ are different for the three catalogs and are
encrypted by a public key from a person designated ``blinder-in-chief''.
Only one of the $dm_2^i$ values is zero. 
These values can be decrypted only by the blinder-in-chief once all the 
conditions for unblinding have been met (see below).

The analysis-level blinding procedure involves blinding of the best-fit values of 
the inferred cosmological constraints. 
All cosmological constraint plots were plotted with shifted values of
cosmological parameters ($\bm{p}$) such that
$\bm{p}_{\mbox{blind}}=\bm{p}-\bm{p}_{\mbox{best}}$, before inspecting
the derived constraints for systematics tests. 

We laid down two conditions for unblinding: 
(1) the passing of sanity checks of the analysis
software and the treatment of systematic effects, and (2)  validation
of analysis choices for cosmic shear TPCFs and studies of their 
impact on the cosmological constraints, which we describe in the
following sections.
After the final unblinding, 
we did not change the analysis setup in any way, 
and we report
the cosmological constraints as at the time of unblinding. 
We unblind in stages; the first analysis-level unblinding was 
removed about a month and a half before the catalog-level unblinding.
Three versions of the paper, corresponding to the analysis 
from each of the three blinded catalogs, were written up prior to 
the catalog-level unblinding.
(Note that this step differs from the unblinding process of
\citealt{2019PASJ...71...43H}, they did the catalog-level unblinding
soon after the first 
analysis-level unblinding, then wrote up the paper based on the true
catalog.)
Then after the catalog-level unblinding and before submission 
to the journal, the paper based on the true catalog underwent 
internal review from the HSC collaboration. 
No change in the results was made at the internal reviewing stage.

It should be noted that although we analyzed the three blind catalogs,
we used the same covariance matrix derived from realistic mock
catalogs which were generated using the true shape catalog (see
Section \ref{sec:covariance} for details).
As a result, derived best-fit $\chi^2$ values for three
blind catalogs were different reflecting the added $dm_1^i$ to each
catalog.
To be specific, for our fiducial analysis setup (see Section
\ref{sec:analyses}), derived best-fit $\chi^2$ values for the true
catalog was found to be 162.0 for the effective degree-of-freedom of
167 (see Section \ref{sec:effective_parameters}), whereas $\chi^2$ for
two false catalogs were 114.0 and 116.0 for $dm_1^i$ values of
0.08491 and 0.08004, respectively (note that those numbers were
generated based on a random number generator and were very close each
other by an accidental chance).
It is true that the $\chi^2$ values were a possible indication of
which was the true catalog, though the true catalog does not
necessarily give the most reasonable $\chi^2$ value.
It is important to note that before unblinding the analysis-level
blinding, we had no idea about inferred cosmological parameter values
as the best-fit values were blinded, and after unblinding the
analysis-level blinding, we did not change any analysis setup.
The catalog-level blinding might not work as
designed because of our use of the same covariance matrix.
Even so, the analysis-level blinding worked to avoid the confirmation
bias. 

%
%
\section{Measurements from the HSC survey data}
\label{sec:measurements}

In this section, we present our measurements of tomographic cosmic shear 
TPCFs from the HSC first-year data.  In addition, we present
measurements of
the auto- and cross-TPCFs of 
the shapes of PSFs and the difference between the shapes of the 
PSF model and of stars, 
which we use to quantify residual 
systematics in our cosmic shear TPCF measurements. 

%
%
\subsection{Cosmic shear TPCFs}
\label{sec:cosmic-shear-TPCS}

We adopt the standard estimates of cosmic shear TPCFs,
$\xi_{\pm}=\langle \gamma_t \gamma_t \rangle \pm \langle \gamma_\times
\gamma_\times \rangle$, where the tangential ($t$) and cross ($\times$)
components of shear are defined with respect to the direction
connecting a pair of galaxies under consideration.
They can be estimated for two tomographic
redshift bins $a$ and $b$ as
\begin{equation}
  \label{eq:xipm}
  \hat{\xi}_\pm^{ab}(\theta)={
    {\sum_{ij} w_i w_j \left[\hat{\gamma}_{i,t}^{a}(\vec{\theta}_i)
        \hat{\gamma}_{j,t}^{b}(\vec{\theta}_j) \pm \hat{\gamma}_{i,\times}^{a}(\vec{\theta}_i)
        \hat{\gamma}_{j,\times}^{b}(\vec{\theta}_j)\right]}
    \over
    {\sum_{ij} w_i w_j }
    },
\end{equation}
where the summation runs over pairs of galaxies with their 
angular separation
$\theta=|\vec{\theta}_i-\vec{\theta}_j|$ within an interval
$\Delta \theta $ around $\theta$.

For the measurement of the TPCFs themselves, we used the public software {\tt
  Athena}\footnote{http://www.cosmostat.org/software/athena}
\citep{2002A&A...396....1S}.
A total of 31 bins with equal logarithmic bin-widths of $\Delta
\log_{10}\theta=0.1$ are chosen with central $\theta$ ranging from
$10^{-0.5}\simeq 0.316$ arcmin to $10^{2.5}\simeq 316$ arcmin,
although only a subset of these angular bins are used 
in our cosmological analyses as described 
in Section~\ref{sec:data_vector}.
As described in Section~\ref{sec:shape-catalog}, 
the HSC first-year shape catalog consists of 6 disjoint fields.
Since gaps between fields are more than 20 degrees, we first 
compute the denominator and numerator of equation~(\ref{eq:xipm}) 
for each field and then sum up each term separately for the final
results. 
Overall, we have non-zero detections in most angular bins between
$\theta \sim 1'$  and $\sim 100'$.

%
%
\subsection{TPCFs of shapes of PSF and residuals}
\label{sec:PSF-TPCS}

The PSF anisotropy induces additional deformation in galaxy shapes, 
which the shear estimation algorithm must correct for
\citep[see][for a review]{2018ARA&A..56..393M}.
However, in the case of the re-Gaussianization PSF correction method, a
small residual in the correction for PSF anisotropy is unavoidable
for two reasons: imperfect measurements and/or 
modeling of PSFs, and the correction error for PSF from galaxy images, 
an effect referred to as PSF leakage.
In fact, systematic tests of the HSC first-year 
shape catalog showed small residual correlations between galaxy 
shears and PSF shapes \citep{2018PASJ...70S..25M}, 
which may bias the cosmic shear TPCFs and our cosmological analysis.

Here we outline our scheme to correct for these systematics.
We follow the simple model used by 
\citet{2019PASJ...71...43H} \citep[see also][]{2018PhRvD..98d3528T},
in which PSF residuals are assumed to be added to the shear linearly
\begin{equation}
  \label{eq:g_psf}
  \gamma^{\mbox{sys}} =\alpha_{\mbox{psf}} \gamma^p + \beta_{\mbox{psf}} \gamma^q,
\end{equation}
where $\gamma^P$ is the shear\footnote{``Shears'' of stars and
  PSFs are converted from the measured distortion using the relation
  between them for intrinsically round objects ($\gamma = e/2$).
  See \citet{2018PASJ...70S..25M} for the definition of distortion of
  star images.} 
of the shape of the model PSF, and $\gamma^q$
is the difference in shears between the PSF model and the true PSF, 
as estimated from the shapes of individual stars, $\gamma^\ast$, i.e.,
$\gamma^q = \gamma^p-\gamma^\ast$.
The first and second terms of the right hand side of 
equation~(\ref{eq:g_psf}) represent the residual
PSF effects from the deconvolution error and the imperfect PSF model 
mentioned above, respectively.
With these terms added to the measured shear $\hat{\gamma}$, the
contributions from these terms to observed TPCFs are written as 
\begin{equation}
  \label{eq:xi_psf}
  \hat{\xi}_{\mbox{psf},\pm}(\theta) = \alpha_{\mbox{psf}}^2 \xi_\pm^{pp}(\theta)
  +2 \alpha_{\mbox{psf}}\beta_{\mbox{psf}} \xi_\pm^{pq}(\theta)
  +\beta_{\mbox{psf}}^2 \xi_\pm^{qq}(\theta),
\end{equation}
where $\xi_\pm^{pp}$ and $\xi_\pm^{qq}$ represent the
auto-TPCFs of $\gamma^p$ and $\gamma^q$, respectively, and
$\xi_\pm^{pq}$ are the cross-TPCFs of $\gamma^p$ and $\gamma^q$.
Those TPCFs are computed using stars that were reserved from the PSF
estimation \citep[see][for details]{2018PASJ...70S...5B}.
In the HSC data reduction pipeline, stars
used for PSF measurement/modeling are selected based on the
distribution of high-$S/N$ objects with stellar sizes. 
About 80\% of selected stars are used for the PSF measurement and
its modeling (those are flagged as {\tt icalib\_psf\_used=True} 
in the HSC 1st-year shape catalog), 
while the remaining 
stars are {\it reserved} for cross-validation of the PSF modeling,
which we use to compute the TPCFs.
The measured TPCFs are presented in
Appendix~\ref{appendix:residual_PSF}. 
An estimation of the proportionality factors $\alpha_{\mbox{psf}}$ and
$\beta_{\mbox{psf}}$ is given in Appendix~\ref{appendix:residual_PSF},
in which we find $\alpha_{\mbox{psf}} \sim 0.03$ and
$\beta_{\mbox{psf}} \sim -1.4$. 
Therefore, given the amplitude of the measured TPCFs, 
$\hat{\xi}_{\mbox{psf},+}$ can be as large as $\sim 10^{-6}$ at
$\theta\sim 10'$.
We correct this effect by adding the term, equation~(\ref{eq:xi_psf}),
to the {\it theoretical model} of the cosmic shear TPCFs (see
Section~\ref{sec:model_correction}). 
Our treatment of this systematic effect in the cosmological analysis
is described in Section~\ref{sec:Systematic_parameters}.
This residual PSF effect on $\xi_-$ is much smaller than
that on $\xi_+$ (see Appendix~\ref{appendix:residual_PSF}), 
so we do not apply that correction to $\xi_-$.

%
%
\section{Models of the cosmic shear TPCFs and covariance matrix}
\label{sec:models}

In this section, we summarize models for the measured cosmic shear
TPCFs, consisting of two major components, the cosmic shear 
arising from the gravitational lensing effect by large-scale 
structures \citep[see][for a review]{2015RPPh...78h6901K} 
and the intrinsic alignment of galaxy shape
\citep[see][for reviews]{2015PhR...558....1T,2015SSRv..193..139K}.
In practice, the measured cosmic shear TPCFs are also affected by
systematics, such as the shear calibration error and residual PSF 
error and/or modeling, which we also summarize in this section.
In addition, we describe our model of covariance matrix used for 
the cosmological analysis.

%
%
\subsection{Cosmic shear TPCFs}
\label{sec:model_cosmicshear}

The cosmic shear TPCFs induced by the gravitational lensing effect 
are related to the cosmic shear power spectra as 
\citep[see e.g.,][and references therein]{2015RPPh...78h6901K}
\begin{equation}
  \label{eq:xi_theory}
  \xi_{\mbox{GG},\pm}^{ab}(\theta) = {1\over {2\pi}} \int d\ell \ell J_{0,4} (\ell
  \theta) P_\kappa^{ab}(\ell),
\end{equation}
where $a$ and $b$ refer to tomographic redshift bins and 
$J_{0,4} (x)$ is the zeroth-order (for $\xi_+$) or fourth-order 
(for $\xi_-$) Bessel function of the first kind. 
We note that in the above expression and in what follows we assume 
no B-mode shear because we find that the B-mode component of cosmic 
shear TPCFs is consistent with zero as shown in 
Appendix~\ref{appendix:B-mode} (see also  
\citealt{2019PASJ...71...43H} from the power spectrum analysis of 
the B-mode shear).
Using the flat-sky and the Limber approximations, the convergence 
power spectrum, $P_\kappa(\ell)$, is computed from the nonlinear 
matter power spectrum, $P_m^{NL}(k)$, as
\begin{equation}
  \label{eq:P_kappa}
  P_\kappa^{ab}(\ell) = \int_0^{\chi_H} d \chi
  {{q^a(\chi)q^b(\chi)} \over f_K^2(\chi)}
  P_m^{NL}\left( {\ell \over {f_K(\chi)}}, \chi \right),
\end{equation}
where $\chi$ is the comoving radial distance, $\chi_H$ is the comoving
horizon distance, and $f_K(\chi)$ is the comoving angular distance.
For the computation of the linear matter power spectrum, we use {\tt CAMB}
\citep{2011PhRvD..84d3516C}.
In order to model the nonlinear matter power spectrum, we 
employ the fitting function by \citet{2012MNRAS.420.2551B}, 
which is based on the {\tt halofit} model
\citep{2003MNRAS.341.1311S,2012ApJ...761..152T} but is modified 
so as to include the effect of non-zero neutrino mass. 
Finally, the lensing efficiency function, $q(\chi)$, is defined 
as
\begin{eqnarray}
  \label{eq:lensingefficiency}
  q^a(\chi)={3 \over 2} \Omega_m \left({{H_0}\over c}\right)^2 \int_\chi^{\chi_H} d \chi'
  p^a(\chi') (1+z){{f_K(\chi)f_K(\chi,\chi')} \over f_K(\chi')},\nonumber\\
\end{eqnarray}
where $p^a(\chi)$ denotes the redshift distribution of source galaxies
in the $a$-th tomographic bin and is normalized so that
$\int d\chi \, p^a(\chi)=1$.

The dependence of cosmological parameters enters the cosmic shear 
TPCFs through the nonlinear matter power spectrum, the 
distance-redshift relation, and the normalization of the 
lensing efficiency function, equation~(\ref{eq:lensingefficiency}). 
Since our cosmological analysis is
limited to the flat $\Lambda$CDM model with non-zero neutrino mass,
the relevant cosmological parameters are the density parameter of CDM
($\Omega_c$), the density parameter of baryons ($\Omega_b$), the Hubble
parameter ($h$), the scalar amplitude of the linear matter power
spectrum on $k=0.05$~Mpc$^{-1}$ ($A_S$), 
the scalar spectrum index ($n_s$),
and the sum of neutrino masses ($\sum m_\nu$).
The cosmological constant parameter is determined under the 
assumption of a flat Universe,
$\Omega_\Lambda = 1 - \Omega_c - \Omega_b -\Omega_\nu$, where
$\Omega_\nu$ is the density parameter corresponding to neutrinos.

%
%
\subsubsection{Effects of baryonic physics on the nonlinear matter power spectrum}
\label{sec:baryon_effect}

It is well known that the evolution of the 
nonlinear  matter power spectrum, especially on small scales, 
is affected by baryon physics such as 
gas cooling, star formation, 
and supernova and active galactic nuclei (AGN) feedbacks
\citep{2010MNRAS.402.1536S,2011MNRAS.415.3649V,2015MNRAS.454.1958M,2016MNRAS.461L..11H,2017MNRAS.465.2936M,2018MNRAS.475..676S,2018MNRAS.480.3962C}.
Quantitative estimates of those effects have not yet converged, due to
uncertainties in the implementation of sub-grid baryon
physics in cosmological hydrodynamical simulations
\citep{2004APh....22..211W,2004ApJ...616L..75Z,2006ApJ...640L.119J,2011MNRAS.417.2020S,2015ApJ...806..186O}. 

We mitigate these effects of baryon physics in our
cosmological analysis by not including the measurements of the 
TPCFs on small scales where the effects are significant (see
Section~\ref{sec:data_vector}).
As a further check, we test their impact using an extreme model, 
the AGN feedback model by \citet{2015MNRAS.450.1212H} that 
is based on the cosmological hydrodynamical simulations of 
\citet{2010MNRAS.402.1536S,2011MNRAS.415.3649V} 
(OverWhelming Large Simulations (OWLS)).
We note that all of other predictions of the baryonic effects based
on other state-of-the-art simulations including the EAGLE simulation
\citep{2016MNRAS.461L..11H}, the IllustrisTNG simulations
\citep{2018MNRAS.475..676S}, and the Horizon set of simulations
\citep{2018MNRAS.480.3962C} have a smaller effect on the matter power
spectrum than the OWLS AGN feedback model we adopt in this study.
However it should be noted that current baryonic simulation results
do not necessarily span all potential real feedback models. We thus
allow to vary the strength of feedback by introducing a parameter.
We follow the methodology of \citet{2017MNRAS.471.4412K}, in which a
modification of the dark matter power spectrum due to the AGN feedback
is modeled by the fitting function derived by
\citet{2015MNRAS.450.1212H}, but an additional parameter ($A_B$) that
controls the strength of the feedback is introduced 
\citep[see Section~5.1.2 of][for the explicit 
expression]{2017MNRAS.471.4412K}. 
We note that
\citet{2019PASJ...71...43H} employed the same methodology.
The case with $A_B=1$ corresponds to the original AGN feedback model
by  \citet{2015MNRAS.450.1212H}, whereas $A_B=0$ corresponds to 
the case of no effect of the baryon physics.
Our treatment of baryon feedback effects in our cosmological
analyses is described in Section~\ref{sec:astrophysical_parameters}.

%
%
\subsection{Intrinsic alignment model}
\label{sec:IAmodel}

The so-called intrinsic alignment (IA) of galaxy shapes is another major
astrophysical systematic in the measurement of the cosmic shear TPCFs 
\citep[see][for recent reviews]{2015SSRv..193..139K,2015PhR...558....1T}.
The IA comes both from the correlation between intrinsic shapes of 
two physically associated galaxies in the same
local field (referred to as the II-term) and from the cross
correlation between lensing shear of background galaxies and the intrinsic
shape of foreground galaxies (referred to as the GI-term).
We employ the standard theoretical model for these terms, namely, the
nonlinear modification of the tidal alignment model
\citep{2004PhRvD..70f3526H,2007NJPh....9..444B,2011A&A...527A..26J}.
In this formalism, TPCFs are given in a similar manner as the cosmic
shear TPCFs, equations~(\ref{eq:xi_theory}), (\ref{eq:P_kappa}), and
(\ref{eq:lensingefficiency}), but with modified power spectra
\begin{equation}
  \label{eq:xi_IA}
  \xi_{\mbox{II/GI},\pm}^{ab}(\theta) = {1\over {2\pi}} \int d\ell \ell J_{0,4} (\ell
  \theta) P_{\mbox{II/GI}}^{ab}(\ell),
\end{equation}
with
\begin{eqnarray}
  \label{eq:P_IA}
  P_{\mbox{II}}^{ab}(\ell) &=& \int_0^{\chi_H} d \chi
  F^2(\chi){{p^a(\chi)p^b(\chi)} \over f_K^2(\chi)}
  P_m^{NL}\left( {\ell \over {f_K(\chi)}}, \chi \right),\\
  P_{\mbox{GI}}^{ab}(\ell) &=& \int_0^{\chi_H} d \chi
  F(\chi){{q^a(\chi)p^b(\chi)+p^a(\chi)q^b(\chi)} \over
    f_K^2(\chi)}\nonumber\\
  &&\times P_m^{NL}\left( {\ell \over {f_K(\chi)}}, \chi \right).
\end{eqnarray}
In the above expressions, $F(\chi)$ represents the correlation strength
between the tidal field and the galaxy shapes, for which we adopt 
the same redshift dependent model as used in \citet{2019PASJ...71...43H}
\begin{equation}
  \label{eq:F_IA}
  F\left[\chi(z)\right] = -A_{\mbox{IA}} C_1 \rho_{\mbox{crit}}
  {{\Omega_m} \over {D_+(z)}}
  \left( {{1+z} \over {1+z_0}} \right)^{\eta_{\mbox{eff}}},
\end{equation}
where $A_{\mbox{IA}}$ is the amplitude parameter, $C_1$ is the fixed
normalization constant ($C_1=5\times 
10^{-14}h^{-2}M_\odot^{-1}$Mpc$^3$), $\rho_{\mbox{crit}}$ is the
critical density at $z=0$, and $D_+(z)$ is the linear growth factor
normalized to unity at $z=0$. We adopt the pivot redshift of $z_0=0.62$,
and treat $A_{\mbox{IA}}$ and the redshift dependence index $\eta_{\mbox{eff}}$ as
nuisance parameters in our cosmological analysis (see
Section~\ref{sec:astrophysical_parameters}). 

%
%
\subsection{Corrections for the redshift-dependent selection bias, 
  PSF related errors, and the constant shear}
\label{sec:model_correction}

The theoretical model for the observed cosmic shear TPCFs is the sum of
three components
\begin{equation}
  \label{eq:xi_sum}
  \xi_{\pm}^{ab}(\theta) = \xi_{\mbox{GG},\pm}^{ab}(\theta) +
  \xi_{\mbox{GI},\pm}^{ab}(\theta) + \xi_{\mbox{II},\pm}^{ab}(\theta).
\end{equation}
In reality, the measured TPCFs are affected by the redshift-dependent
shear calibration bias (Section~\ref{sec:selection-bias}) and 
the residual PSF and PSF modeling error (Section~\ref{sec:PSF-TPCS},
equation~\ref{eq:xi_psf}).
In addition, $\xi_+$ components may be biased by the constant shear over
a field arising from systematics (Appendix~\ref{sec:mean_shear}).
We note that the constant shear arising from the gravitational
lensing effect on scales larger than a survey field is taken 
into account properly in our analysis, as our model for the
covariance matrix includes the super-survey mode.
We apply these corrections to $\xi_+$ as
\begin{eqnarray}
  \label{eq:model_corrections+}
  \xi_{+}^{ab}(\theta) &\rightarrow&
  (1+m_{\mbox{sel}}^a+m_R^a)(1+m_{\mbox{sel}}^b+m_R^b)
  \xi_{+}^{ab}(\theta)\nonumber\\
  &&+\alpha_{\mbox{psf}}^2 \xi_+^{pp}(\theta)
  +2 \alpha_{\mbox{psf}}\beta_{\mbox{psf}} \xi_+^{pq}(\theta)
  +\beta_{\mbox{psf}}^2 \xi_+^{qq}(\theta)\nonumber\\
  &&+\bar{\gamma}^2,
\end{eqnarray}
where $\bar{\gamma}$ is the redshift-independent constant shear term
that we treat as a nuisance parameter (see
Section~\ref{sec:Systematic_parameters}).
Since the PSF-related corrections to $\xi_-$ are found to be
very small (see Appendix~\ref{appendix:residual_PSF}), we do not 
apply these corrections to $\xi_-$. As a result, the corrected 
expression for $\xi_-$ is
\begin{eqnarray}
  \label{eq:model_corrections-}
  \xi_{-}^{ab}(\theta) &\rightarrow&
  (1+m_{\mbox{sel}}^a+m_R^a)(1+m_{\mbox{sel}}^b+m_R^b)
  \xi_{-}^{ab}(\theta).
\end{eqnarray}
The values of $(m_{\mbox{sel}}^a+m_R^a)$ are taken from Table~3 of
\citet{2019PASJ...71...43H}; 
from the lowest to highest redshift bins, they are 0.0086, 0.0099,
0.0241, and 0.0391\footnote{While deriving the covariance, we also
account for $m_R$ in the mocks, although not the selection bias. 
This can cause an at most 2 percent difference in the 
covariance matrix.}.
In our cosmological analysis, we treat $\alpha_{\mbox{psf}}$ and
$\beta_{\mbox{psf}}$ as nuisance parameters (see
Section~\ref{sec:Systematic_parameters} for our choice of prior
ranges).

%
%
\subsection{Covariance}
\label{sec:covariance}

We derive a covariance matrix of the TPCF measurement using 2268
realizations of mock HSC shape 
catalogs. See Appendix~\ref{appendix:mock} for a brief description of
the mock catalogs, which are described in detail in
\citet{2019MNRAS.486...52S}.
We measure the cosmic shear TPCFs for all 2268 mock catalogs in
exactly the same manner as the real cosmic shear measurement.
Since the HSC mock catalogs are constructed based on full-sky lensing
simulation data with galaxy positions, intrinsic shape noise,
and measurement noise taken from the real HSC shape catalog, the mock data
naturally have the same survey geometry and the same noise properties as
the real catalog, and include super-survey cosmic shear signal
from these full-sky lensing simulations. 
In addition, the effects of nonlinear structure formation
on the lensing shear field are included in the mock data. 
Therefore the covariance matrix computed from the mock catalogs
automatically includes all the contributions, namely, Gaussian,
non-Gaussian, super-survey covariance and the
survey geometry are naturally taken into account.
\citet{2019MNRAS.486...52S} found that in the case of the HSC 1st-year 
data we adopt in this study, the shape noise covariance dominates the
covariance at the smallest angular bin, while the cosmological Gaussian
covariance is prominent at the largest angular bin.

The accuracy of the covariance matrix from the mocks was studied in 
detail by \citet{2019MNRAS.486...52S}. 
They investigated the impact of photo-$z$ errors and 
field-to-field variation among the six separate HSC fields on the 
covariance estimation.  
They found that the change in the variance due to the different 
photo-$z$ methods can yield a $5-10\%$ difference in 
signal-to-noise ratio of the cosmic shear TPCFs, whereas 
the field variation can change the covariance
estimation by $3-5\%$.
\citet{2019MNRAS.486...52S} also addressed the effect of the
multiplicative bias on the covariance estimation. 
They found that multiplicative bias of 10\% can change shape 
noise covariance at the $\sim20\%$ level.
We already included the effect by assuming the fiducial value of
multiplicative bias. 
A 1\% level uncertainty in the multiplicative bias was confirmed in
\citet{2018MNRAS.481.3170M}, leading to less than 2\% uncertainty in our
estimation of shape noise covariance. 

Overall, we expect the covariance matrix estimated from mocks to be
calibrated with $<10\%$ accuracy against various systematic effects in
the cosmic shear analysis, 
if the cosmological model in the mock catalogs is correct. 

One drawback of this approach is that we are not able to include 
the cosmology dependence of the covariance, because the HSC mock 
catalogs are
based on a set of full-sky gravitational lensing ray-tracing
simulations that adopt a specific flat $\Lambda$CDM cosmology (see
Appendix~\ref{appendix:mock}). 
This is in contrast to
\citet{2019PASJ...71...43H} who used a halo-model-based analytic
model of covariance matrix (which was tested against the HSC mock 
catalogs) in their cosmological analysis.
In the case of the TPCF, the halo-model-based analytic covariance
matrix was formulated
\citep{2001ApJ...554...56C,2009MNRAS.395.2065T,2013PhRvD..87l3504T}.
However, it is found in \citet{2019MNRAS.486...52S} that in order to
derive an accurate covariance, the survey geometry must be properly
taken into account which requires $N_g^2$ operations ($N_g$ is a
total number of galaxies) and is computationally very expensive.
\citet{2019PASJ...71...43H} also studied the effect of the 
cosmology dependence of the covariance on the cosmological analysis 
in their cosmic shear power spectrum study, by comparing 
cosmological constraints derived using the cosmology-dependent
covariance (which is their fiducial model) with those derived 
using a cosmology-independent one (fixed to the best-fit
cosmological model). They found that the best-fit $\Omega_m$ 
and $S_8(=\sigma_8 (\Omega_m/0.3)^\alpha$ with 
$\alpha = 0.45$ or $0.5$) values agree with each other 
within 20\% of the statistical uncertainty.
It is therefore reasonable to assume that the cosmology dependence 
of the covariance matrix does not significantly impact our 
cosmological analysis.
We refer the readers to
\citet{2009A&A...502..721E,2019A&A...631A.160H,2019OJAp....2E...3K}
for dependence of the covariance on cosmology and its impact on 
cosmological parameter constraints.

%
%
\section{Cosmological analyses}
\label{sec:analyses}

We employ the standard Bayesian likelihood analysis for the cosmological
inference of measured cosmic shear TPCFs. The log-likelihood is given
by
\begin{equation}
  \label{eq:log-like}
  -2 \ln {\cal L}(\bm{p}) = \sum_{i,j} \left(d_i-m_i(\bm{p})\right) {\mbox{Cov}}_{ij}^{-1}
  \left(d_j-m_j(\bm{p})\right), 
\end{equation}
where $d_i$ is the data vector that is detailed in
Section~\ref{sec:data_vector}, $m_i(\bm{p})$ is the theoretical model with
$\bm{p}$ is a set of parameters detailed in
Section~\ref{sec:model_parameters}, and ${\mbox{Cov}}_{ij}$ is the
covariance matrix that is described in Section~\ref{sec:covariance}.
Since our covariance matrix is constructed from 2268 mock 
realizations, its inverse covariance is known to be biased high
\citep[see][and references
  therein]{Anderson2003, 2007A&A...464..399H}.
When calculating the inverse covariance, we therefore include the so-called
Anderson-Hartlap correction factor $\alpha =
(N_{\mbox{mock}}-N_d-2)/(N_{\mbox{mock}}-1)$, where 
$N_{\mbox{mock}}=2268$ is the number of mock realizations
(see Appendix \ref{appendix:mock}) and $N_d=170$ 
(for our fiducial choice, see Section~\ref{sec:data_vector})
is the length of our data vector. 

In order to sample the likelihood efficiently, we employ the multimodal
nested sampling algorithm 
\citep{2008MNRAS.384..449F,2009MNRAS.398.1601F,2013arXiv1306.2144F},
as implemented in the public software {\tt MultiNest} (version 3.11).

%
%
\subsection{Data vector}
\label{sec:data_vector}

The data vector, $d_i$, is constructed from ten tomographic 
combinations of cosmic shear TPCFs $\hat{\xi}_+^{ab}$ and
$\hat{\xi}_-^{ab}$ presented in Figure~\ref{fig:xi_pm_best}.
Although TPCFs are detected with a good signal-to-noise ratio 
over a wide angular range as shown in the Figure, we limit
angular ranges for our cosmological analysis for the 
following reasons. 

First, we remove the angular range where the uncertainty in 
the theoretical model of cosmic shear TPCFs due to baryon physics 
is not negligible. We employ the AGN feedback model considered in
\citet{2015MNRAS.450.1212H} as an extreme case, and deduce from
Figure~5 of their paper that scales where the AGN feedback effect
becomes less than 5\% for $\xi_+$ and $\xi_-$ are $\theta>4'$ and
$\theta>20'$, respectively. Since their results were obtained 
assuming the galaxy redshift distribution with the mean redshift
of $\langle z \rangle \sim 0.75$,
the feedback effect may have a larger impact on larger scale 
signals for the lower source redshift sample.
Considering the lower mean redshift of our lowest-$z$ 
tomographic sample, we conservatively adopt about 50\% larger scales 
than the scales mentioned above as our threshold scales. 

Second, we remove the angular range where the extra shape 
correlations due to PSF leakage and PSF model error are not 
negligible.
The effects of these errors on cosmic shear TPCFs are examined in
Appendix~\ref{appendix:residual_PSF}. It is found that their total
contribution to $\xi_+$ is about $10^{-6}$ on scales $5'<\theta<60'$.
Since this estimate is based on a simple model for PSF errors and 
the associated errors are large, $(0.4-1)\times10^{-6}$,
the above value should be considered as a rough estimate.
Comparing this estimate with the measured signals and 
errors, we set an upper limit of $\theta \lesssim 60'$ for $\xi_+$.
Since the contribution of this systematic to $\xi_-$ is found to be very
small,  
about $10^{-8}$ even at around 1 degree scale, no upper limit is
set to $\xi_-$ from this condition.

Third, we remove the angular range where the signal-to-noise ratio
including the cosmic variance for individual angular bins becomes
$\lesssim 1$. This condition sets the upper limit 
$\theta\lesssim 200'$ for $\xi_-$. 

Taking these three points into consideration, 
we adopt angular bins $\theta_i =
10^{0.1\times i}$ arcmin with $9\leq i \leq 17$ for $\xi_+$, and $15\leq
i \leq 22$ for $\xi_-$. The corresponding angular ranges of galaxy-pair
separation are $7\farcm08<\theta<56\farcm2$ 
and $28\farcm2<\theta<178'$ for $\xi_+$ and $\xi_-$, 
respectively. The total length of the data vector for our 
fiducial choice is $N_d=(9+8)\times 10=170$.

%
%
\subsubsection{Signal-to-noise ratio}
\label{sec:SNR}

Using the fiducial data vector described in Section~\ref{sec:data_vector} 
and the covariance matrix described in
Section~\ref{sec:covariance}, the total signal-to-noise 
ratio is found to be 18.7.
The value of the signal-to-noise ratio depends on the assumed cosmological 
model through the covariance matrix.
Our covariance matrix is based on the mock catalogs assuming {\it WMAP9}
cosmology.
\citet{2019PASJ...71...43H} evaluated the total signal-to-noise of HSC
tomographic cosmic shear power spectra using
a covariance matrix based on the {\it Planck} cosmology, and found
$SN=15.6$ for their fiducial multipole range $300< \ell <1900$.
The difference between these signal-to-noise ratio values is 
mostly accounted for by the different angular ranges 
adopted in these two studies 
(see also Appendix~\ref{appendix:powerspectrum}), 
and by the different cosmological models assumed 
for the covariance matrices.

%
%
\subsubsection{Effective angular scale of angular bins and bin-averaged TPCFs}
\label{sec:effective_angular_scale}

We determine the pair-weighted effective mean center of each angular bin
as follows.
In our TPCF measurements, we adopt a regular log-interval binning
with the bin width of $\Delta \log \theta =0.1$. For $i$-th bin, the
minimum and maximum angular scales are given by $\theta_{min}=10^{0.1(i-0.5)}$ and
$\theta_{max}=10^{0.1(i+0.5)}$, respectively, with the simple bin center of
$\theta_c=10^{0.1i}$.
Assuming the number of galaxy pairs scales with separation as
$n_p(<\theta)\propto \theta^2$
(here we ignore the irregular survey geometry), 
the pair-number weighted mean separation for each bin is given by 
\begin{equation}
  \label{eq:effective_scale}
  \bar{\theta}=
      {
        {\int_{\theta_{min}}^{\theta_{max}} d \theta ~ \theta n_p(\theta)}
        \over
        {\int_{\theta_{min}}^{\theta_{max}} d \theta ~
          n_p(\theta)}
        }.
\end{equation}
For our bin width of $\Delta \log \theta =0.1$, we find
$\bar{\theta}=1.011\times \theta_c$.

The same bin-averaged effect should be taken into account in the
computation of the theoretical model of the cosmic shear TPCFs.
The exact integration over the bin-width
would be computationally expensive. 
Instead, we adopt an approximate estimate based on the following consideration 
(for other approximate estimates, see \citealt{2019A&A...624A.134A} and
references therein).  
Assuming a power-law form for the cosmic shear TPCF within a bin-width, that
is, $\xi(\theta)\propto \theta^\mu$ ($-1\lesssim \mu \lesssim -0.5$ for
the cosmic shear TPCFs on scales of our interest) and ignoring the
irregular survey  geometry, the bin-averaged TPCF is given by 
\begin{equation}
  \label{eq:bin_averaged_TPCF}
  \bar{\xi}=
      {
        {\int_{\theta_{min}}^{\theta_{max}} d \theta ~ \xi(\theta) n_p(\theta)}
        \over
        {\int_{\theta_{min}}^{\theta_{max}} d \theta ~
          n_p(\theta)}
        }.
\end{equation}
In the case of $\mu =-1(-0.5)$, we find
$\bar{\xi} = 0.993(0.996)\times \xi(\theta_c)$,
which is very close to
the value evaluated at
an angular scale of $\hat{\theta}=1.007\times \theta_c$.
Specifically, 
$\xi(\hat{\theta})= \xi(\theta_c) \times (\hat{\theta}/\theta_c)^\mu =
0.993(0.997)\times \xi(\theta_c)$, for $\mu=-1(-0.5)$.
On these grounds, we decide to adopt the  TPCFs at
$\hat{\theta}$ as our 
estimate of the bin-averaged cosmic shear TPCFs\footnote{In the original
version of this paper, $\bar{\theta}$ was incorrectly evaluated and was
adopted for the angular scale to estimate the bin-averaged cosmic shear
TPCFs, though its effect on results are very small.
In this revised version, all the theoretical predictions are evaluated
at $\hat{\theta}$}.

%
%
\subsection{Model parameters and prior ranges}
\label{sec:model_parameters}

In this subsection, we summarize model parameters and their prior
ranges used in our cosmological analysis.
Prior ranges and choice of parameter set for systematic tests are
summarized in Table~\ref{table:parameters}.

%
%
\begin{table*}
\tbl{Summary of cosmological, astrophysical, and systematics parameters
  used in our cosmological analysis.
  ``flat[$x_1$, $x_2$]'' means a flat prior between $x_1$ and
  $x_2$, whereas ``Gauss($\bar{x}$, $\sigma$)'' means a Gaussian
  prior with the mean $\bar{x}$ and the standard deviation
  $\sigma$. For detail descriptions of parameters, 
  see section \ref{sec:cosmological_parameters} for the cosmological
  parameters, section \ref{sec:astrophysical_parameters} for the
  astrophysical nuisance parameters, and section
  \ref{sec:Systematic_parameters} for the systematics nuisance
  parameters. \label{table:parameters}}
{
\begin{tabular}{lllll}
\hline
  Parameter & \multicolumn{3}{l}{~~~~~~~~~~~~~~~~~~~~~~Prior range} & Section \\
  {}        & Fiducial $\Lambda$CDM & $w$CDM & Systematics tests & {} \\
\hline
  Cosmological    & {} & {} & {} &  \ref{sec:cosmological_parameters} \\
  $\Omega_c$      & flat[0.01, 0.9] & {} & {} & {} \\
  $\log(A_s \times 10^9)$ & flat[-1.5, 2.0] & {} & {} & {} \\
  $\Omega_b $   & flat[0.038, 0.053] & {} & {} & {} \\
  $n_s$   & flat[0.87, 1.07] & {} & {} & {} \\
  $h$   & flat[0.64, 0.82] & {} & {} & {} \\
  $\sum m_\nu$ [eV]  & fixed to 0.06  & & {} flat[0, 0.5] for ``$\sum m_\nu$ varied`` & {} \\
  $w$  & fixed to -1  & flat[$-2$, $-1/3$] & {} & {}  \\
  \hline
  Astrophysical    & {} & {} & {} & \ref{sec:astrophysical_parameters} \\
  $A_{\mbox{IA}}$  & flat[-5, 5] & {} & fixed to 0 for ``w/o IA'' & {} \\
  $\eta_{\mbox{IA}}$  & flat[-5, 5] & {} & fixed to 3 for ``IA $\eta_{\mbox{IA}}=3$'' & {} \\
  $A_B$  & fixed to 0 & {} & fixed to 1 for ``$A_B=1$'' or flat[-5, 5] for ``$A_B$ varied'' & {} \\
  \hline
  Systematics    & {} & {} & {} & \ref{sec:Systematic_parameters} \\
  $\alpha_{\mbox{psf}}$  & Gauss(0.029, 0.010) & {} & fixed to 0 for ``w/o PSF error'' & {} \\
  $\beta_{\mbox{psf}}$  & Gauss(-1.42, 1.11) & {} & fixed to 0 for ``w/o PSF error'' & {} \\  
  $\Delta m$           & Gauss(0, 0.01) & {} & fixed to 0 for ``w/o $\Delta m$'' & {} \\
  $\Delta z_1$           & Gauss(0, 0.0374) & {} & fixed to 0 for ``w/o $p(z)$ error'' & {} \\
  $\Delta z_2$           & Gauss(0, 0.0124) & {} & fixed to 0 for ``w/o $p(z)$ error'' & {} \\
  $\Delta z_3$           & Gauss(0, 0.0326) & {} & fixed to 0 for ``w/o $p(z)$ error'' & {} \\
  $\Delta z_4$           & Gauss(0, 0.0343) & {} & fixed to 0 for ``w/o $p(z)$ error'' & {} \\
  $\bar{\gamma}$  & fixed to 0 & {} & flat[0, $5\times 10^{-3}$] for ``w/ const-$\gamma$'' & {} \\
  \hline
\end{tabular}
}
\begin{tabnote}
{}
\end{tabnote}
\end{table*}

%
%
\subsubsection{Cosmological parameters}
\label{sec:cosmological_parameters}

We focus on the flat $\Lambda$CDM cosmological model characterized
by six parameters ($\Omega_c$, $A_s$, $\Omega_b$, $n_s$, $h$, and
$\sum m_\nu$, see Section~\ref{sec:model_cosmicshear}).
Among thees parameters, the cosmic shear TPCFs are most sensitive to
$\Omega_c$ and $A_s$, or the derived parameter $\sigma_8$.
Thus we adopt prior ranges that are sufficiently wide 
for these parameters (see Table~\ref{table:parameters}).
For ($\Omega_b$, $n_s$, and $h$), which are only 
weakly constrained with cosmic shear TPCFs, we set prior 
ranges which largely bracket allowed values from external 
experiments (see Table~\ref{table:parameters}). 
For the sum of neutrino mass, we take $\sum m_\nu =0.06$~eV from the
lower bound indicated by
the neutrino oscillation experiments \citep[e.g.,][for a
  review]{2013neco.book.....L} for our fiducial choice.
As a systematics test, we check the impact of neutrino mass on our
conclusions by varying $\sum m_\nu$.

In addition to the fiducial $\Lambda$CDM model, we consider an extended
model by including the time-independent equation-of-state parameter for
the dark energy ($w$), referred to as the $w$CDM model.
We take a flat prior with $-2<w<-1/3$, which excludes the
non-accelerating expansion of the present day Universe, 
and brackets allowed values from external experiments.

%
%
\subsubsection{Astrophysical nuisance parameters}
\label{sec:astrophysical_parameters}

Our fiducial model for the TPCFs includes the contribution of the intrinsic
alignment of galaxy shapes as described in Section~\ref{sec:IAmodel}.
The nonlinear alignment model we employed has two 
parameters, the amplitude
parameter $A_{\mbox{IA}}$ and the redshift dependence parameter
$\eta_{\mbox{IA}}$ which represents the effective redshift evolution of 
the IA amplitude beyond the redshift evolution of the matter distribution 
due to a possible intrinsic redshift evolution and/or the change of the 
galaxy population as a function of redshift.
Following recent cosmic shear studies
e.g., \citet{2017MNRAS.465.1454H}, \citet{2018PhRvD..98d3528T}, and
\citet{2019PASJ...71...43H}, we adopt very wide prior ranges for these
parameters.

The effect of baryon physics on the nonlinear matter power spectrum
(see Section~\ref{sec:baryon_effect}) is another possible astrophysical
systematic effect on the cosmological analysis.
Nevertheless, since we restrict the angular ranges of cosmic shear 
TPCFs conservatively so that the baryon effects do not have a 
significant impact  on our analysis (see Section~\ref{sec:data_vector}), 
we do not include the baryon effect in our fiducial model, 
but check its impact in our systematics tests, employing the 
AGN feedback model by \citet{2015MNRAS.450.1212H} by adding a parameter
$A_B$ which controls the amplitude of the baryon effect. We carry out
two tests; one fixing $A_B=1$ that corresponds to 
the original AGN feedback model, and the other in which $A_B$ is a free
parameter.

%
%
\subsubsection{Systematics nuisance parameters}
\label{sec:Systematic_parameters}

To summarize, in our fiducial model we account for systematic 
effects from PSF leakage and PSF modeling errors, the uncertainty 
in the shear multiplicative bias correction, and 
uncertainties in the source galaxy redshift distributions.
In our cosmological analysis, we include these effects by modeling 
them with nuisance parameters which are marginalized over in the 
final cosmological inference.
In addition, in systematics tests we check the impact of the uncertainty of 
the constant shear over fields.
Below we summarize our choices for prior ranges on nuisance 
parameters in these models.

Our models for the PSF leakage and PSF modeling errors are described in
Section~\ref{sec:PSF-TPCS}. We apply the correction for these systematics by
equation~(\ref{eq:model_corrections+}).
The model parameters are estimated in Appendix~\ref{appendix:residual_PSF}, 
in which we find
$\alpha_{\mbox{psf}}=0.029\pm 0.010$ and $\beta_{\mbox{psf}}=-1.42\pm
1.11$.
We adopt Gaussian priors for these parameters and include them in our fiducial
model.

Regarding the uncertainty in the shear multiplicative bias correction,
we follow \citet{2019PASJ...71...43H} to
introduce the nuisance parameter $\Delta m$, which represents the
residual multiplicative bias, and modifies the {\it theoretical
  prediction} for the cosmic shear TPCFs to
\begin{equation}
  \label{eq:delta_m}
  \xi_{\pm}^{ab}(\theta) \rightarrow (1+\Delta m)^2 \xi_{\pm}^{ab}(\theta).
\end{equation}
The prior range of $\Delta m$ is taken to be Gaussian with zero mean
and the standard deviation of 0.01. This is based on the calibration 
of the HSC first-year shear catalog done with image simulations 
\citep{2018MNRAS.481.3170M}, in which it is confirmed that the
multiplicative bias is controlled at the 1\% level, leaving a 1\%
uncertainly on the residual bias.

Regarding uncertainties in the redshift distributions of source
galaxies, we again follow the methodology of \citet{2019PASJ...71...43H}
\citep[see also][]{2018PhRvD..98d3528T}, in which uncertainties 
for each tomographic bin are assumed to be represented by 
a single parameter $\Delta z_a$. The source redshift distribution,
which is derived by the COSMOS re-weighted method 
(see Section~\ref{sec:pz}), is then shifted by
\begin{equation}
  \label{eq:pz_shift}
  p^a(z) \rightarrow  p^a(z+\Delta z_a).
\end{equation}
The prior ranges for the shift parameters are estimated by comparing the
COSMOS re-weighted $p^a(z)$ with ones derived from stacked-PDFs
following the method described in Section~5.8 of \citet{2019PASJ...71...43H}.
The derived prior ranges, which are summarized in Table
\ref{table:parameters}, are in reasonable agreement with those 
found in \citet{2019PASJ...71...43H} with the largest difference of 24\%. 

Finally, as discussed in Appendix~\ref{sec:mean_shear}, 
we introduce the single parameter $\bar{\gamma}$, which
represents the redshift-independent constant shear arising from
systematics, when checking the impact of the uncertainty 
in the constant shear over fields. The constant shear is 
added to the theoretical model of $\xi_+$ as shown in
equation~(\ref{eq:model_corrections+}). 
Given that we have not found a strong evidence of the existence of 
the residual constant shear (see Appendix~\ref{sec:mean_shear}), we do
not include it in our fiducial model, but check its impact as a
systematics test, in which we treat $\hat{\gamma}$ as a 
nuisance parameter with a flat prior for a wide range 
$0<\hat{\gamma}<5\times 10^{-3}$.
We constrain $\hat{\gamma}$ to be positive, because 
only the square of $\hat{\gamma}$ enters $\xi_+$.

%
%
\subsubsection{Effective number of {\it free} parameters}
\label{sec:effective_parameters}

It should be noted that not all the model parameters should be considered to
be free as more than half of them are tightly
constrained by priors. In other words, posteriors of those parameters
are not driven by data but are dominated by priors, and fixing those
parameters does not significantly change the cosmological constraints.
In fact, as will be found in the following sections,
although the total number of model parameters is
14 for our fiducial case (5 cosmological, 2 astrophysical, and 7
systematics parameters, see Table~\ref{table:parameters}), only 
three of them
($\Omega_c$, $A_s$, and $A_{\mbox{IA}}$) are constrained by the data 
with much
narrower posterior distributions than with priors.
Therefore, the standard definition of degree-of-freedom (d.o.f.)
$N_d-N_p$($=170-14$ for our fiducial case) is likely to be an underestimation.
A conservative choice of the effective number of free parameters
($N_p^{\rm eff}$) should account for only these three
parameters\footnote{See \citet{2019PhRvD..99d3506R} and 
Section~6.1 of \citet{2019PASJ...71...43H} 
for a more mathematically robust way to define the effective number of 
free parameters.}. 

%
%
\section{Results}
\label{sec:results}

In this section we first present cosmological constraints from our
cosmic shear analysis. We then discuss the robustness of the results
against various systematics, and finally we perform internal 
consistency checks 
among different choices of angular ranges and of tomographic redshift bins. 

%
%
\begin{figure}[t]
\begin{center}
  \includegraphics[width=82mm]{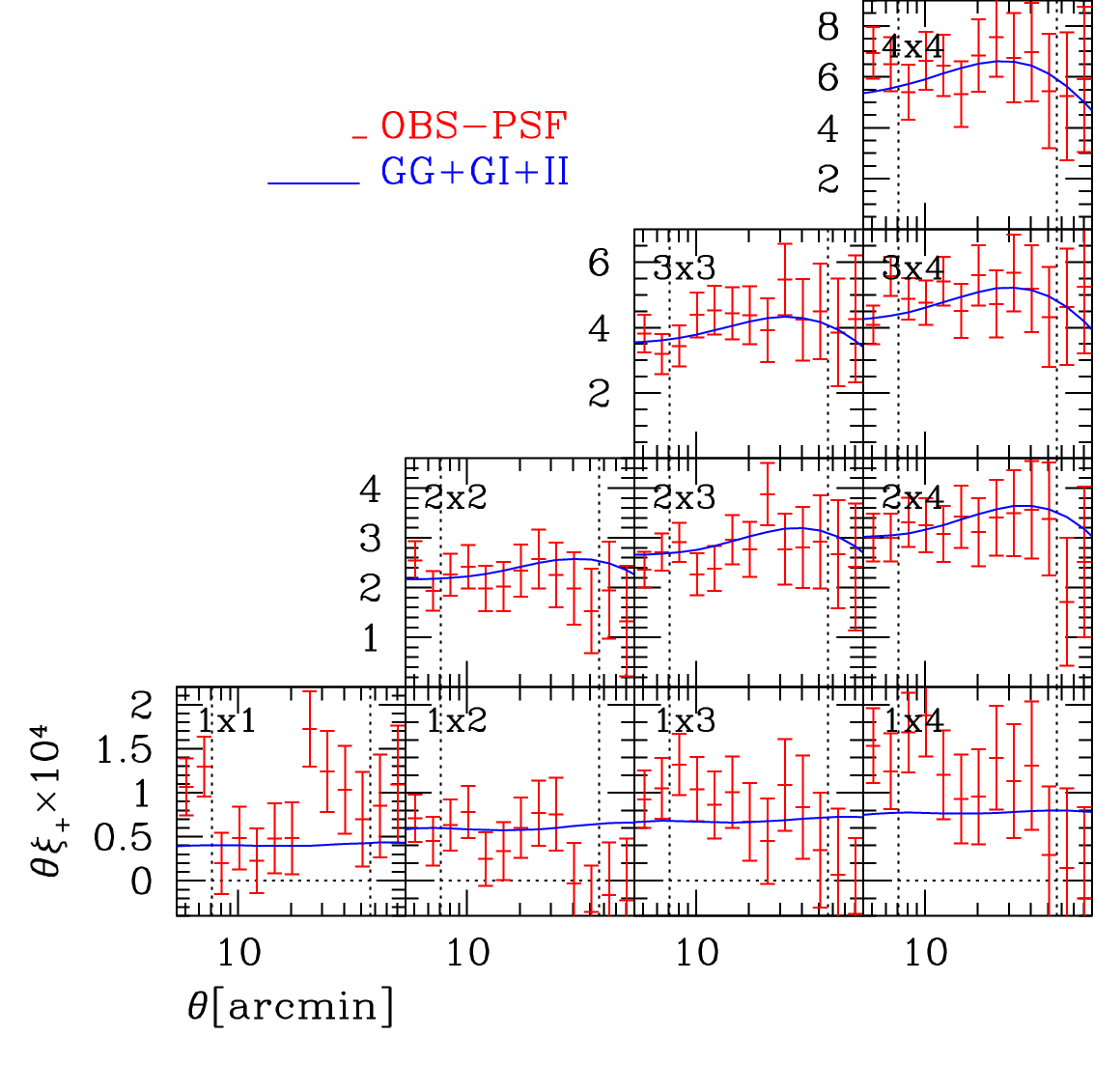}
  \includegraphics[width=82mm]{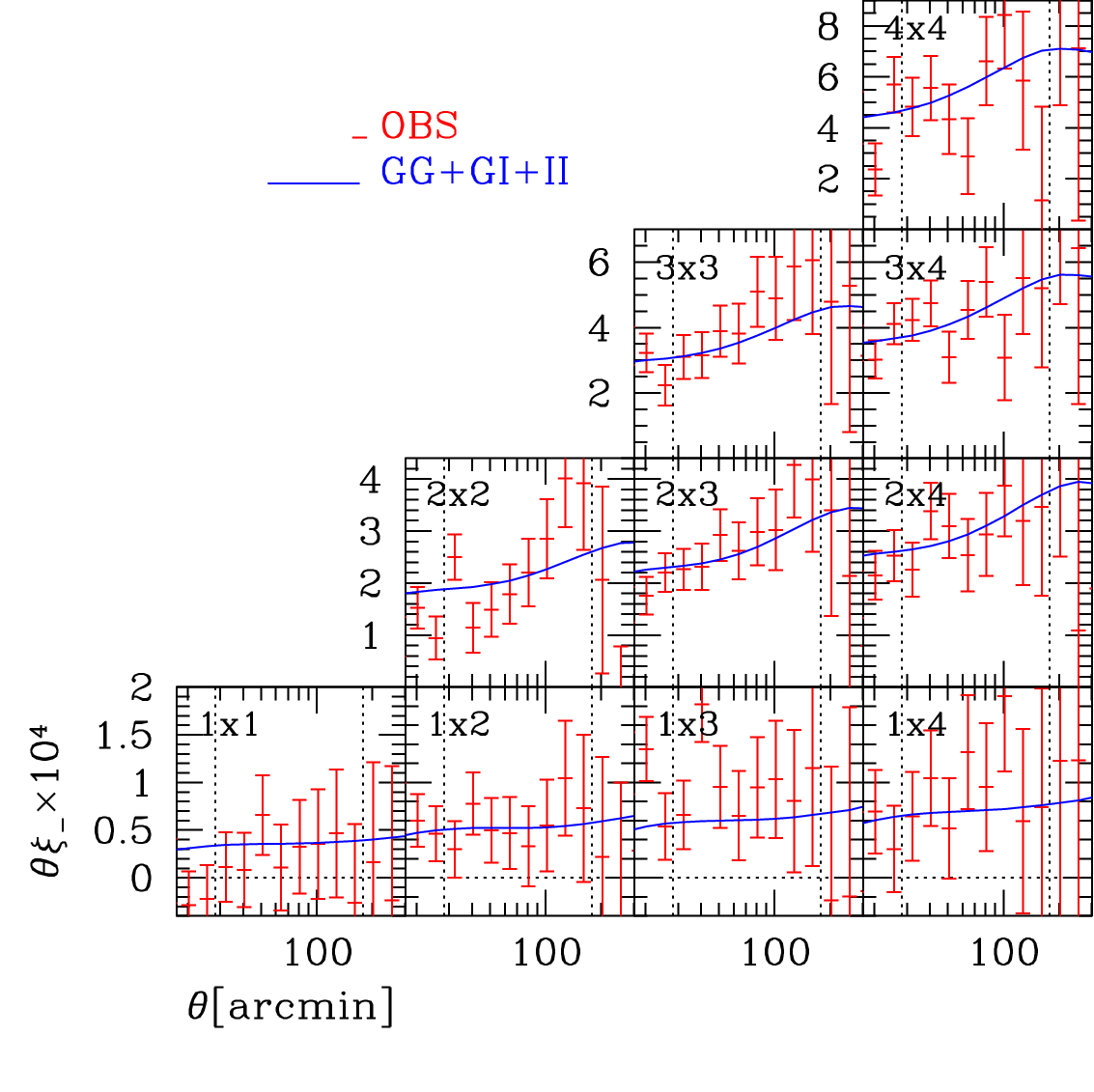}
\end{center}
\caption{Comparison of the HSC tomographic cosmic shear TPCFs with
  the best-fitting theoretical model for the fiducial
  flat $\Lambda$CDM model. Upper and lower triangular-tiled panels show
  $\xi_+$ and $\xi_-$, respectively. The measured $\xi_+$ are corrected
  for the PSF leakage and PSF modeling errors.  Error bars represent the
  square-root of the diagonal elements of the covariance matrix. The solid line
  corresponds to the best-fit (maximum likelihood) fiducial model
  including the residual multiplicative bias correction shown in 
  equation~(\ref{eq:delta_m}).
  Vertical dotted lines show the angular ranges used for the likelihood
  analysis.
  \label{fig:xi_pm_best}}
\end{figure}

%
%
\subsection{Cosmological constraints in the fiducial flat $\Lambda$CDM model}
\label{sec:fiducial_cosmology}

First we compare the HSC tomographic cosmic shear TPCFs 
with the theoretical model with best-fit parameter values for the fiducial
flat $\Lambda$CDM model in Figure~\ref{fig:xi_pm_best}, in which the measured
$\xi_+$ are corrected for the PSF leakage and PSF modeling errors with
equation~(\ref{eq:xi_psf}). 
In these plots, error bars represent the square-root of the diagonal 
elements of the covariance matrix. 
We find that our model with the fiducial parameter setup reproduces the
observed tomographic cosmic shear TPCFs quite well.
The $\chi^2$ value for the best-fit parameter set is
$\chi^2=162.0$ for the {\it effective} d.o.f. of $170-3=167$,
resulting in a $p$-value of 0.595. 

%
\begin{figure}
\begin{center}
  \includegraphics[height=82mm,angle=-90]{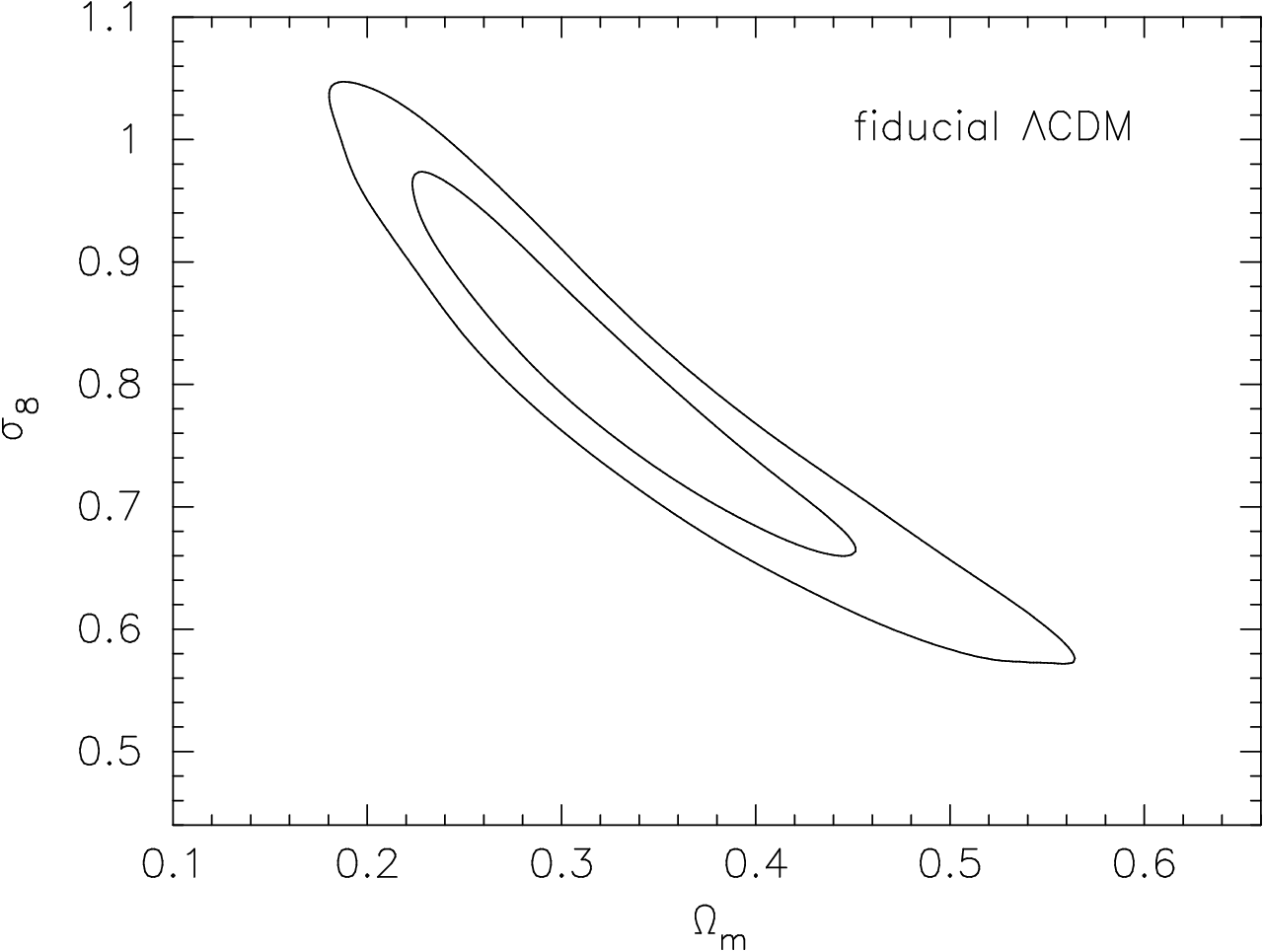}
  \includegraphics[height=82mm,angle=-90]{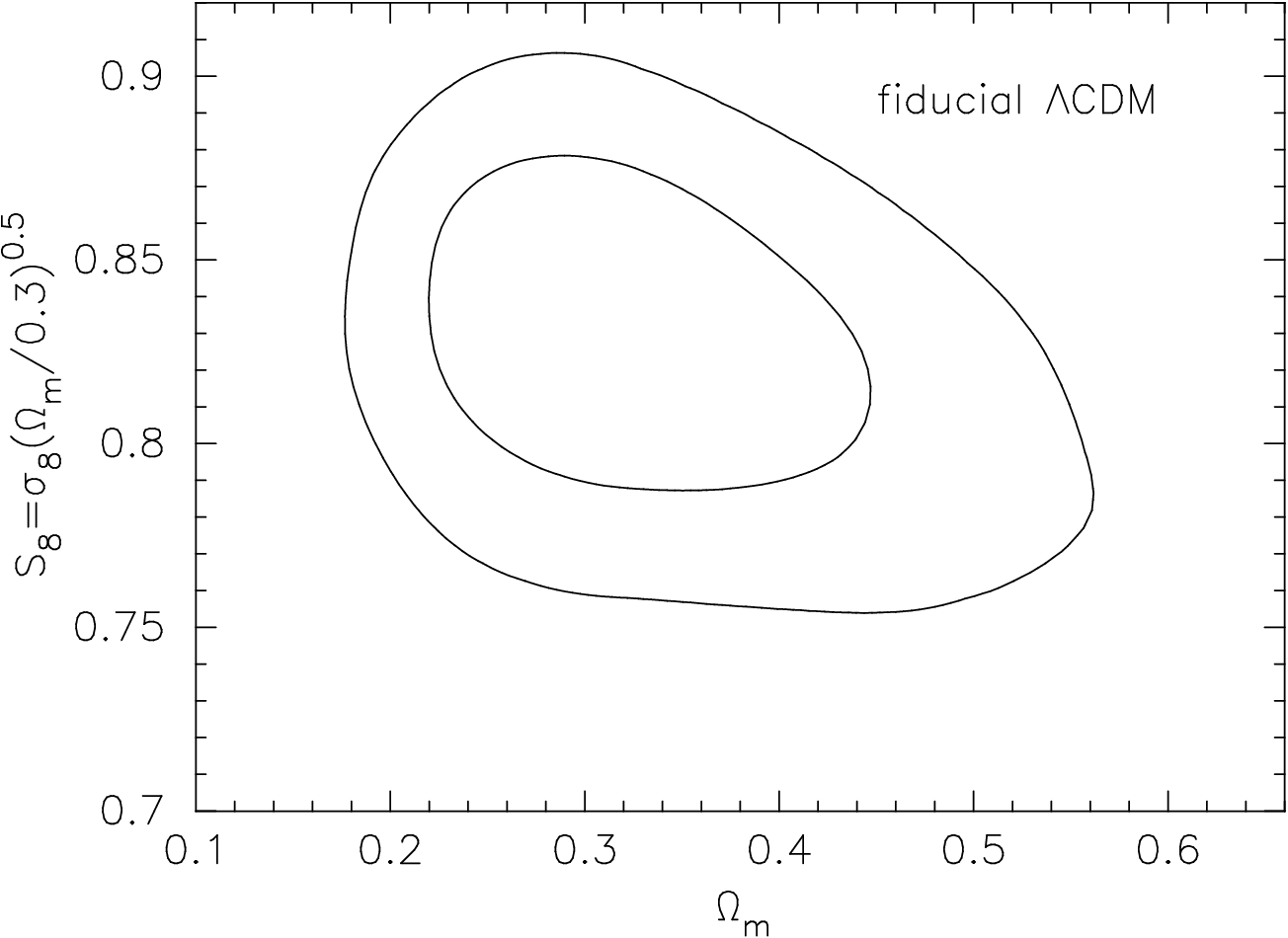}
\end{center}
\caption{Marginalized posterior contours (68\% and 95\%
    confidence levels) in the
    $\Omega_m$-$\sigma_8$ plane (top panel) and in the $\Omega_m$-$S_8$
    plane (bottom panel), where $S_8=\sigma_8\sqrt{\Omega_m/0.3}$ in the fiducial
    flat $\Lambda$CDM model.
  \label{fig:om_sig8_S8}}
\end{figure}

%
%
\begin{figure*}
\begin{center}
  \includegraphics[height=120mm,angle=-90]{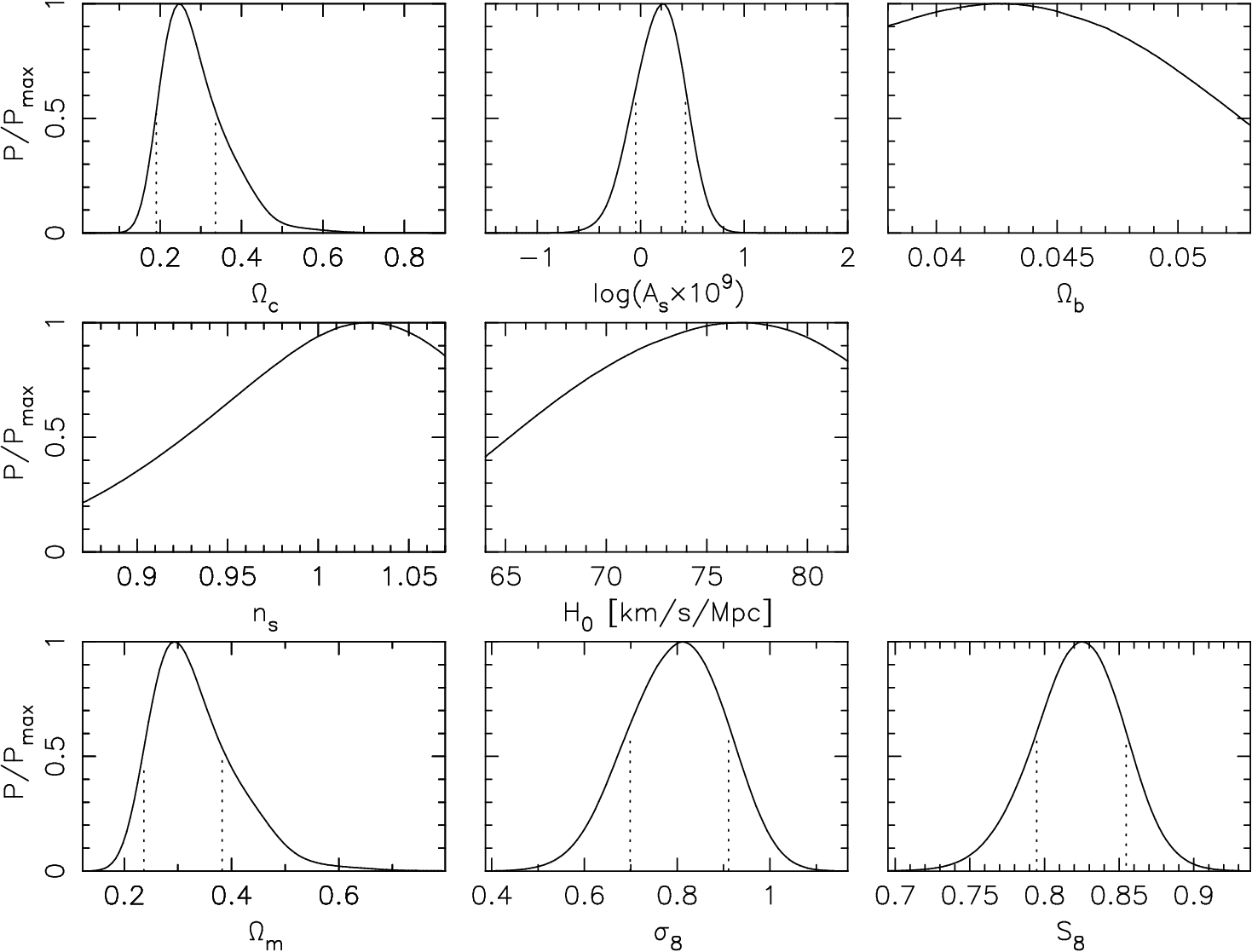}
\end{center}
\caption{Marginalized one-dimensional posterior distributions of different 
  cosmological parameters in the fiducial flat $\Lambda$CDM model. 
  The upper five 
  panels show the posterior distributions for the model parameters, whereas 
  the bottom three panels are for derived parameters. 
  For the five top panels, the plotted range of the horizontal-axis 
  indicates its flat prior range.
  Dotted vertical lines represent the approximate 68\% confidence 
  intervals, which are not shown 
  for poorly constrained parameters.
  \label{fig:posterior_cosmo}}
\end{figure*}

We marginalize over a total of 14 model parameters (5 cosmological, 2
astrophysical, and 7 systematics parameters, see Table~\ref{table:parameters}) 
in our fiducial flat $\Lambda$CDM model to derive marginalized posterior 
contours in the $\Omega_m$-$\sigma_8$ and $\Omega_m$-$S_8$ planes, which 
are presented in Figure~\ref{fig:om_sig8_S8}.
We also show marginalized one-dimensional posterior distributions 
of cosmological parameters in Figure~\ref{fig:posterior_cosmo}.
We find marginalized 68\% confidence intervals of
$0.237<\Omega_m<0.383$, 
$0.699<\sigma_8<0.911$, and
$0.795<S_8<0.855$. 
From the posterior distributions shown in
Figure~\ref{fig:posterior_cosmo}, it can be seen that the current 
HSC cosmic shear TPCFs alone cannot place useful constraints on the 
Hubble constant
($H_0$), the baryon density parameter ($\Omega_b$), and the 
spectral index ($n_s$).
We have confirmed that the constraint on $S_8$ is not strongly
affected by uncertainties in these parameters 
as long as they are restricted 
within the prior ranges considered in this paper.

%
%
\subsubsection{Neutrino mass}
\label{sec:impact_neutrino}

%
%
\begin{figure*}
\begin{center}
  \includegraphics[height=120mm,angle=-90]{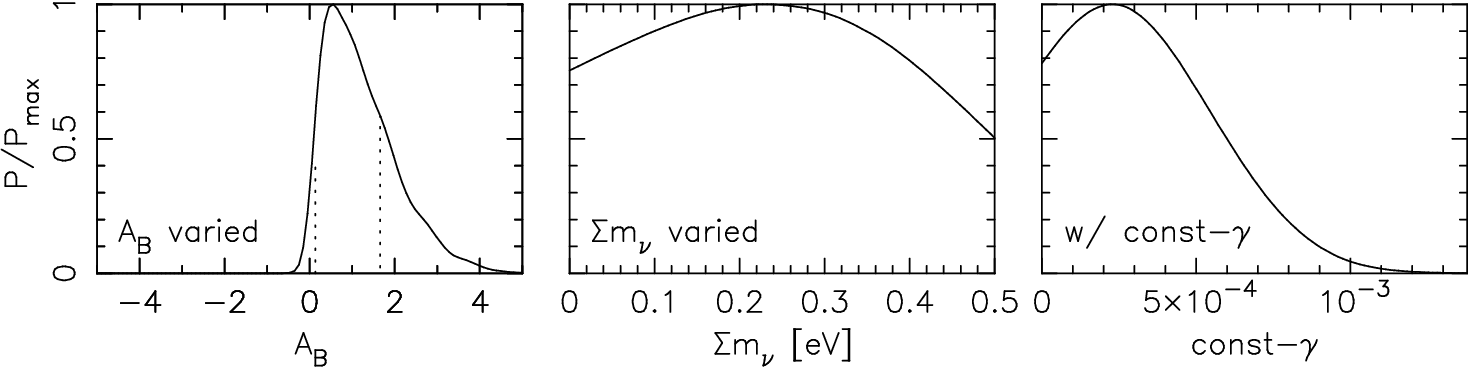}
\end{center}
\caption{Marginalized one-dimensional posterior distributions of nuisance
  parameters derived from non-fiducial models. From left to
  right, we show the baryon feedback model parameter from the ``$A_B$ varied''
  setup, the neutrino mass from the ``$\sum m_{\nu}$ varied'' setup, and the
  residual constant shear $\bar{\gamma}$ from the ``w/ const-$\gamma$'' setup.
  \label{fig:posterior_3addedpar}}
\end{figure*}

%
%
\begin{figure}[t]
\begin{center}
  \includegraphics[height=82mm,angle=-90]{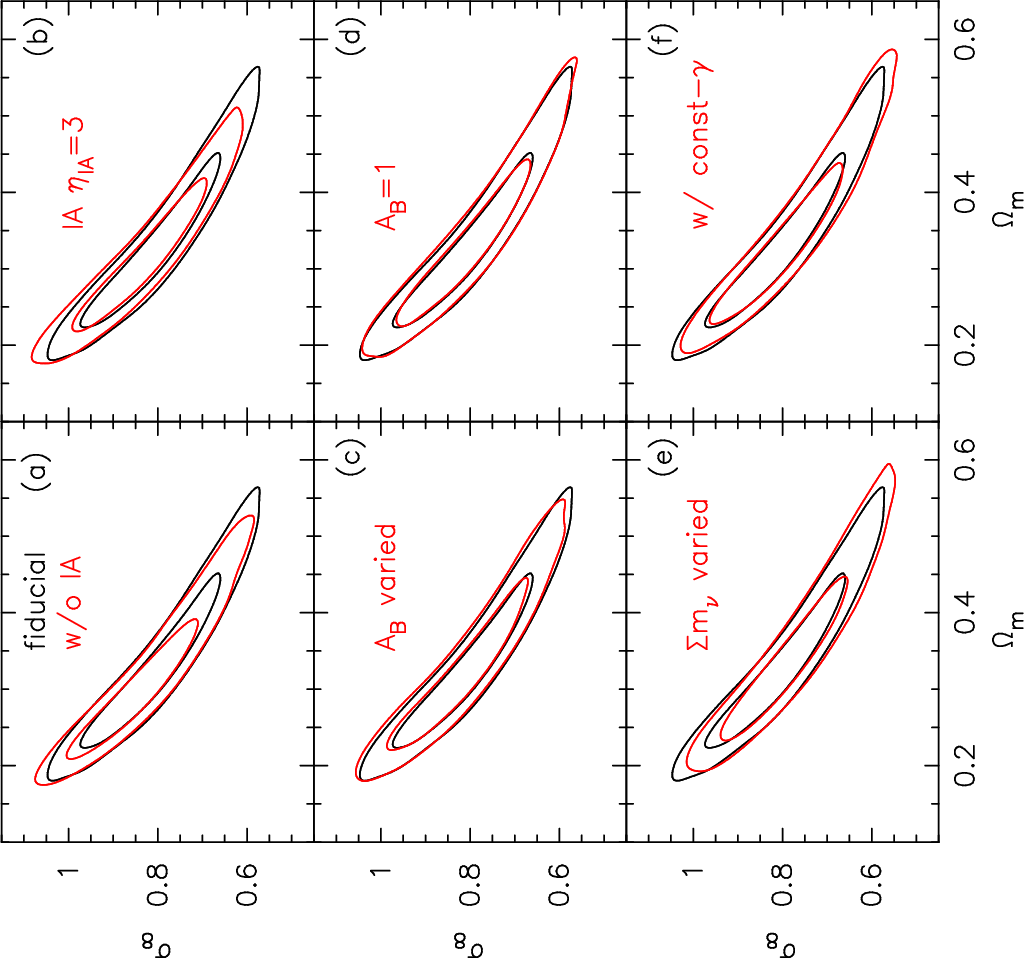}
\end{center}
\caption{Comparison of constraints in the
  $\Omega_m$-$\sigma_8$ plane between the fiducial setup (black
  contours) and different assumptions, as described in the text 
  (red contours showing 68\% and 95\% confidence levels).
  \label{fig:om_sig8_sys1}}
\end{figure}

\begin{figure}
\begin{center}
  \includegraphics[height=82mm,angle=-90]{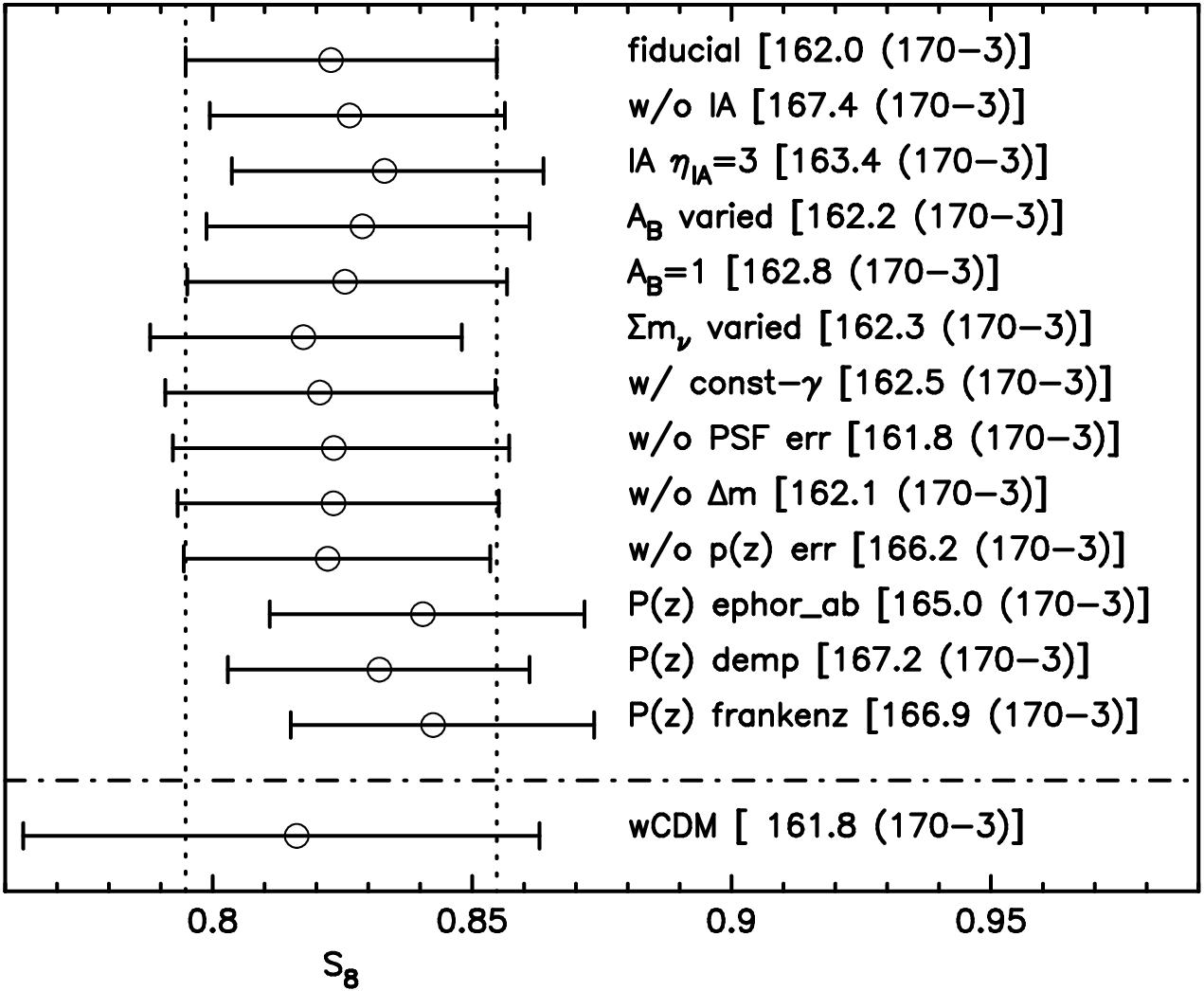}
\end{center}
\caption{Means and 68\% confidence intervals of
  marginalized one-dimensional constraints on $S_8=\sigma_8\sqrt{\Omega_m/0.3}$.
  The fiducial case (top) is compared with different setups to check the 
  robustness of our result.
  Vertical dotted lines show the 68\% confidence interval of the
  fiducial case. The numbers in the bracket after the setup name indicate
  [$\chi_{\rm min}^2$ ($N_d-N_p^{\rm eff}$)].
  \label{fig:s8ranges}}
\end{figure}

\begin{figure}[t]
  \begin{center}   
  \includegraphics[height=82mm,angle=-90]{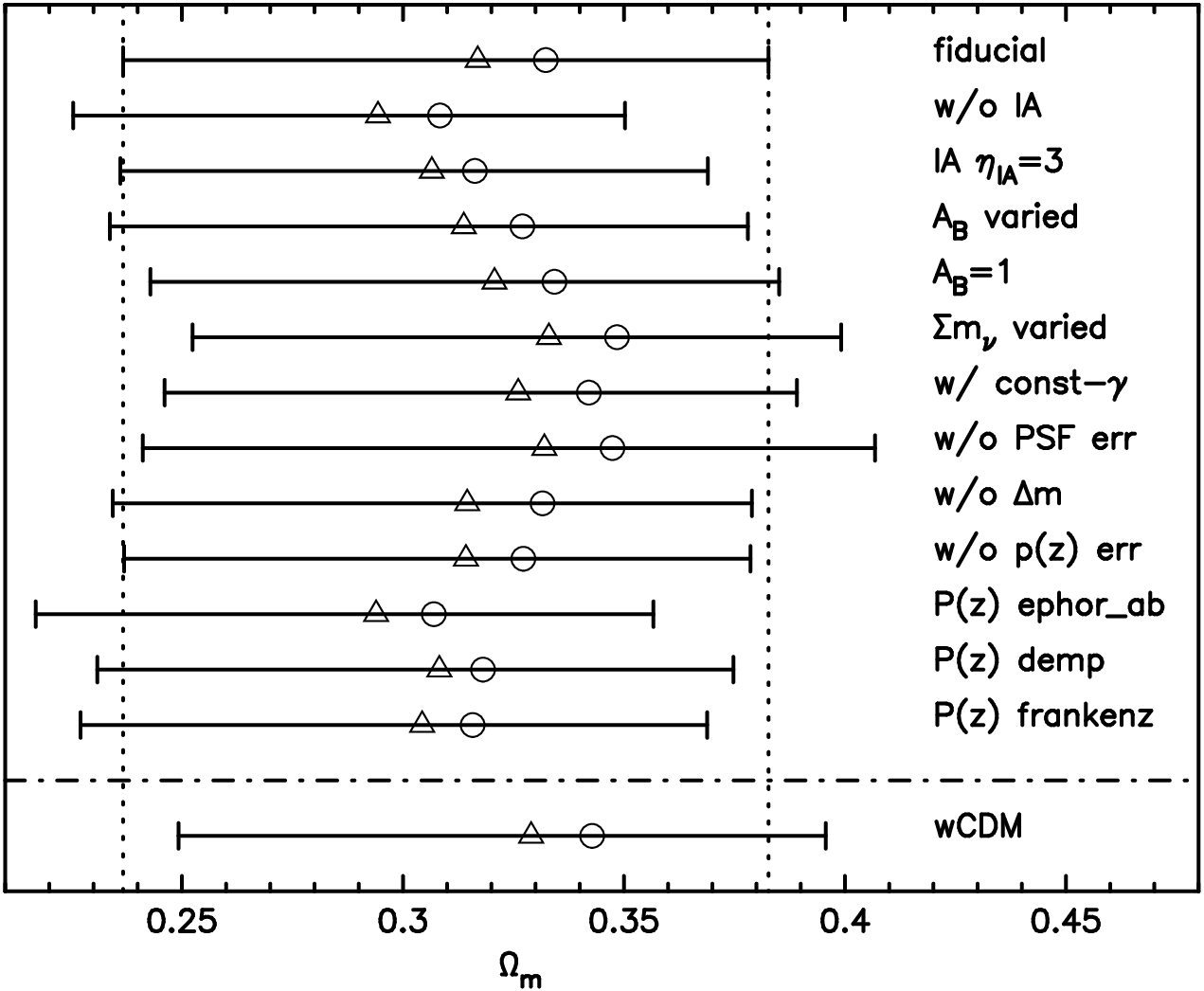}
\end{center}
\caption{Same as Fig~\ref{fig:s8ranges}, but for 
  marginalized one-dimensional constraints on $\Omega_m$.
  Open circles and open triangles show the means and medians of the
  marginalized posterior distributions, respectively.
  We note that the means of the marginalized posterior
  distributions are preferentially located on the right side of the 68\%
  confidence  
  intervals, because their posterior distributions are skewed toward high 
  $\Omega_m$ values (as shown in Figure~\ref{fig:posterior_cosmo}).
  \label{fig:omranges}}
\end{figure}

\begin{figure}
\begin{center}
  \includegraphics[height=82mm,angle=-90]{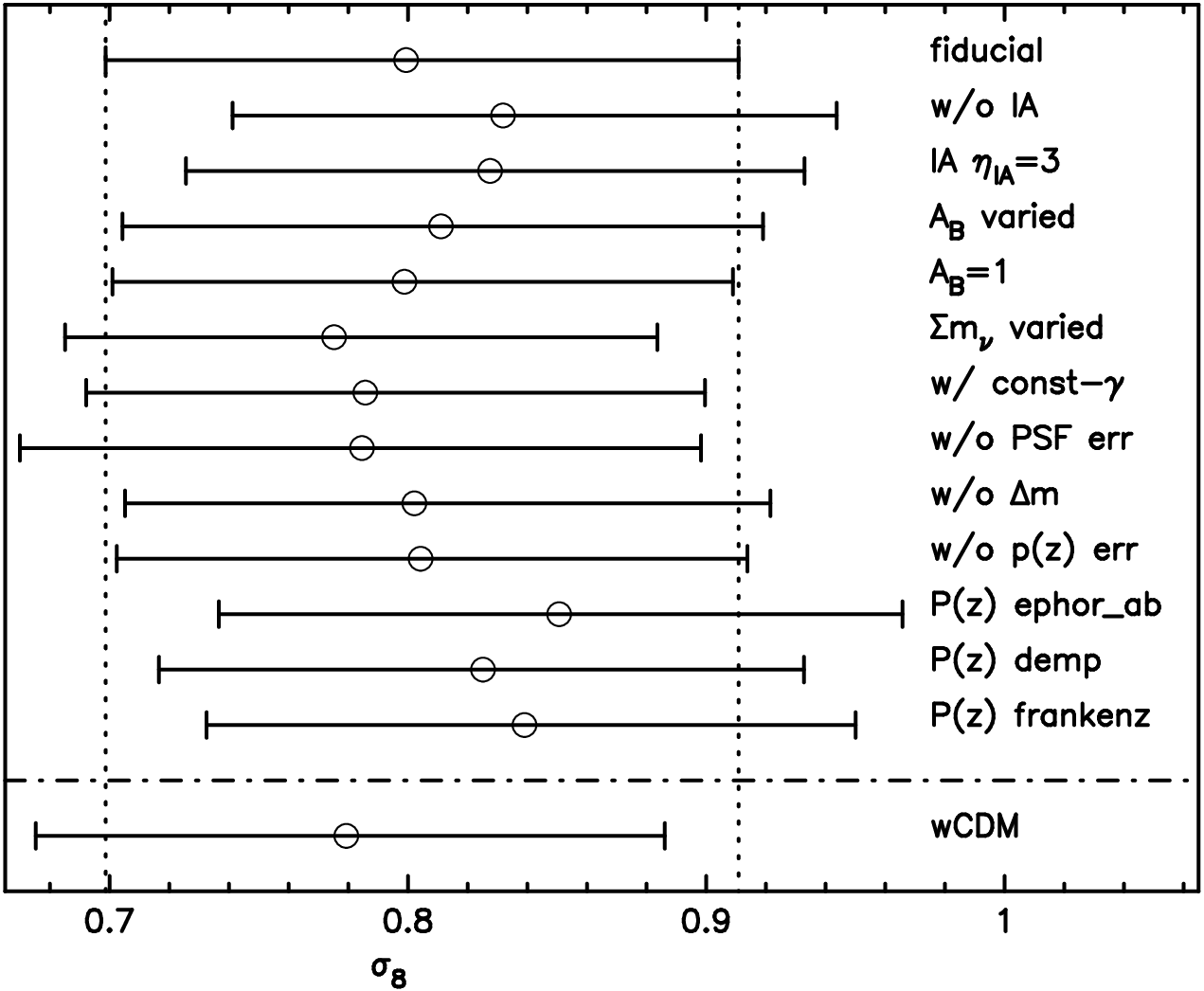}
\end{center}
\caption{Same as Fig~\ref{fig:s8ranges}, but for 
  marginalized one-dimensional constraints on $\sigma_8$. 
  \label{fig:sig8ranges}}
\end{figure}

Since the non-zero neutrino mass leads to a redshift-dependent suppression 
of the matter power spectrum at small scales, it has, in principle, an impact 
on the cosmological inference.
In our fiducial setup, the neutrino mass is fixed at $\sum m_{\nu}=0.06$~eV;
the current measurement precision 
of the cosmic shear TPCFs is expected to be insufficient to place a
useful constraint on the neutrino mass, 
especially given the fact that we exclude small scales from our analysis.
We check this expectation with a setup in which the neutrino mass is allowed
to vary with a flat prior in the range $0<\sum m_{\nu}<0.5$~eV.
Figure~\ref{fig:posterior_3addedpar} shows the one-dimensional posterior 
distribution of $\sum m_{\nu}$, from which it is indeed found that the 
current HSC cosmic shear TPCFs do not place a useful constraint on 
the neutrino mass. 
The derived marginalized posterior contours in the $\Omega_m$-$\sigma_8$
plane are compared with the fiducial case in panel (e) of
Figure~\ref{fig:om_sig8_sys1}\footnote{At first look it may seem
strange that the 68 percent confidence contours corresponding to the
posterior distribution marginalized over neutrino masses is smaller
than the case where we assume a fixed mass for neutrinos equal to 
0.06~eV. This happens because the probability distribution is 
peaked at a value for $\sum m_\nu>0.06 $~eV where the posterior 
volume in $\Omega_m$-$\sigma_8$ plane is smaller.}. 
Confidence intervals on $S_8$, $\Omega_m$, and $\sigma_8$ are 
compared with the fiducial case in Figures~\ref{fig:s8ranges}, 
\ref{fig:omranges}, and \ref{fig:sig8ranges}, respectively.
These comparisons indicate that the non-zero neutrino mass 
indeed has little impact on our cosmological constraints.
It is also found that the neutrino
mass constraint does not correlate with any of $\Omega_m$, $\sigma_8$, or $S_8$.
These findings confirm the validity of our treatment of the neutrino mass in 
our fiducial cosmological inference.

%
%

%
%

%
%

%
%
\begin{figure*}
  \begin{center}
    \includegraphics[height=120mm,angle=-90]{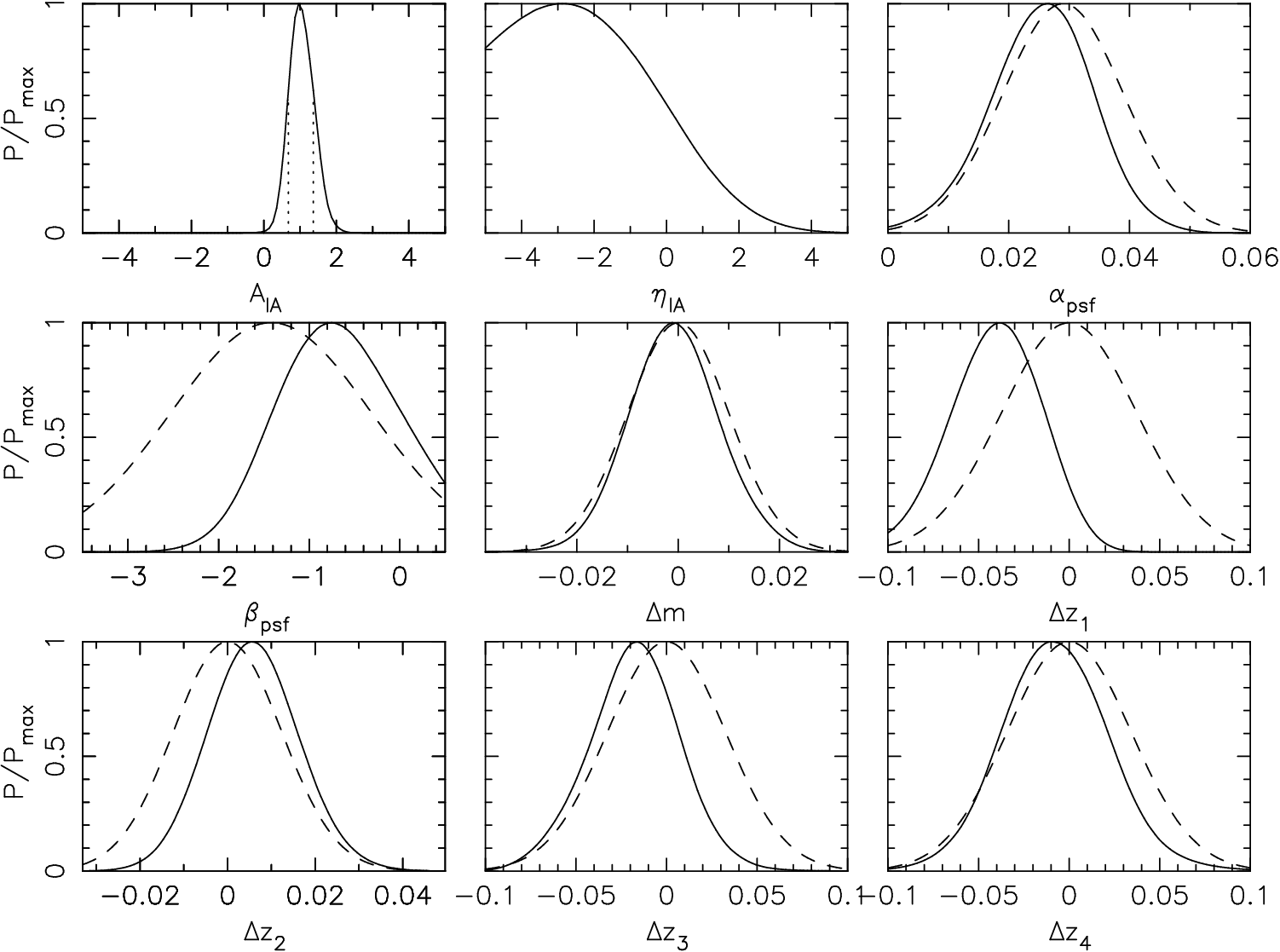}
  \end{center}
  \caption{Marginalized one-dimensional posterior distributions of
    astrophysical
    and systematics parameters in the fiducial flat $\Lambda$CDM model.
    For the cases of $A_{\mbox{IA}}$ and $\eta_{\mbox{IA}}$,
    the horizontal axis range corresponds to the flat prior range
    ($-5<x<5$), whereas for the other cases Gaussian priors are shown 
    by the dashed curves.
    In the top left panel, vertical lines represent
    the approximate 68\% confidence interval of $A_{\mbox{IA}}$.
  \label{fig:posterior_nuisance}}
\end{figure*}

%
%
\subsubsection{Posteriors of nuisance parameters}
\label{sec:posterior_nuisance}

The marginalized one-dimensional posterior distributions of astrophysical 
and systematics parameters in the fiducial flat $\Lambda$CDM model are shown
in Figure~\ref{fig:posterior_nuisance}. It is found that, 
except for $A_{\mbox{IA}}$, the posteriors are dominated by priors.
Below, we discuss effects of these nuisance parameters on the 
cosmological inference by changing the parameter setup.
Comparisons of the one-dimensional constraints on $S_8$, $\Omega_m$,
and $\sigma_8$ between the fiducial case and cases with different setups 
are summarized in  Figures~\ref{fig:s8ranges}, \ref{fig:omranges},
and \ref{fig:sig8ranges}, respectively.

%
%
\subsection{Impact of astrophysical uncertainties}
\label{sec:impact_astro}

%
%
\subsubsection{Intrinsic galaxy alignment}
\label{sec:impact_IA}

We find that the marginalized one-dimensional constraint on $A_{\mbox{IA}}$ is 
$A_{\mbox{IA}}=1.04_{-0.37}^{+0.32}$, which is consistent with the result from the HSC cosmic
shear power spectrum analysis by \citet{2019PASJ...71...43H}. 
They found $A_{\mbox{IA}}=0.38\pm 0.70$ for their fiducial setup.
The 1$\sigma$ error on $A_{\mbox{IA}}$ from our analysis is
smaller than one from the power spectrum analysis. The reason for this
is currently not known. A possible reason would be different angular
ranges adopted in the two analyses (see Appendix~\ref{appendix:powerspectrum}).
On the other hand, our constraint on $\eta_{\mbox{IA}}$ is $-2.0\pm1.8$,
which is consistent with the shear power spectrum analysis.
As discussed in section 5.4 of \citet{2019PASJ...71...43H}, a plausible
value of $\eta_{\mbox{IA}}$ from available observations is
$\eta_{\mbox{IA}}=3\pm 0.75$ which would be about 2$\sigma$ higher
compared to our derived value. 
Given this, we will examine the impact of the IA modeling on our
cosmological inference below.

In order to test the robustness of the cosmological constraints against
the uncertainty of the intrinsic galaxy alignment, we perform two
cosmological inferences with different IA modeling. In one case, 
the IA contribution is completely ignored i.e., $A_{\mbox{IA}}$ is fixed to 0, 
and in the other case $\eta_{\mbox{IA}}$ is fixed to 3 
(See section 5.4 of \citealt{2019PASJ...71...43H}) while 
$A_{\mbox{IA}}$ is treated as a free parameter.
The results from these settings are compared with the fiducial ones in
Figure~\ref{fig:om_sig8_sys1} (panels (a) and (b)) and 
Figure~\ref{fig:s8ranges}.
We find that the corresponding changes in cosmological constraints 
are not significant.
For instance, the shift of the mean $S_8$ value is found to be
0.33$\sigma$ for the ``IA $\eta_{\mbox{IA}}=3$'' case.

Finally, we examine how the IA contribution affects the constraints in the 
$\Omega_m$-$\sigma_8$ plane.
As shown in panel (a) of Figure~\ref{fig:om_sig8_sys1}, the inclusion 
of the IA contribution moves the posterior contour toward higher
$\Omega_m$ and lower $\sigma_8$, and as we have seen, slightly reduces $S_8$.
This behavior may appear somewhat 
counter-intuitive, because the IA contribution, mostly given a 
negative GI term, suppresses TPCFs, leading to a larger $S_8$ to compensate.
A plausible explanation for this is as follows.
Since the negative redshift dependence of IA contribution, which is
preferred as seen in Figure~\ref{fig:posterior_nuisance}, suppresses TPCFs
at lower redshifts more strongly than at higher redshifts, larger matter
fluctuations at lower redshifts are required to compensate the
redshift-dependent suppression.
This requires more rapid growth of matter fluctuations at lower
redshifts, leading to the higher $\Omega_m$ along with the lower 
$\sigma_8$ to adjust the overall amplitude of tomographic TPCFs.

%
%
\subsubsection{Baryonic feedback}
\label{sec:impact_baryon}

In our fiducial setup, we do not include the effect of the baryonic 
feedback, but instead remove the angular scales where its impact is 
not negligible (see Section~\ref{sec:data_vector}). 
It is therefore expected that the baryonic feedback effect does not
strongly affect our cosmological constraints. We check this 
expectation explicitly by employing an empirical 
``AGN feedback model'' by \citet{2015MNRAS.450.1212H} 
(as described in Section~\ref{sec:baryon_effect}).
Specifically we consider two cases; the original 
AGN feedback model by \citet{2015MNRAS.450.1212H}, which corresponds 
to fixing the baryon feedback parameter $A_B=1$, and 
a more flexible model in which $A_B$ is allowed to vary with a flat prior 
in the range $-5<A_B<5$.

Since the baryonic feedback suppresses the amplitude of the matter power
spectrum on scales we are probing, it leads to a higher values of $S_8$
to compensate. This is indeed seen in the ``$A_B=1$'' case, 
as shown in Figure~\ref{fig:s8ranges}.
However the shift of the mean $S_8$ value is not significant, 
0.07$\sigma$, as expected.

In the ``$A_B$ varied'' case, 
Figure~\ref{fig:posterior_3addedpar} shows that the constraint on $A_B$ is 
weak with the marginalized posterior of 
(its mean and 68\% confidence interval) $A_B=1.19_{-1.06}^{+0.47}$. The expected correlation 
between $A_B$ and $S_8$ is confirmed. 
Again, it is found from panel (c) of Figure~\ref{fig:om_sig8_sys1} and 
Figure~\ref{fig:s8ranges} that its impact on cosmological constraints 
is not significant.
We conclude that the effect of baryonic feedback on our fiducial 
cosmological constraints is insignificant given the size of our 
statistical errors.

%
%
\subsection{Impact of systematics}
\label{sec:impact_systematics}

%
%
\subsubsection{Residual constant shear}
\label{sec:impact_const_shear}

In the fiducial model, we do not include the correction for the residual
constant shear, because the statistical significance of its existence is
found to be marginal (see Appendix~\ref{sec:mean_shear}).
In order to check the robustness of our fiducial cosmological constraints
against the residual constant shear, we test the same setup as the
fiducial case but including a single parameter $\bar{\gamma}$ that models the
residual constant shear as equation~(\ref{eq:model_corrections+}).
We adopt a flat prior in the range $0<\bar{\gamma}<5\times10^{-3}$.
The derived constraints are compared with the fiducial case in panel (f)
of Figure~\ref{fig:om_sig8_sys1} and Figure~\ref{fig:s8ranges}.
We find that the resulting changes in the cosmological constraints are very small.
The marginalized one-dimensional posterior distribution of $\bar{\gamma}$ 
is shown in Fig~\ref{fig:posterior_3addedpar}.
The derived 1$\sigma$ upper limit is
found to $4.5\times10^{-4}$, which is smaller than the constant shear
expected from the cosmic shear that is coherent over the field 
(see Appendix~\ref{appendix:cov_mean_shear}).

%
%
\begin{figure}
\begin{center}
  \includegraphics[height=82mm,angle=-90]{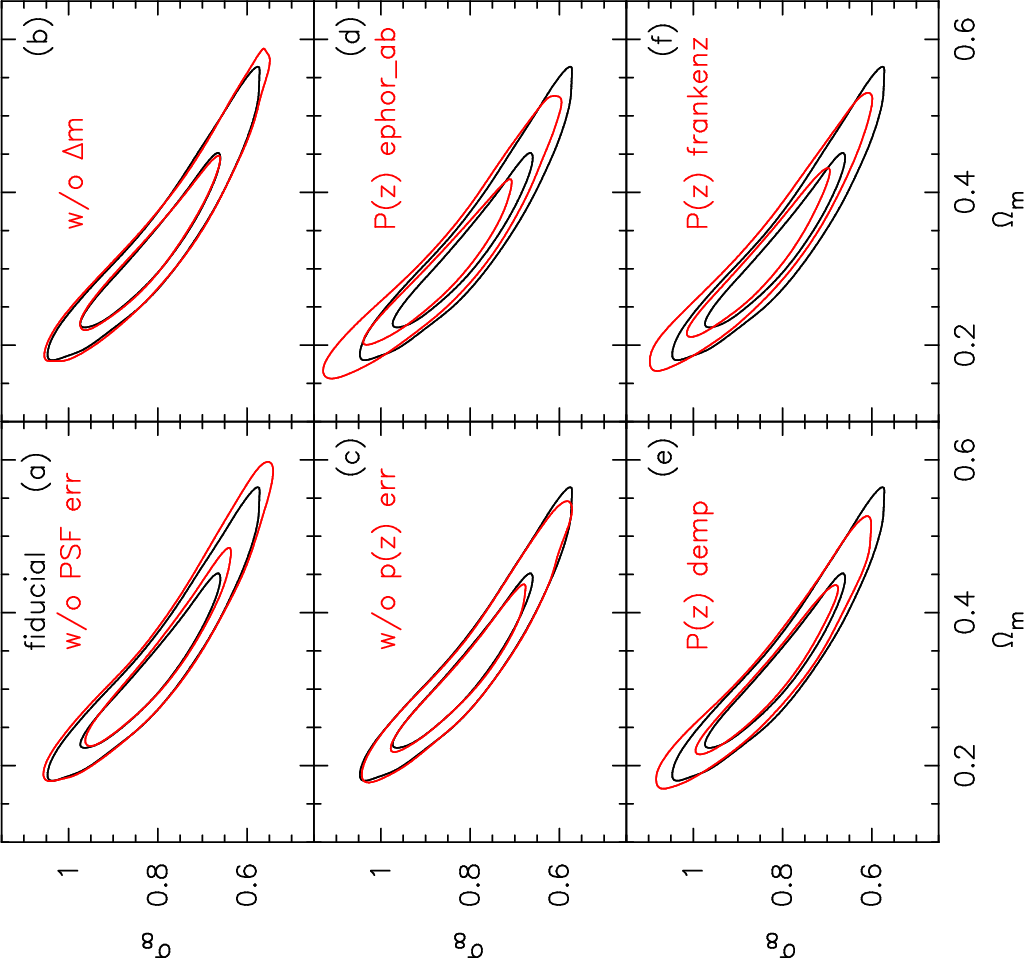}
\end{center}
\caption{Same as Figure~\ref{fig:om_sig8_sys1}, but for other setups for
  systematics tests. 
  \label{fig:om_sig8_sys2}}
\end{figure}

%
%
\subsubsection{PSF leakage and PSF modeling errors}
\label{sec:impact_psf}

In this paper we employ a simple model for the PSF leakage and PSF
modeling errors given by equation~(\ref{eq:g_psf}), and apply the
correction to the cosmic shear TPCFs as described in
equation~(\ref{eq:model_corrections+}). 
The priors for the model parameters $\alpha_{\mbox{psf}}$ and 
$\beta_{\mbox{psf}}$ are derived in Appendix~\ref{appendix:residual_PSF}.
Marginalized one-dimensional posterior distributions of these parameters 
from our fiducial analysis are shown in Figure~\ref{fig:posterior_nuisance}.
We found that the posteriors are largely determined by the priors.
We also find that the marginalized constraints on these parameters are not
strongly correlated with either $\Omega_m$, $\sigma_8$, or $S_8$.

In order to check the robustness of our cosmological constraints against
these systematics, we test the same setup as the fiducial case but
ignoring these parameters i.e., setting $\alpha_{\mbox{psf}} =
\beta_{\mbox{psf}} =0$.
The results are shown in panel (a) of Figure~\ref{fig:om_sig8_sys2} and
Figure~\ref{fig:s8ranges}. We find that the changes in the cosmological
constraints are very small. This is expected, as the corrections
due to PSF leakage and PSF modeling errors small compared 
with the current size of errors on the HSC cosmic shear TPCFs.

%
%
\subsubsection{Shear calibration error}
\label{sec:impact_deltam}

In our fiducial analysis we also take account of the uncertainty 
in the shear multiplicative bias correction using a simple model, 
equation~(\ref{eq:delta_m}), with a Gaussian prior corresponding to 
a 1\% uncertainty (see Section~\ref{sec:Systematic_parameters}).
The marginalized one-dimensional posterior distribution of the model 
parameter $\Delta m$ from our fiducial analysis is shown in 
Figure~\ref{fig:posterior_nuisance}, which indicates that the posterior 
is dominated by the prior.

In order to check the effect of this residual calibration bias on our
cosmological constraints, we test the same setup as the fiducial case but
ignoring the nuisance parameter i.e., setting $\Delta m =0$.
The results are shown in panel (b) of Figure~\ref{fig:om_sig8_sys2} and
Figure~\ref{fig:s8ranges}. We find that the changes in the cosmological
constraints are very small.

%
%
\subsubsection{Source redshift distribution errors}
\label{sec:impact_dz}

We take account of uncertainties in the redshift distributions of
source galaxies by introducing parameters $\Delta z_a$, which represent 
a shift of the source redshift distributions as defined in 
equation~(\ref{eq:pz_shift}). We consider independent shifts for the four 
tomographic bins, leading to four nuisance parameters. Priors on these 
parameters are determined based on differences of source redshift 
distributions from different approaches (see
Section~\ref{sec:Systematic_parameters}), and we marginalize over 
these nuisance parameters in our fiducial setup.
Marginalized one-dimensional posterior distributions of 
these parameters from our fiducial analysis are shown in 
Figure~\ref{fig:posterior_nuisance}.
Although peak positions of these posteriors show shifts from the
peak the prior distributions, the sizes of the shifts are reasonably
within the the Gaussian priors. In the case of the lowest redshift bin
which shows the largest shift, the peak shift is 1.0$\sigma$ of the
Gaussian priors, and thus is not statistically significant.
However, notice that it may indicate an unknown bias in
estimation of the source redshift distribution that is not captured in
the prior knowledge.

In order to check the robustness of our cosmological constraints against
these uncertainties, we test the same setup as the fiducial analysis but
ignoring these parameters.
The results are shown in panel (c) of Figure~\ref{fig:om_sig8_sys2} and
Figure~\ref{fig:s8ranges}. We find that the changes in the cosmological
constraints are small, with the shift of the mean $S_8$ value being
$-0.02\sigma$.

In addition, we also check for possible 
systematic effects coming from the uncertainty of the redshift 
distributions due to photo-$z$ methodology. 
We explore this by replacing
the default COSMOS re-weighted $p^a(z)$ with ones derived from
stacked PDFs. For this purpose we adopt three different 
photo-$z$ methods, {\tt DEmP}, Ephor AB, and FRANKEN-Z 
(see Section~\ref{sec:pz}), for which stacked PDFs are shown in
Figure~\ref{fig:pz}. This is a rather empirical test, as each photo-$z$ 
method has its own bias and errors \citep{2018PASJ...70S...9T}, 
thus this test should be considered as a sensitivity check.
The results are shown in Figure~\ref{fig:om_sig8_sys2} (panels (d), (e),
and (f)) and Figure~\ref{fig:s8ranges}. Again, we find that the changes in the
cosmological constraints are not significant.
Thus we conclude that no additional systematics are identified from
this test.

%
%
\begin{figure}
\begin{center}
  \includegraphics[height=82mm,angle=-90]{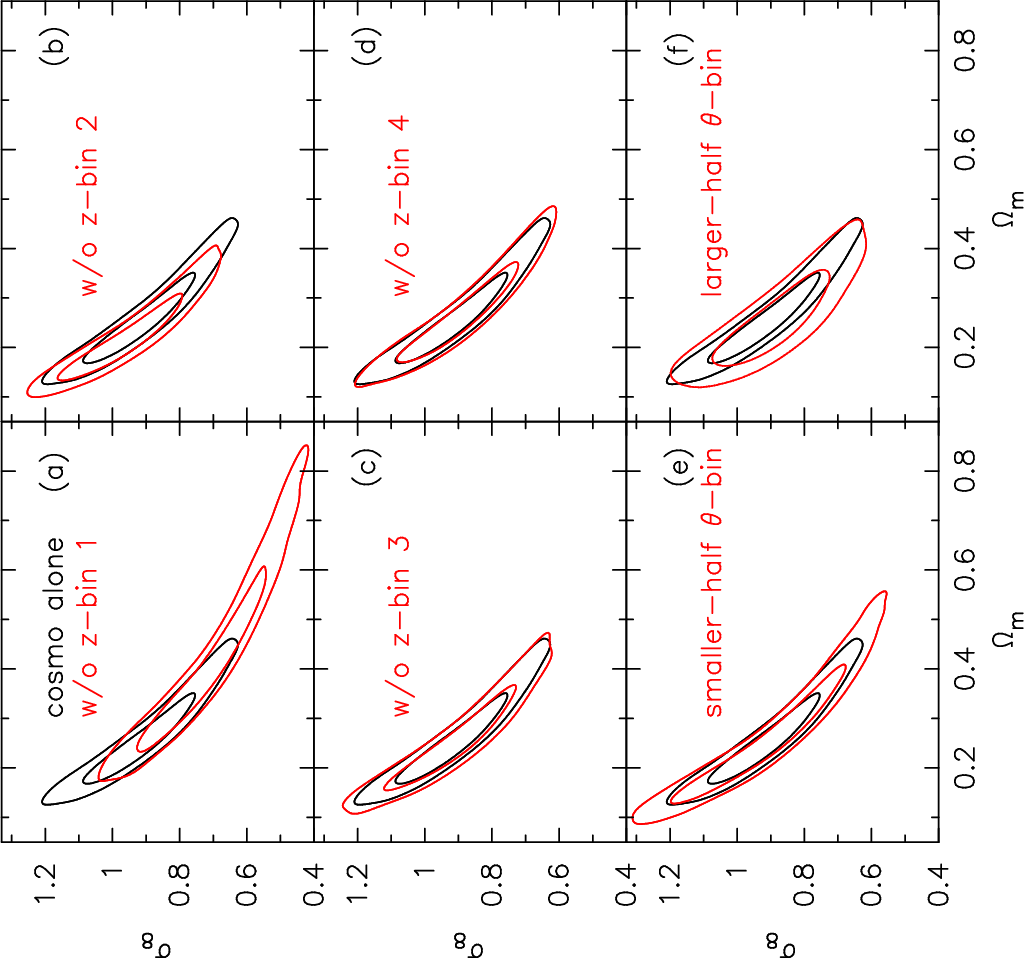}
\end{center}
\caption{Comparison of constraints in the
  $\Omega_m$-$\sigma_8$ plane from the cosmology alone setup (black
  contours) with different setups for internal consistency checks 
  (red contours showing 68\% and 95\% confidence levels).
  \label{fig:om_sig8_sys3}}
\end{figure}

%
%
\begin{figure}
\begin{center}
  \includegraphics[height=82mm,angle=-90]{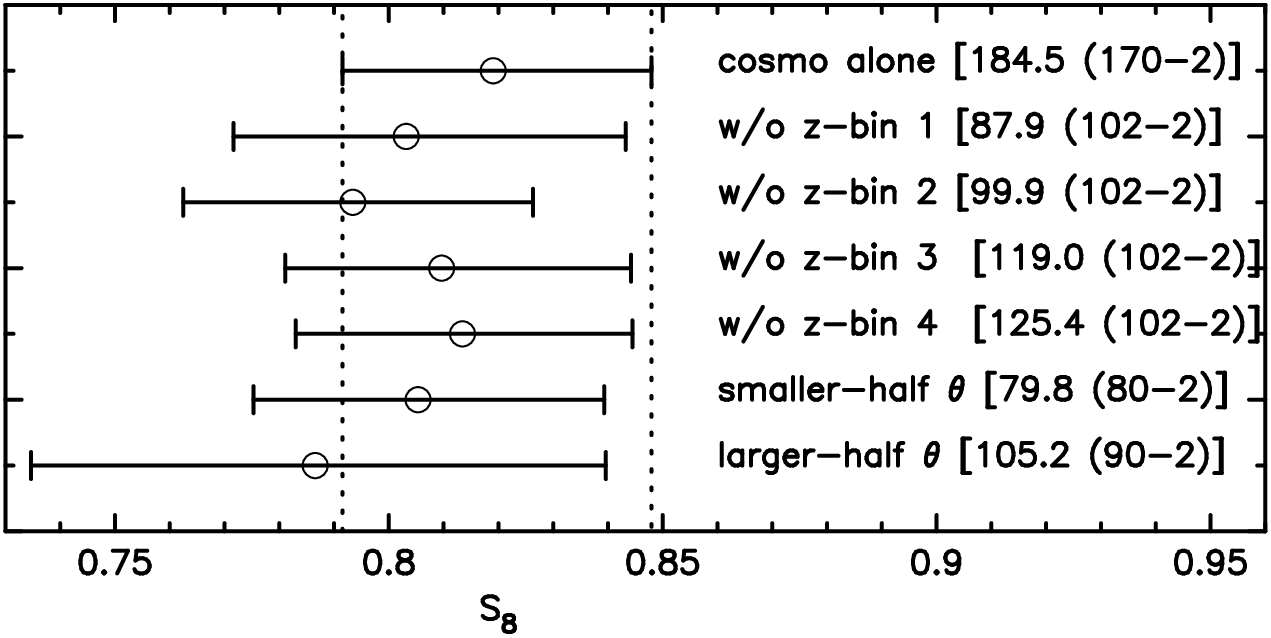}
\end{center}
\caption{Means and 68\% confidence intervals of
  marginalized one-dimensional constraints on $S_8$.
  The ``cosmology alone'' case (top) is compared with different setups 
  for internal consistency checks.
  Vertical dotted lines show the 68\% confidence interval of
  the cosmology alone case.
  \label{fig:s8ranges_ic}}
\end{figure}

%
%
\subsection{Internal consistency}
\label{sec:internal_consistency}

Here we present results of internal consistency checks in which we
derive cosmological constraints from subsets of the data vector and compare
the results with ones from a reference setup.
In doing so, we do not use the fiducial results as the reference, but 
instead we adopt the results from the ``cosmology alone'' setup in which 
we include neither systematics nor astrophysical parameters 
but only five cosmological parameters are included as a baseline 
for comparison. 
The reason for this choice is to avoid undesirable changes in nuisance
parameters, especially redshift-dependent parameters such as the 
redshift dependence parameter of the IA $\eta_{\mbox{IA}}$ and 
photo-$z$ error parameters $\Delta z_i$, which may add or cancel 
out shifts in parameter constraints.
Of course, this has the side effect that the reference setup does not
provide the best cosmological constraints, although the difference 
from the fiducial case is not significant.
In fact, the differences in the marginalized cosmological constraints
between the fiducial setup and ``cosmology alone'' setup is about the level
of these between the fiducial setup and ``w/o IA'' setup, as ignoring IA
contribution has the largest effect. 
To summarize, considering the facts that our aim here is to carry out 
an {\it internal}
consistency check and that the side effect is not significant, we 
adopt ``cosmology alone'' setup as the reference.

%
%
\subsubsection{Tomographic redshift bins}
\label{sec:tomographic_bins}

First, we exclude one of the four redshift bins and perform
the cosmological inference with three tomographic bins.
The resulting cosmological constraints are shown in
Figure~\ref{fig:om_sig8_sys3} (panels (a)-(d)), and the derived 68\%
confidence intervals of $S_8$ are compared in
Figure~\ref{fig:s8ranges_ic}.
We find that constraints on $S_8$ from these setups are consistent within 
1$\sigma$ of the reference result.
Figure~\ref{fig:s8ranges_ic} may look odd in the sense that all the setups
have a lower mean value of $S_8$ than that of the reference setup. This
is a result of changes of the posterior distributions in
the $\Omega_m$-$\sigma_8$ plane in different directions, leading to 
a smaller $S_8$ by chance.
Also Figure~\ref{fig:om_sig8_sys3} shows that 68\% confidence
contours in the $\Omega_m$-$\sigma_8$ plane in these cases largely overlap
with the reference contour.  
Thus we conclude that the no significant internal inconsistency is found
from this test.

It may be worth noting that excluding one redshift bin leads to relatively
large shifts in $\Omega_m$ constraints, as shown in
Figure~\ref{fig:om_sig8_sys3}.
This is due the fact that the constraint on $\Omega_m$ is mainly driven by 
the relative amplitudes of cosmic shear TPCFs in different tomographic bins, 
as was discussed in \citet{2019PASJ...71...43H}. 

%
%
\subsubsection{Angular ranges}
\label{sec:angular_ranges}

Next, we check the internal consistency among different angular
ranges by splitting the fiducial angular bins in half.
To be specific, the 9(8) angular bins of $\xi_+$($\xi_-$) are split 
into 4(4) smaller  $\theta$ bins and 5(4) larger $\theta$ bins.
The resulting cosmological constraints, in comparison with the
``cosmology alone'' case, are shown in 
Figure~\ref{fig:om_sig8_sys3} (panels (e) and (f)) and
Figure~\ref{fig:s8ranges_ic}.
It is found that for the smaller-half bins case, the constraint on $S_8$
shifts to smaller value by 0.48$\sigma$, with the
posterior contours on $\Omega_m$-$\sigma_8$ plane being elongated along
the $\Omega_m$-$\sigma_8$ degeneracy direction.
On the other hand, the constraint on $S_8$ from the larger-half bins
case shifts slightly more than $1\sigma$ of the reference result.
However, the 68\% confidence interval of this case is about two times larger
than that of the reference case.
In addition, the confidence contours in the $\Omega_m$-$\sigma_8$ plane largely
overlap with those of reference cases.
Thus no strong evidence of internal
inconsistency is found by this test.

%
%
\subsection{$w$CDM model}
\label{sec:wcdm}

%
%
\begin{figure}
\begin{center}
  \includegraphics[height=82mm,angle=-90]{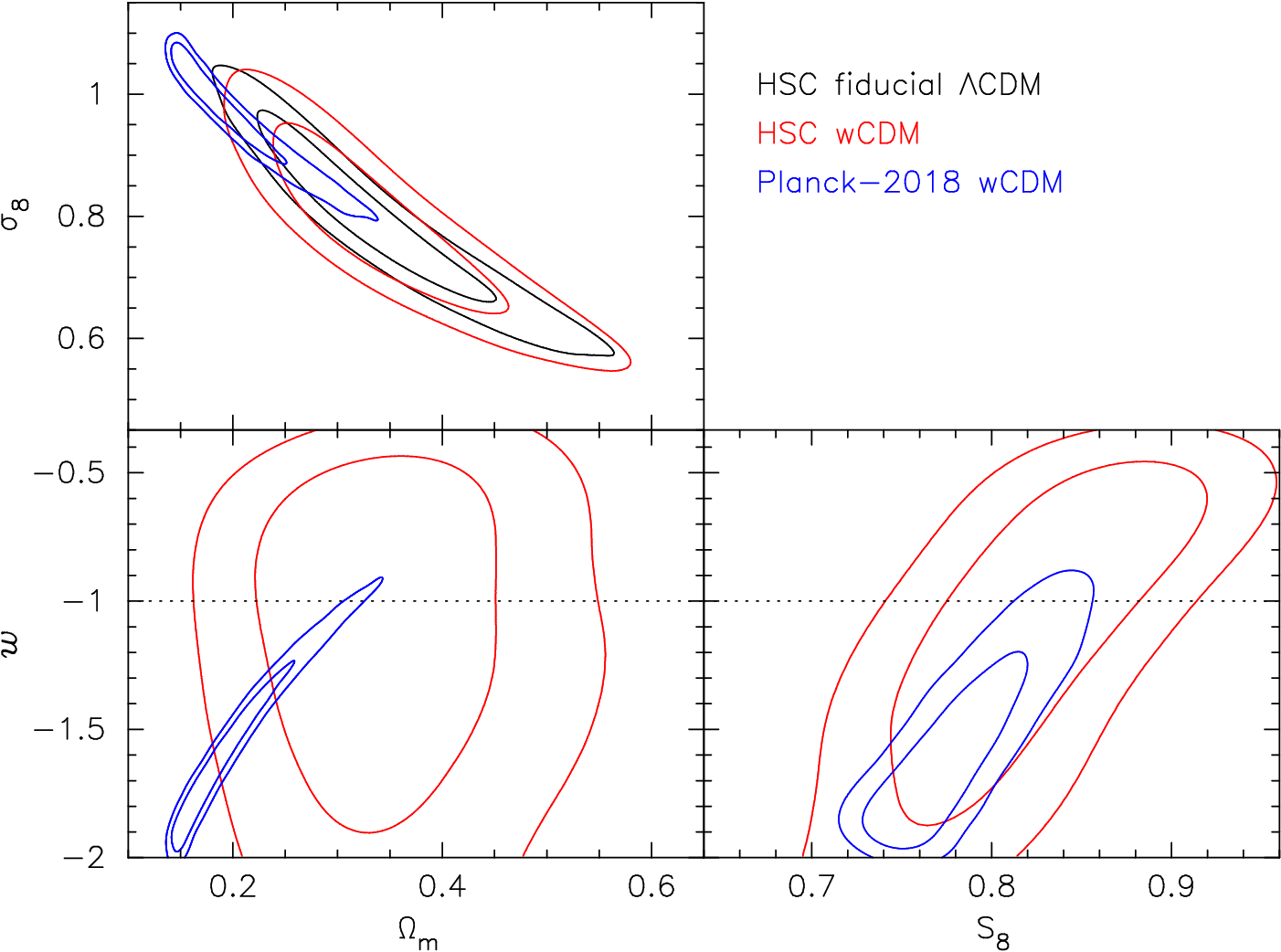}
\end{center}
\caption{Marginalized posterior contours (68\% and 95\%
    confidence levels) in the
    $\Omega_m$-$\sigma_8$ plane (top), the $\Omega_m$-$w$ plane (bottom left)
    and the $S_8$-$w$ plane (bottom right) in the $w$CDM model are shown by 
    red contours.
    Constraints from the fiducial $\Lambda$CDM model are shown by the black
    contours, and 
    {\it Planck} 2018 results for the $w$CDM model
    \citep[][TT+TE+EE+lowE]{2020A&A...641A...6P} are also shown by blue
    contours. 
    \label{fig:om_sig8_w}}
\end{figure}

In addition to the fiducial $\Lambda$CDM model, we test one extension
model by including the time-independent dark energy equation of state
parameter $w$. We allow $w$ to vary with a flat prior in the range
$-2<w<-1/3$. The setup of the other parameters are same as the
fiducial $\Lambda$CDM model.

The marginalized constraints in the $\Omega_m$-$\sigma_8$, $\Omega_m$-$w$, and 
$S_8$-$w$ planes are shown
in Figure~\ref{fig:om_sig8_w}, along with constraints from the
fiducial $\Lambda$CDM model and the {\it Planck} 2018 results for the
$w$CDM model \citep[][TT+TE+EE+lowE]{2020A&A...641A...6P}. 
Marginalized one-dimensional constraint ranges of $\Omega_m$, $\sigma_8$, 
and $S_8$ are shown in Figures~\ref{fig:omranges}, \ref{fig:sig8ranges}, and
\ref{fig:s8ranges}, respectively.
It is found that adding $w$ as a model parameter degrades constraints on
cosmological parameters, and that the current HSC cosmic shear TPCFs
alone cannot place a useful constraint on $w$.
This is quantitatively very similar to the result found in the HSC
cosmic shear power spectrum analysis by \citet{2019PASJ...71...43H}. 

%
%
\subsection{Comparison to other constraints from the literature}
\label{sec:comparison}

%
%
\begin{figure}[t]
\begin{center}
  \includegraphics[height=82mm,angle=-90]{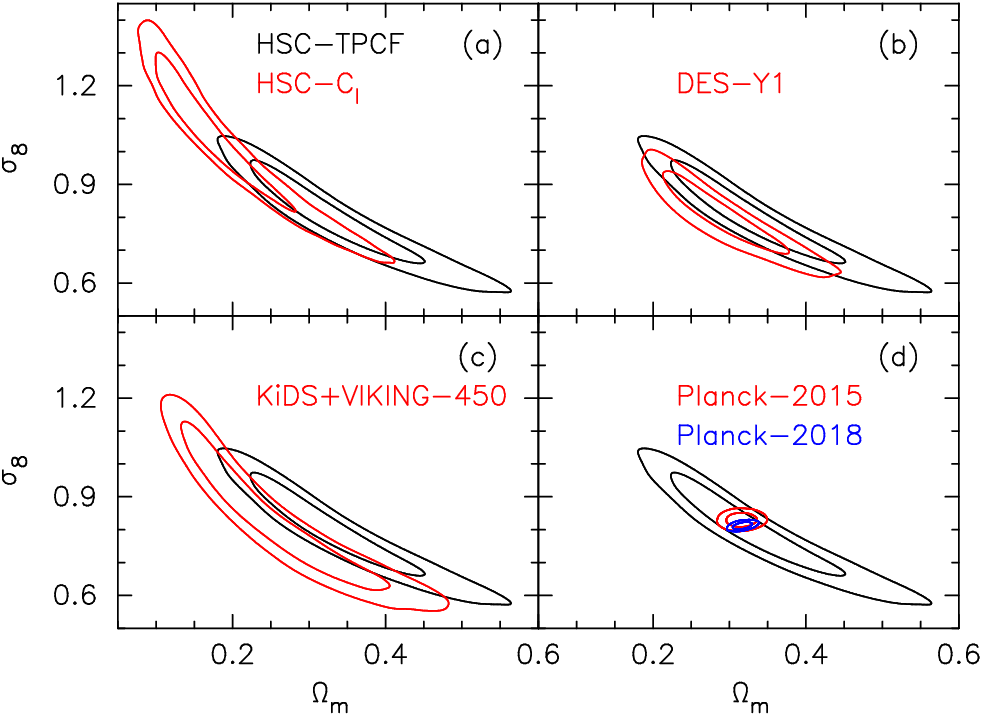}  
\end{center}
\caption{Marginalized posterior contours (68\% and 95\%
  confidence levels) in the $\Omega_m$-$\sigma_8$ plane.
  Our result from the fiducial $\Lambda$CDM model (black contours) is
  compared with results in the literature (red-line contours):
  Note that although different studies adopt different priors and
  different modeling choices, we
  do not adjust them to our fiducial setup, but rather use their
  original setups.
  Therefore, part of the difference in the posteriors may be due to
  the different choice of priors and modeling.
  (A) HSC first-year cosmic shear power spectrum result 
  \citep{2019PASJ...71...43H}. (b) Dark Energy Survey Year 1 (DES-Y1)
  cosmic shear TPCF result \citep{2018PhRvD..98d3528T}. (c)
  KiDS+VIKING-450 cosmic shear TPCF result
  \citep{2020A&A...633A..69H}.
  (d) {\it Planck} 2018 CMB result without
  CMB lensing \citep[][TT+TE+EE+lowE]{2020A&A...641A...6P}  
  (red lines) and {\it Planck} 2015 CMB result without
  CMB lensing \citep[][TT+lowP]{2016A&A...594A..13P} (blue lines).
  \label{fig:om_sig8_others}}
\end{figure}

%
%
\begin{figure}
\begin{center}
  \includegraphics[height=82mm,angle=-90]{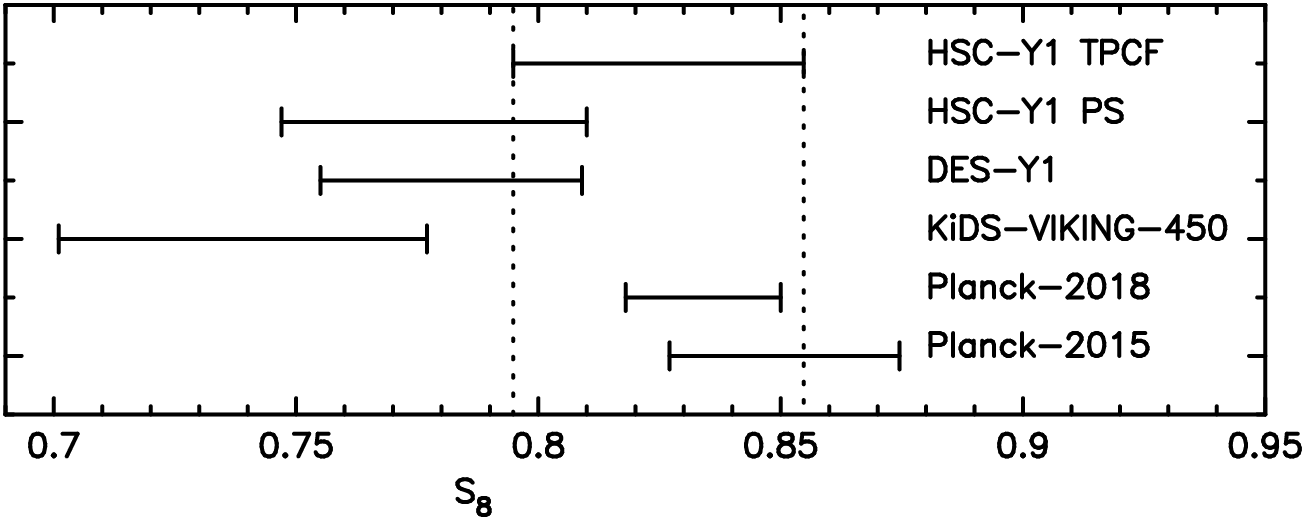}
\end{center}
\caption{68\% confidence intervals of marginalized posterior distributions
  of $S_8=\sigma_8\sqrt{\Omega_m/0.3}$. Our result from the fiducial
  $\Lambda$CDM model is compared with other results in the 
  literature, HSC first year (HSC-Y1)
  cosmic shear power spectra \citep{2019PASJ...71...43H}, DES-Y1
  cosmic shear TPCFs \citep{2018PhRvD..98d3528T}, KiDS+VIKING-450 cosmic
  shear TPCFs \citep{2020A&A...633A..69H}, and {\it Planck} 2018 CMB
  \citep[][TT+TE+EE+lowE]{2020A&A...641A...6P}, and {\it
    Planck} 2015 CMB \citep[][TT+lowP without
    lensing]{2016A&A...594A..13P}.  
  Since different studies adopt different definitions of the
  central values (mean, median, or peak of the posterior
  distribution), central values are not shown to avoid possible 
  misunderstanding.
  \label{fig:s8ranges_others}}
\end{figure}

Finally, we compare the cosmological constraints from our fiducial
$\Lambda$CDM model with other results in the literature.
Comparison plots in the $\Omega_m$-$\sigma_8$ plane are presented in
Figure~\ref{fig:om_sig8_others}, where constraints from other studies are
derived from publicly available chains.
Note that although different studies adopt different priors, we
do not adjust them to our fiducial setup, but rather use their
original priors.
Also, different studies adopt different modeling choices, for example,
Dark Energy Survey Year 1 \citep[DES-Y1;][]{2018PhRvD..98d3528T} adopts
the uniform sampling of $A_S$, instead of the logarithmic sampling that
adopted in KiDS+VIKING-450 \citep{2020A&A...633A..69H} and this study.
Therefore, part of the difference in the posteriors
may be due to the different choices of priors and modeling.
Figure~\ref{fig:s8ranges_others} compares the 68\% confidence
intervals of $S_8=\sigma_8\sqrt{\Omega_m/0.3}$, where results of other studies 
are taken from the literature.

DES-Y1 covers much larger area (1321~deg$^2$) than the HSC first year
data, yielding slightly tighter constraints than our fiducial results. 
The confidence contours of DES-Y1 in the $\Omega_m$-$\sigma_8$ plane 
largely overlap with our results, although our confidence regions are 
roughly $1.3$ times larger than theirs.
However, the two constraints are slightly misaligned in the direction
perpendicular to the $\Omega_m$-$\sigma_8$ degeneracy direction.
This results in about 1$\sigma$ difference in best-fit $S_8$ values, 
as seen in Figure~\ref{fig:s8ranges_others}.

KiDS+VIKING-450 covers 341.3~deg$^2$.
A large part of our survey fields are included in their survey fields.
Their total number of galaxies is $\sim 12$ million, about 30\%
larger than our sample.
The redshift range of galaxies they used in their cosmological analysis is
$0.1<z<1.2$, which is lower than the redshift range adopted in our analysis, 
$0.3<z<1.5$.
As is found in Figure~\ref{fig:om_sig8_others}, compared with our
posterior contours, contours from KiDS+VIKING-450 are located on the
lower $\Omega_m$ side, and are slightly elongated to the higher-$\sigma_8$
direction.
Their best-fit $S_8$ value is about 2$\sigma$ lower than ours.

It is found from Figure~\ref{fig:om_sig8_others} that the confidence
contours in the $\Omega_m$-$\sigma_8$ plane from the 
{\it Planck} 2018 CMB result
\citep[][TT+TE+EE+lowE without CMB lensing]{2020A&A...641A...6P}
as well as {\it Planck} 2015 CMB result
\citep[][TT+lowP without CMB lensing]{2016A&A...594A..13P}
overlap well with our confidence contours from the HSC first year 
TPCF analysis.
The 68\% confidence intervals of $S_8$ from {\it Planck} 2015 and 
2018 are also consistent with our result.
We therefore conclude that there is no tension between
{\it Planck} 2015 and 2018 constraints and our cosmic 
shear constraints.
The concordance between our HSC cosmic shear TPCF result and the 
{\it Planck}
CMB result in the flat $\Lambda$CDM model will place useful constraints on
extended models such as the $w$CDM model, although a combined cosmological
inference with {\it Planck} data is beyond the scope of this study.
In fact, a comparison between those constraints shown in
Figure~\ref{fig:om_sig8_w} implies that a tighter lower limit
on $w$ may be obtained by such a combined analysis.

%
%
\subsection{Comparison with HSC first year
  cosmic shear power spectrum result}
\label{sec:comparison_Hikage}

%
%
\begin{figure}
\begin{center}
  \includegraphics[height=82mm]{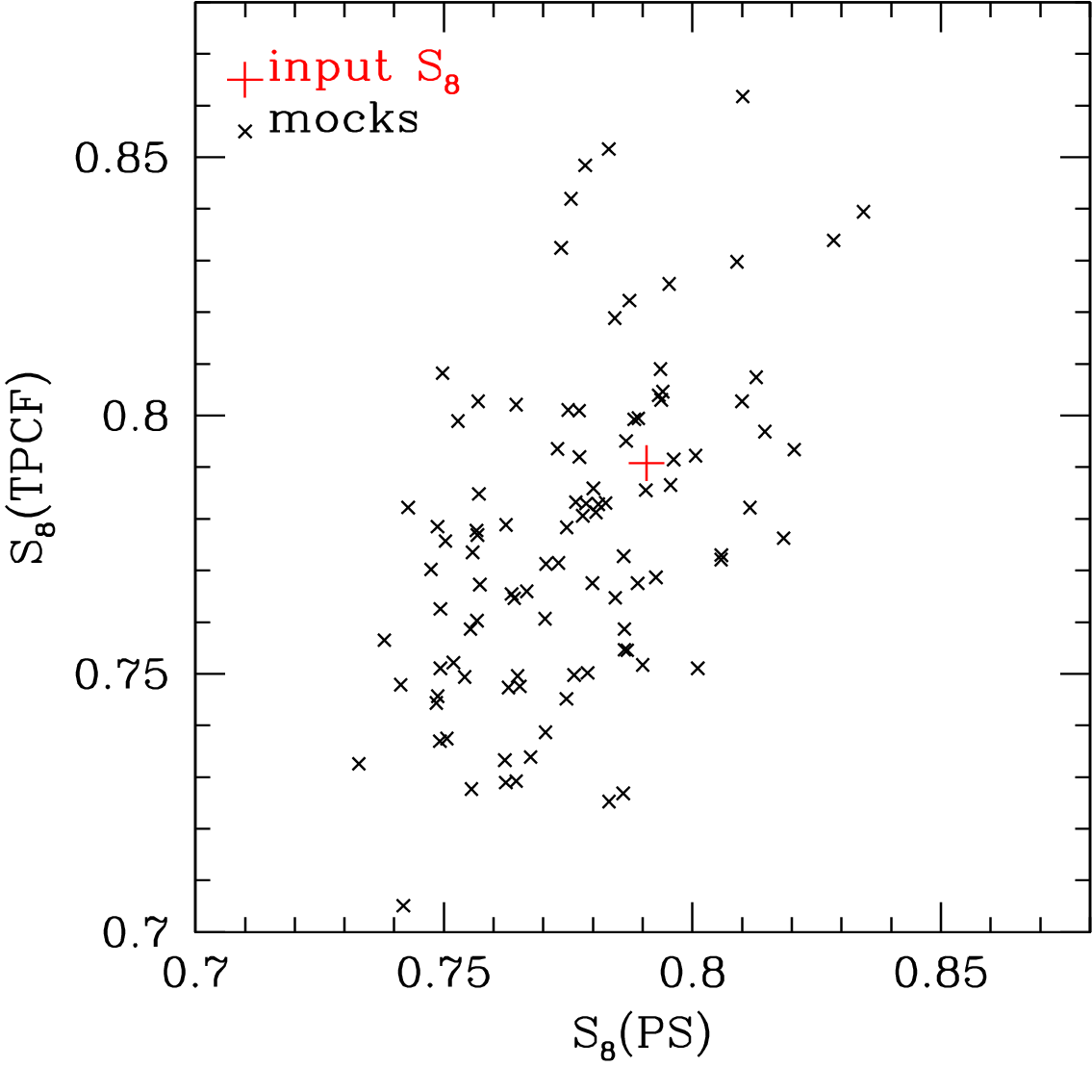}  
\end{center}
\caption{Scatter plot showing median values of marginalized 
  one-dimensional posterior distributions of $S_8$ derived from 
  cosmological analyses on 100 mock catalogs. 
  Results from the power spectrum analysis by
  \citet{2019PASJ...71...43H} are compared with ones from the TPCF
  analysis in this study. The red cross shows the value of $S_8$ 
  adopted in generating the mock catalogs. 
  \label{fig:S8r_S8f}}
\end{figure}

Figure~\ref{fig:om_sig8_others} indicates that the 68\% confidence
contours from the cosmic shear power spectrum analysis by
\citet{2019PASJ...71...43H} and from this study
overlap only mildly, even though they share
the same HSC first year weak lensing shape catalog 
\citep{2018PASJ...70S..25M} and 
adopt a similar analysis setup, including the definition 
of tomographic bins and the treatment of the IA and systematics 
parameters. The 68\% marginalized one-dimensional confidence 
intervals of $\Omega_m$ and $\sigma_8$ from these two studies also 
overlap only slightly. For instance, Figure~\ref{fig:s8ranges_others} 
indicates that there is $\sim 1 \sigma$ difference in the $S_8$ constraints 
between these two studies.  The differences between the median values of 
$S_8$ and $\Omega_m$ are  $-0.042$ and $-0.16$, respectively, where the
standard deviations of those parameters found in this study is $0.030$ and 
$0.082$, respectively.
These differences could be indicative of unknown systematic 
errors in either or both of the analyses and/or originate from 
different angular scales used in those two cosmological analyses, 
and therefore we will examine this carefully below.

We use realistic HSC mock catalogs to check whether these differences 
can be explained simply by a statistical fluctuation. 
The mock catalogs used in this analysis are the ones described in 
\citet{2018PASJ...70S..26O} and adopted in \citet{2019PASJ...71...43H}.
These differ slightly from the mock catalogs used in this paper
to derive the covariance matrix in Appendix~\ref{appendix:mock}, although
we note that these two sets of mock catalogs are generated by almost 
the same methodology and therefore are very similar. We perform the
cosmological inference on the 100 mock catalogs using the same parameter 
setup as the fiducial setup except that we fix the PSF modeling errors 
$\alpha_{\mbox{psf}}$ and $\beta_{\mbox{psf}}$ to zero 
because no PSF modeling error is added in the mock data.
\citet{2019PASJ...71...43H} also performed their power-spectrum based
cosmological inference on the same mock catalogs adopting their fiducial
setup. 
From these analyses on the mocks, we can determine the covariance of 
best-fit cosmological parameters between the cosmic shear power 
spectrum analysis in \citet{2019PASJ...71...43H}  and our cosmic 
shear TPCF analysis.

We present the scatter plot comparing $S_8$ values from these two 
cosmological analyses on the same mock catalogs in
Figure~\ref{fig:S8r_S8f}.
We find that $S_8$ values from these two analyses are only weakly
correlated. 
We find that the correlation is even weaker for $\Omega_m$.
We find that, for $S_8$, 8 out of 100 cases have a difference 
$\Delta S_8$ less than the observed value of $-0.042$,
and for $\Omega_m$, 17 out of 100 cases have a difference
$\Delta\Omega_m$ less than the observed difference of $-0.16$.
If we take the two-side estimate, we find that 
for $S_8$($\Omega_m$), 12(19) out of 100 cases have an absolute
difference of $|\Delta S_8| > 0.042$ ($|\Delta\Omega_m|>0.16$).
These mean that these differences can be explained by a statistical 
fluctuation at the $\sim 1.6\sigma$ level.

To quantify the covariance of best-fit cosmological parameters further, 
we compute the correlation coefficient
\begin{equation}
  \label{eq:covariance_cosmopar}
  r(q)={ {\mbox{Cov}(q_{,R},q_{,F})} \over
    { {\mbox{Cov}(q_{,R},q_{,R})^{1/2}}
      {\mbox{Cov}(q_{,F},q_{,F})^{1/2}} }},
\end{equation}
where $q$ is either $S_8$ or $\Omega_m$, and the subscripts $R$ and $F$ stand
for the real-space TPCF and Fourier-space power spectrum, respectively.
We find $r(S_8)= 0.51$ and $r(\Omega_m)=0.17$, which confirms that 
the correlation between derived cosmological constraints from the two 
analyses is weak, especially for $\Omega_m$.
The main reason for this weak correlation is the different multipole 
ranges probed in these two analyses.
\citet{2019PASJ...71...43H} adopted the multipole range $300<\ell<1900$,
whereas in Appendix~\ref{appendix:powerspectrum}, we examine  
the contribution to $\xi_{\pm}$ from different $\ell$-ranges to show
that a large part of the contribution to $\xi_{\pm}$ on scales adopted
in this study comes from $\ell<300$.
This indicates that in deriving cosmological constraints,
these two studies utilize fairly different and complementary information. 

%
%
\section{Summary and conclusions}
\label{sec:summary}

We have presented a cosmological analysis of the cosmic shear TPCFs
measured from the HSC first year data, covering 136.9~deg$^2$ and
including 9.6 million 
galaxies to $i\sim 24.5$ AB mag.
We used the HSC first year shape catalog \citep{2018PASJ...70S..25M},
which is based on the re-Gaussianization PSF correction method
\citep{2003MNRAS.343..459H} and is calibrated with image simulations
\citep{2018MNRAS.481.3170M}.
In order to examine the impact of residual
PSF errors on cosmic shear TPCFs, we utilized the HSC star catalog which
contains information on both the star shapes and PSF models. 
Photometric redshifts derived from the HSC five-band photometry are
adopted to divide galaxies into four tomographic redshift bins ranging
from $z=0.3$ to $1.5$ with equal widths of $\Delta z =0.3$.
The unweighted galaxy number densities for each tomographic bin are
(from the lowest to highest redshift) 6.1, 6.1, 4.6, and 2.7 arcmin$^{-2}$.

In addition to the HSC data set, we utilized HSC mock shape catalogs
constructed based on full-sky 
gravitational lensing ray-tracing simulations \citep{2017ApJ...850...24T}.
The mock catalogs have the same survey geometry and shape noise
properties as the real data \citep{2019MNRAS.486...52S}.
We derived the covariance matrix adopted in our cosmological analysis
from 2268 mock realizations.
The mock catalogs are also used to assess the statistical significance of
some of our results.

Ten combinations of auto and cross tomographic TPCFs were measured with
high signal-to-noise ratio over a wide angular range.
The total signal-to-noise ratio computed over the angular ranges that we
adopted in our cosmological analysis ($7'<\theta<56'$ for $\xi_+$
and $28'<\theta<178'$ for $\xi_-$) was $S/N=18.7$, although 
a caveat is that this estimate depends on the cosmological model used to
derive the covariance matrix; we adopt the WMAP9 cosmology.
We also examined the E/B-mode decomposition of the cosmic shear TPCFs
to test our assumption in the cosmological analysis that the cosmic
shear field is B-mode free.
In appendix \ref{appendix:B-mode}, we evaluated the standard 
$\chi^2$ value for B-mode TPCFs with the shape
noise covariance, and found $\chi^2=84.7$ for $N_d=90$.
We thus conclude that no evidence of significant B-mode shear is found.

We performed a standard Bayesian likelihood analysis for the
cosmological inference of the measured cosmic shear TPCFs.
Our fiducial $\Lambda$CDM model consists of five cosmological
parameters and includes contributions from intrinsic alignment of
galaxies as well as seven nuisance parameters (2 for PSF errors,
1 for shear calibration error, and 4 for source redshift
distribution errors).
We found that our model fits the measured TPCFs very well with a
minimum $\chi^2$ of 162.0 for 167 effective degrees-of-freedom.
Marginalized one-dimensional constraints are (mean and 68\% confidence interval)
$S_8=\sigma_8\sqrt{\Omega_m/0.3}=0.823_{-0.028}^{+0.032}$,  
$\Omega_m=0.332_{-0.096}^{+0.050}$, and
$\sigma_8=0.799_{-0.101}^{+0.112}$.
Although we fixed the neutrino mass of $\sum m_\nu=0.06$~eV in the
fiducial model, we found that varying the neutrino mass has little
effect on the cosmological constraints.
We also tested $w$CDM model to find that allowing the dark energy
equation of state parameter $w$ to vary degrades the $S_8$ constraint to
$S_8=0.816_{-0.053}^{+0.047}$.
We have found that the current HSC cosmic shear
TPCFs alone cannot place a useful constraint on $w$. 

We have carefully checked the robustness of our cosmological results
against astrophysical uncertainties in modeling and systematics
uncertainties in measurements.
The former includes the intrinsic alignment of galaxies and the baryonic
feedback effect on the nonlinear matter power spectrum, and the 
latter includes PSF errors, shear calibration error, errors in the 
estimation of source redshift distributions, and a residual constant 
shear over fields. 
We have tested the validity of our treatment of those uncertainties by
changing parameter setups or by adopting extreme models for them.
We have found that none of these uncertainties has a significant impact
on the cosmological constraints. Specifically, different setups yield
shifts in best-fit $S_8$ values of $\sim 0.6\sigma$ of the statistical
error at most.
We have also confirmed the internal consistency of our results among
different redshift and angular bins.

Our constraint contours in the $\Omega_m$-$\sigma_8$ plane largely 
overlap with those of DES-Y1 \citep{2018PhRvD..98d3528T}, although 
the two contours are slightly misaligned, resulting in about a
$1\sigma$
difference in the best-fit $S_8$ value; our best-fit $S_8$ is 
higher than that from DES-Y1.
A larger difference was found between KiDS+VIKING-450
\citep{2020A&A...633A..69H} and our result.
In fact, the best-fit $S_8$ value from KiDS+VIKING-450 is
$\sim 2\sigma$
lower than our result.
We have found that the $S_8$ constraint from  {\it Planck}
\citep{2020A&A...641A...6P} is consistent with our result within
$1\sigma$ level.
We found that the 68\% confidence
contour in the $\Omega_m$-$\sigma_8$ plane from {\it Planck} nicely
overlaps with our result. 

\citet{2019PASJ...71...43H} used the same HSC first year weak lensing
shape catalog but adopted the cosmic shear power spectra to derive 
cosmological constraints.
We have found about a 1$\sigma$ level difference in $S_8$ constraints 
between the cosmic shear power spectrum analysis and our comic shear
TPCF analysis, even though they share the same shape catalog.
We used mock catalogs to examine the statistical significance of
the difference. We have found that the difference can be explained 
by a statistical fluctuation at about the $1.6\sigma$ level.
We also used the mock catalog to examine the correlation in derived 
cosmological constraints between these two studies, and have found 
the cross-correlation coefficients of $r(S_8)=0.51$ and 
$r(\Omega_m)=0.17$.
The reason for these weak correlations, especially for $\Omega_m$, 
is the different multipole ranges probed in these two analyses.
\citet{2019PASJ...71...43H} adopted the multipole range $300<\ell<1900$,
whereas a large part of the contribution to
$\xi_{\pm}$ over angular ranges adopted in this study comes from
$\ell<300$, indicating that two studies utilize fairly different
and complementary information in deriving cosmological constraints.

In summary, our $S_8$ constraint is located on the high side among 
recent cosmic shear studies and is fully consistent with the latest 
{\it Planck} CMB result.
Among the recent studies mentioned above, only the KiDS+VIKING-450
result is inconsistent with our result at $\sim 2\sigma$ level.
Since the KiDS survey fields largely overlap with HSC survey
fields, it is worth analyzing their public shape catalog with our
methodology to understand its origin, which we leave for future 
work.

This paper presents cosmological results based on the 
HSC 1st-year data.
When the HSC survey is completed, we will have about seven times 
more area, which will improve both the statistical error and 
the cosmic variance.
In addition to this, improvement efforts on several analysis 
techniques are underway, including PSF measurement and modeling 
\citep{2019PASJ...71..114A}, photo-$z$ estimations, 
and shear measurements.
In future, it would be important
to explore other missing redshift-dependent selection biases by using
techniques such as metacalibration
\citep{2017arXiv170202600H,2017ApJ...841...24S}, and it would be also
important to implement more advanced methods of accounting for baryonic
feedback effects such as one proposed by \citet{2015MNRAS.454.2451E}.

Recently, \citet{2020A&A...638L...1J} argued
that the systematic uncertainties in the redshift distribution of
galaxies derived by the 
re-weighted method bases on the COSMOS 30-band photo-$z$
\citep{2009ApJ...690.1236I} might be underestimated and could lead to
a bias in the cosmological constraints due to outliers in the COSMOS
30-band catalog. They showed that $S_8$ constraints from both KiDS-VIKING 450
and DES-Y1 inferred adopting the redshift distributions based on
spectroscopic samples are lower than ones based on COSMOS 30-band
photo-$z$ sample.
A plausible reason for these differences could be the systematic
uncertainties in the COSMOS 30-band photo-$z$, though further close
examination of the redshift distribution is
needed to reach a firm conclusion.
One might deduce from their finding that a similar
bias may exist in our analysis.
Take the case of DES-Y1 for example, the difference between $S_8$
values inferred adopting the two redshift distributions is $|\Delta
S_8|=0.030$ \citep{2020A&A...638L...1J}, which corresponds to
$\sim1\sigma$ of our $S_8$ constraint.
Thus, this is indeed one important issue
to be explored in a future study
(see Appendix
\ref{appendix:photoz_uncertainties} for a related discussion).
However, in our case it is not feasible to use spec-z samples for a
reference sample, because a spec-z sample that reaches the depth of
HSC weak lensing catalog is not available now. A possible method to
calibrate the photo-$z$ without relying on COSMOS 30-band photo-$z$ is
a cross-correlation method \citep{2008ApJ...684...88N} which we will
adopt in a future work.

%
%
\begin{ack}
We would like to thank the anonymous referee for constructive comments
on the earlier manuscript which improved the paper.
We would like to thank R.~Takahashi for useful discussions and for
making full-sky gravitational lensing simulation data publicity
available. 
We would like to thank Martin Kilbinger for making the software {\tt
  Athena} publicly available, Antony Lewis and Anthony Challinor for
making the software {\tt CAMB} publicly available, {\tt MultiNest}
developers for {\tt MultiNest} publicly available, and HEALPix team for
HEALPix software publicity available.

This work was supported in part by World Premier International
Research Center Initiative (WPI Initiative), MEXT, Japan, JSPS KAKENHI
Grant Number JP15H05887, JP15H05892, JP15H05893, 
17H06600, JP17K05457, 18H04350, 18H04358, and JP18K03693.
A portion of this research was carried out at the Jet
Propulsion Laboratory, California Institute of Technology, under a contract
with the National Aeronautics and Space Administration.
RMa is supported by the Department of Energy Cosmic Frontier program,
grant DE-SC0010118.

Data analysis were in part carried out on PC cluster at Center for
Computational Astrophysics, National Astronomical Observatory of
Japan. Numerical computations were in part carried out on Cray XC30 and
XC50 at Center for Computational Astrophysics, National Astronomical
Observatory of Japan, and also on Cray XC40 at YITP in Kyoto
University.

The Hyper Suprime-Cam (HSC) collaboration includes the astronomical communities 
of Japan and Taiwan, and Princeton University.  
The HSC instrumentation and software were developed by the National Astronomical 
Observatory of Japan (NAOJ), the Kavli Institute for the Physics and Mathematics 
of the Universe (Kavli IPMU), the University of Tokyo, the High Energy Accelerator 
Research Organization (KEK), the Academia Sinica Institute for Astronomy and 
Astrophysics in Taiwan (ASIAA), and Princeton University.  
Funding was contributed by the FIRST program from Japanese Cabinet Office, 
the Ministry of Education, Culture, Sports, Science and Technology (MEXT), 
the Japan Society for the Promotion of Science (JSPS),  Japan Science
and Technology Agency (JST),  the Toray Science  Foundation, NAOJ, Kavli
IPMU, KEK, ASIAA, and Princeton University.
This paper makes use of software developed for the Large Synoptic Survey
Telescope. We thank the LSST Project for making their code available as
free software at \url{http://dm.lsst.org}

The Pan-STARRS1 Surveys (PS1) have been made possible through 
contributions of the Institute for Astronomy, the University of Hawaii, 
the Pan-STARRS Project Office, the Max-Planck Society and its 
participating institutes, the Max Planck Institute for Astronomy, 
Heidelberg and the Max Planck Institute for Extraterrestrial Physics, 
Garching, The Johns Hopkins University, Durham University, the 
University of Edinburgh, Queen's University Belfast, the 
Harvard-Smithsonian Center for Astrophysics, the Las Cumbres Observatory 
Global Telescope Network Incorporated, the National Central University 
of Taiwan, the Space Telescope Science Institute, the National 
Aeronautics and Space Administration under Grant No. NNX08AR22G issued 
through the Planetary Science Division of the NASA Science Mission 
Directorate, the National Science Foundation under Grant No. AST-1238877, 
the University of Maryland, and Eotvos Lorand
University (ELTE) and the Los Alamos National Laboratory.

Based in part on data collected at the Subaru Telescope and retrieved
from the HSC data archive system, which is operated by Subaru Telescope
and Astronomy Data Center at National Astronomical Observatory of Japan.
\end{ack}

\bibliographystyle{apj}

\appendix

%
%
\section{Mean shear values over fields}
\label{sec:mean_shear}

%
%
\begin{figure}[t]
\begin{center}
 \includegraphics[width=82mm]{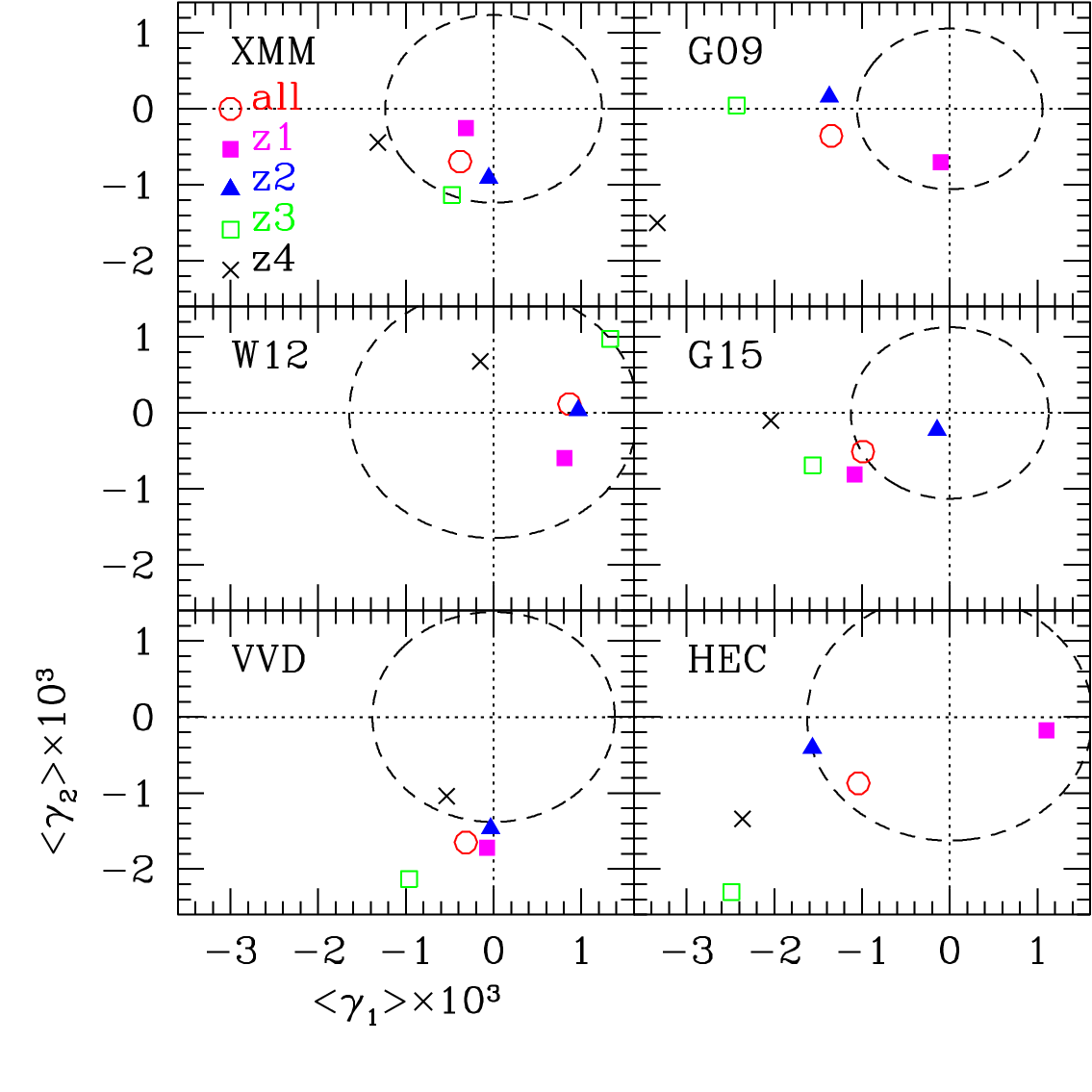}
\end{center}
\caption{Measured mean shear values over each field are plotted for each
  tomographic sample (magenta filled squares, blue triangles, green open
  squares, and black crosses from the lowest to
  highest redshift tomographic bins, respectively),
  and for the combined sample of all four 
  tomographic bins (red open circles).
  The dashed circle shows the 68.3\% enclosing mean cosmic shear value for
  the combined sample, as estimated from mock catalogs (see
  Appendix~\ref{sec:mean_shear}).  
  The 68.3\% enclosing mean shear values for each tomographic 
  bin are about 0.77, 1.1, 1.4, and 1.7 times larger than those for
  the combined sample for the lowest to
  highest redshift tomographic bins, respectively.
  \label{fig:conste}}
\end{figure}

%
%
\begin{table}
\caption{Summary of $\chi^2$ and $p$-values of the mean shear over each
  field, where there are 8 degrees of freedom (2 shear components
  multiplied by 4 tomographic bins), and the covariance matrix is
  derived from mock catalogs (see
  Appendix~\ref{appendix:mock}). \label{table:conste}}
\begin{center}
  \begin{tabular}{lcc}
    \hline
    Field & $\chi^2$ & $p$-value \\
    \hline
    XMM      & 4.6  & 0.80 \\
    GAMA09H  & 17.2 & 0.028 \\
    WIDE12H  & 5.2  & 0.73 \\
    GAMA15H  & 13.7 & 0.089 \\
    VVDS     & 9.1  & 0.33 \\
    HECTOMAP & 14.8 & 0.063 \\
    \hline
  \end{tabular}
\end{center}
\begin{tabnote}
  {}
\end{tabnote}
\end{table}

The value of the shear averaged over a field is not expected to be zero
due to the presence of the cosmic shear signal on scales larger than a field. 
However, it could also be non-zero due to residual systematics in the shear 
estimation and/or data reduction process.
The latter, if present, may bias the cosmological inference.
While systematic tests on the HSC 1st-year shape catalog
\citep{2018PASJ...70S..25M} found no evidence of a mean shear above 
that expected from large-scale cosmic shear, we closely reexamine this
question here
because the shear correlation function, especially $\xi_+$, is directly
affected by the residual mean shear.

The measured mean shear values over each field are shown in
Figure~\ref{fig:conste} 
for each tomographic sample, as well as for the combined sample of 
the four tomographic bins. From those plots, we find that 
mean shear values for each field are about $|\gamma | \sim 10^{-3}$.
In order to estimate the amplitude of the mean shear caused by the
cosmic shear signal on scales larger than a field, we use a set of 2268 mock
catalogs described in Appendix~\ref{appendix:mock}. For each field and
for each tomographic sample from a mock catalog, we measure mean 
shear values. We repeat this measurement for each of the 2268 mock 
catalogs, and sort the mean shear values to find a 68.3\% enclosing mean
shear value below which 1549 mock samples are enclosed.
The results for the combined sample of the four
tomographic bins are shown in Figure~\ref{fig:conste} as 
the dashed-line circle for each field, with slightly different circle sizes 
for mean shear values of individual tomographic bins (see the caption 
of Figure~\ref{fig:conste} for more details).
We note that the mean shear value expected from the intrinsic shape 
noise is $\sigma_e /\sqrt{N_g} \sim 0.3/\sqrt{10^{6-7}} \sim
O(10^{-4})$, where $\sigma_e$ is the root-mean-square value of the
intrinsic galaxy distortion (in shear units) and $N_g$ is the number of
galaxies. This value is much smaller than the mean shear from the
mock catalogs, suggesting that the mean shear value is indeed 
dominated by cosmic shear that is coherent over the field.
We find from Figure~\ref{fig:conste} that most of the measured values
are located within the 68.3\% enclosing circle, which is consistent 
with the finding in \citet{2018PASJ...70S..25M}. 
In fact, only the highest redshift tomographic bin of GAMA09H field 
has a mean shear beyond the 95.5\% range 
($|\gamma|=3.0\times 10^{-3}$ for this case).

In addition to the above test, we estimate a statistical significance of
the measured mean shears against the null hypothesis that they 
arise solely from the cosmic shear as follows.
Using the data set of mean shears measured from the tomographic mock
catalogs, we derive the covariance matrix  (see
Appendix~\ref{appendix:cov_mean_shear} for details),
$\mbox{Cov}(d_i,d_j)$, where $d_i$ is the data vector consisting of 
mean values of the two shear components in each of the four tomographic
bin, namely 
$d_i=(\bar{\gamma}_1^1,\bar{\gamma}_1^2,\bar{\gamma}_1^3,\bar{\gamma}_1^4,
\bar{\gamma}_2^1,\bar{\gamma}_2^2,\bar{\gamma}_2^3,\bar{\gamma}_2^4)$.
Given this covariance matrix for each field, we compute $\chi^2$ of the
data relative to the null hypothesis;
the results are summarized in Table~\ref{table:conste} along
with the corresponding $p$-values. 
The $p$-values are reasonable for all fields
except the GAMA09H field, in which the $p$-value is slightly
smaller than the conventional criterion of 0.05.
However, since we measured the mean shear independently in six fields, 
the chances of getting one field with a $p$ value less than 0.028 is 
$1 - 0.972^6  = 0.16$. 
Thus we conclude that the measured mean shears are consistent with 
that expected from large-scale cosmic shear. 

Although we have found no clear evidence of additive 
shear bias arising from residual systematics, 
we check the impact of such a possible residual shear on our cosmological analysis 
by modeling it as a redshift-independent constant shear, which we 
denote as $\bar{\gamma}$.
We expect that the redshift-independent
constant shear is a reasonable assumption for the following
reason.
The redshift-dependence of shape measurements may arise
from the difference of galaxy properties such as sizes between
different redshifts, which are estimated as the selection biases,
$m_{\rm sel}^a$ and $m_R^a$ (see Section
\ref{sec:selection-bias} and Section 5.7 of
\citealt{2019PASJ...71...43H}).  
We find that the variation in the redshift-dependent selection biases
(to be specific $1+m_{\rm sel}^a+m_R^a$) among four tomographic bins is
$\sim2\%$ at largest (see Table 3 of \citealt{2019PASJ...71...43H}).
In our systematics tests in Sec.~\ref{sec:Systematic_parameters}, 
we add $\bar{\gamma}$ to 
the {\it theoretical model} of $\xi_+$, and than marginalize over $\bar{\gamma}$
to see how cosmological constraints change.
We note that $\xi_-$ is unaffected by this constant shear due to
the cancellation between $\langle \gamma_t \gamma_t\rangle$ and 
$\langle \gamma_\times \gamma_\times \rangle$.

%
%
\section{PSF leakage and residual PSF model errors}
\label{appendix:residual_PSF}

%
%
\begin{figure}
  \begin{center}
    \includegraphics[width=82mm]{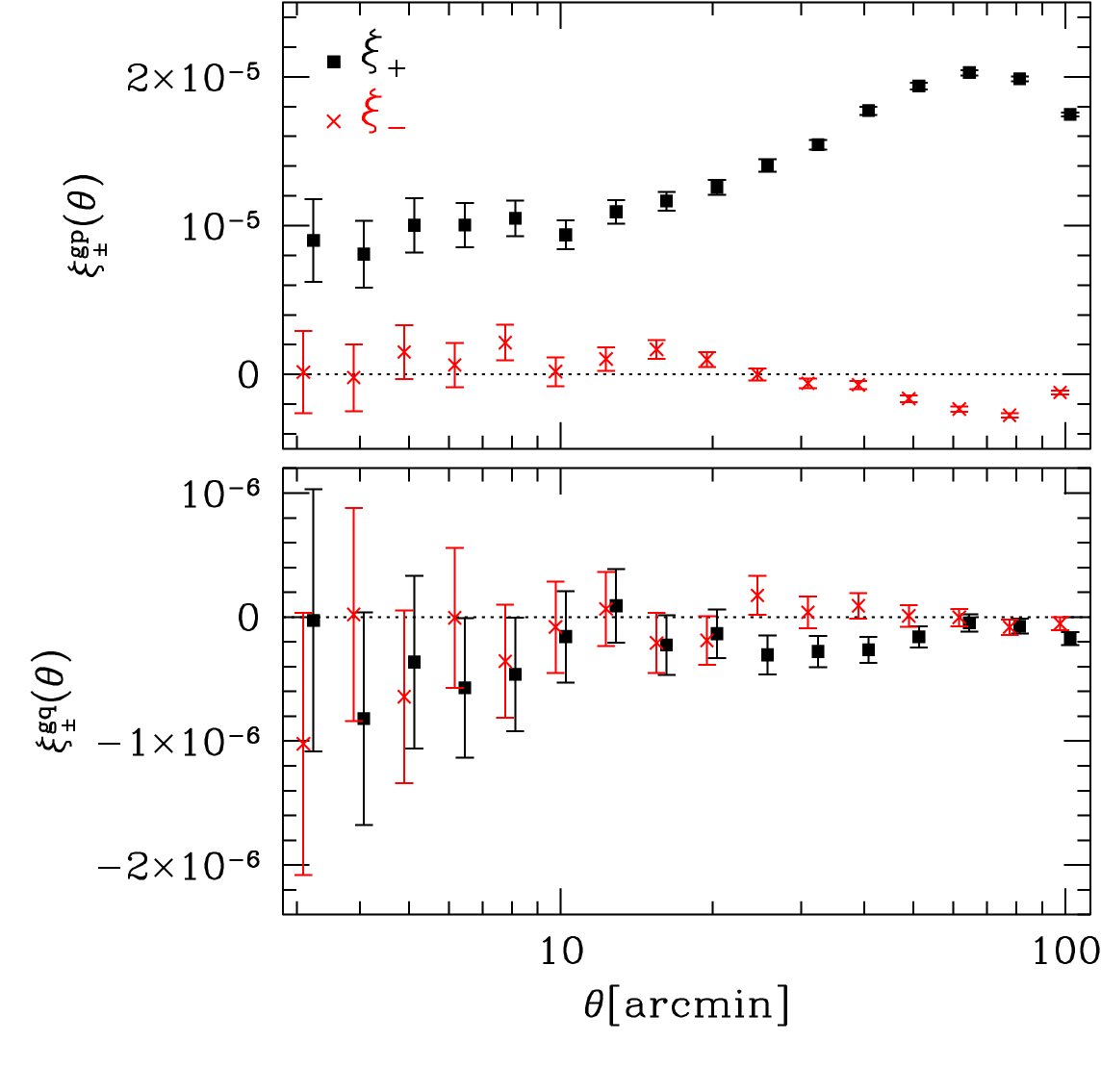}
  \end{center}
  \caption{Upper (bottom) panel shows the cross correlation function
    between galaxy shears and PSF shapes  (difference in shapes between the
    PSF and stars) converted into shear. Filled squares and red crosses
    are for $\xi_+$ and $\xi_-$, respectively.
    In the bottom panel, points are horizontally shifted slightly for clarity.
    In measuring these signals, the combined catalog of the four tomographic
    redshift bins is used for the galaxy shear sample, and reserved
    stars (as described in Section~\ref{sec:PSF-TPCS}) are used for the
    PSF sample. 
    \label{fig:xipm_gp_gq}}
\end{figure}

%
%
\begin{figure}
  \begin{center}
    \includegraphics[width=82mm]{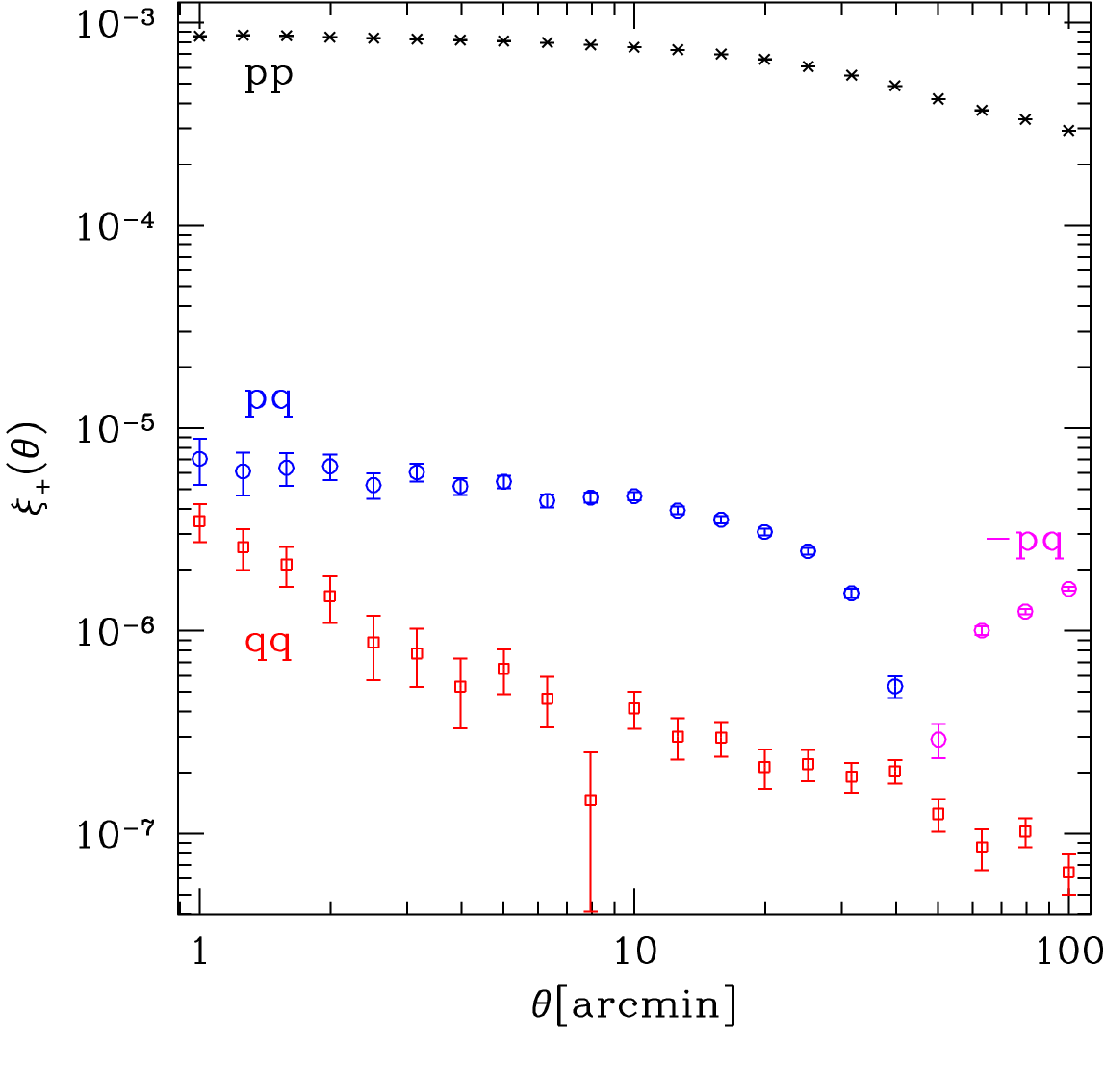}
  \end{center}
  \caption{The auto- and cross-correlation functions between PSF shapes
    ($\gamma^p$) and the difference between PSF and star shapes
    ($\gamma^q$), i.e., $\xi_+^{pp}$ (black crosses), $\xi_+^{qq}$ (red
    squares), and $\xi_+^{pq}$ (blue bars for $\xi_+^{pq}>0$, and
    magenta bars for $\xi_+^{pq}<0$ plotted as $-\xi_+^{pq}$).
    \label{fig:xip_pp_qq_pq}}
\end{figure}

In this Appendix, we examine the impact of PSF leakage and residual PSF
model error on the measurement of shear correlation functions, employing
the simple linear model as described in equations~(\ref{eq:g_psf}) and 
(\ref{eq:xi_psf}). The model parameters, $\alpha_{\mbox{psf}}$ and
$\beta_{\mbox{psf}}$, can be estimated by the cross correlation functions between
$\gamma^{p,q}$ and galaxy shears, $\xi^{gp,gq}=\langle \hat{\gamma}
\gamma^{p,q} \rangle$, which are related to $\xi_\pm^{pp,pq,qq}$ as
\begin{eqnarray}
  \label{eq:xi_gp_gq}
  \xi_{\pm}^{gp}&=&\alpha_{\mbox{psf}}\xi_\pm^{pp}+\beta_{\mbox{psf}}\xi_\pm^{pq},\\
  \label{eq:xi_gp_gq2}
  \xi_{\pm}^{gq}&=&\alpha_{\mbox{psf}}\xi_\pm^{pq}+\beta_{\mbox{psf}}\xi_\pm^{qq}.
\end{eqnarray}
In measuring these quantities, we use reserved stars, which are described 
in more detail in Section~\ref{sec:PSF-TPCS}, for the PSF sample, and for
the galaxy shear sample, we use the combined catalog of the four
tomographic redshift bins, because the measurement of
$\xi_-^{gp,gq}$ is very noisy as shown below. 
As a consequence, we do not take into account possible redshift dependence of
$\alpha_{\mbox{psf}}$ and $\beta_{\mbox{psf}}$.
See Section~4.2 of \citet{2019PASJ...71...43H} for further discussions on
this point.

%
%
\begin{figure}
  \begin{center}
    \includegraphics[width=82mm]{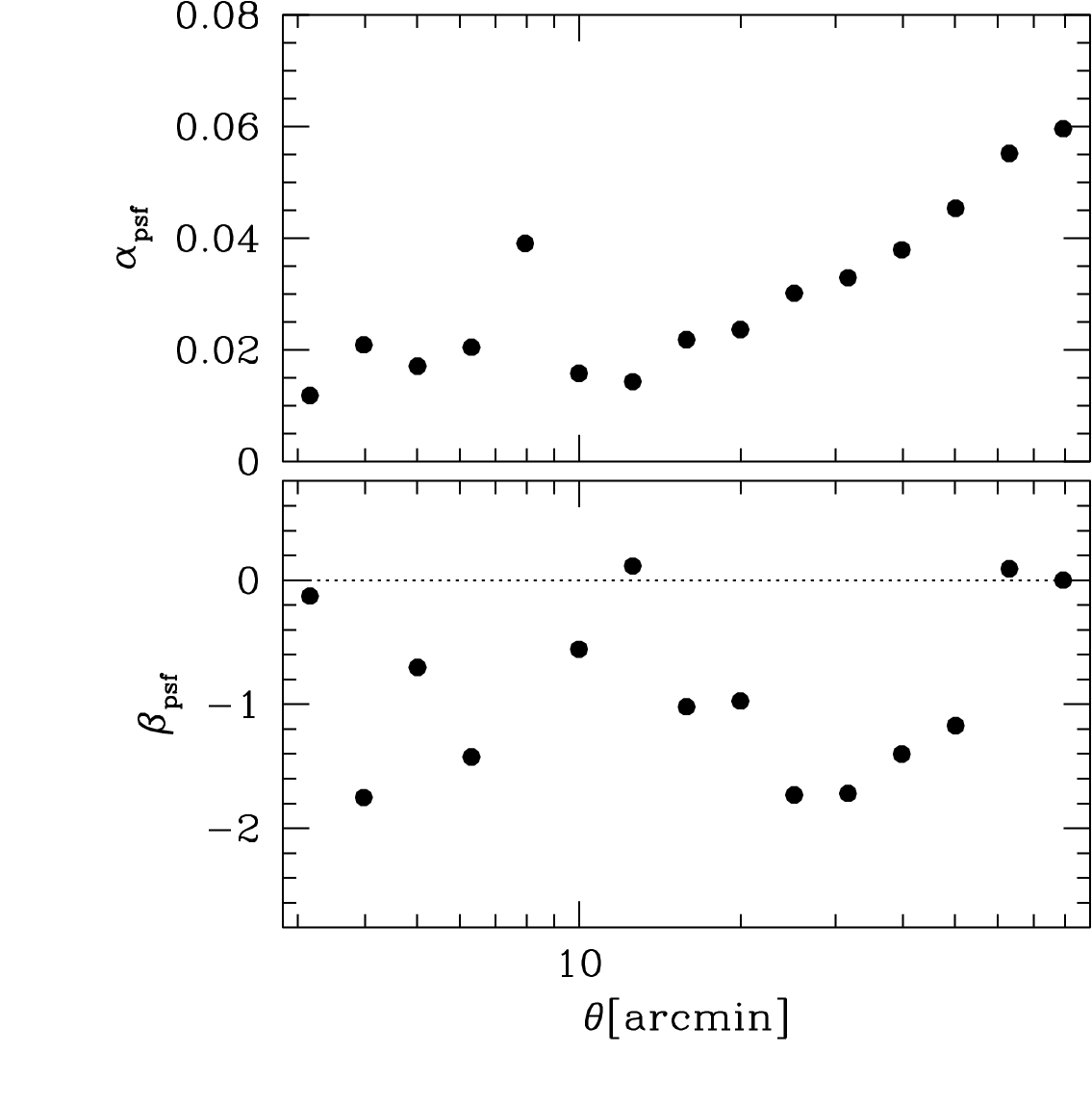}
  \end{center}
  \caption{Model parameters in PSF leakage and residual PSF model
    derived for $\xi_+$ using equations~(\ref{eq:xi_gp_gq}) and
    (\ref{eq:xi_gp_gq2}).
    Error bars, which largely come from errors on $\xi_+^{gq}$ (see
    the lower panel of Figure~\ref{fig:xipm_gp_gq}), are
    not shown.
    \label{fig:psf_alp_beta_xip_allz}}
\end{figure}

We first consider the $\xi_+$ component.
The measured $\xi_\pm^{gp,gq}$ are shown in Fig \ref{fig:xipm_gp_gq}, 
where the error bars represent the shape noise.
As shown in the upper panel, we obtain high signal-to-noise ratio detections
for $\xi_+^{gp}$ over a wide angular range.  
The signal-to-noise ratios for $\xi_-^{gq}$ are marginal, but there is a clear 
trend toward negative values.
Using these measured values, together with $\xi_+^{pp,pq,qq}$ shown in
Figure~\ref{fig:xip_pp_qq_pq}, we derive $\alpha_{\mbox{psf}}$ and
$\beta_{\mbox{psf}}$ with equations~(\ref{eq:xi_gp_gq}) and (\ref{eq:xi_gp_gq2}).
The results are shown in Figure~\ref{fig:psf_alp_beta_xip_allz}, where
we omit error bars which are dominated by errors on $\xi_+^{gq}$
(see the lower panel of Figure~\ref{fig:xipm_gp_gq}). 
Taking the simple average and standard deviation of the 9 points in the 
angular range from $\sim 8'$ to $\sim 50'$, the range is used in the
cosmological analysis in this study, we find
$\alpha_{\mbox{psf}}=0.029\pm 0.010$ and $\beta_{\mbox{psf}}=-1.42\pm
1.11$, which we adopt as the prior ranges of these parameters.
\citet{2019PASJ...71...43H} derived the same quantities with the same
data set but in the power spectrum analysis, and found
$\alpha_{\mbox{psf}}=0.057\pm 0.018$ and $\beta_{\mbox{psf}}=-1.22\pm 0.74$. 
There is a difference in the central values of $\alpha_{\mbox{psf}}$, 
although they are marginally consistent with each other.
This difference might reflect the different angular ranges between 
the two studies (see Appendix \ref{appendix:powerspectrum}).

%
%
\begin{figure}
  \begin{center}
    \includegraphics[width=82mm]{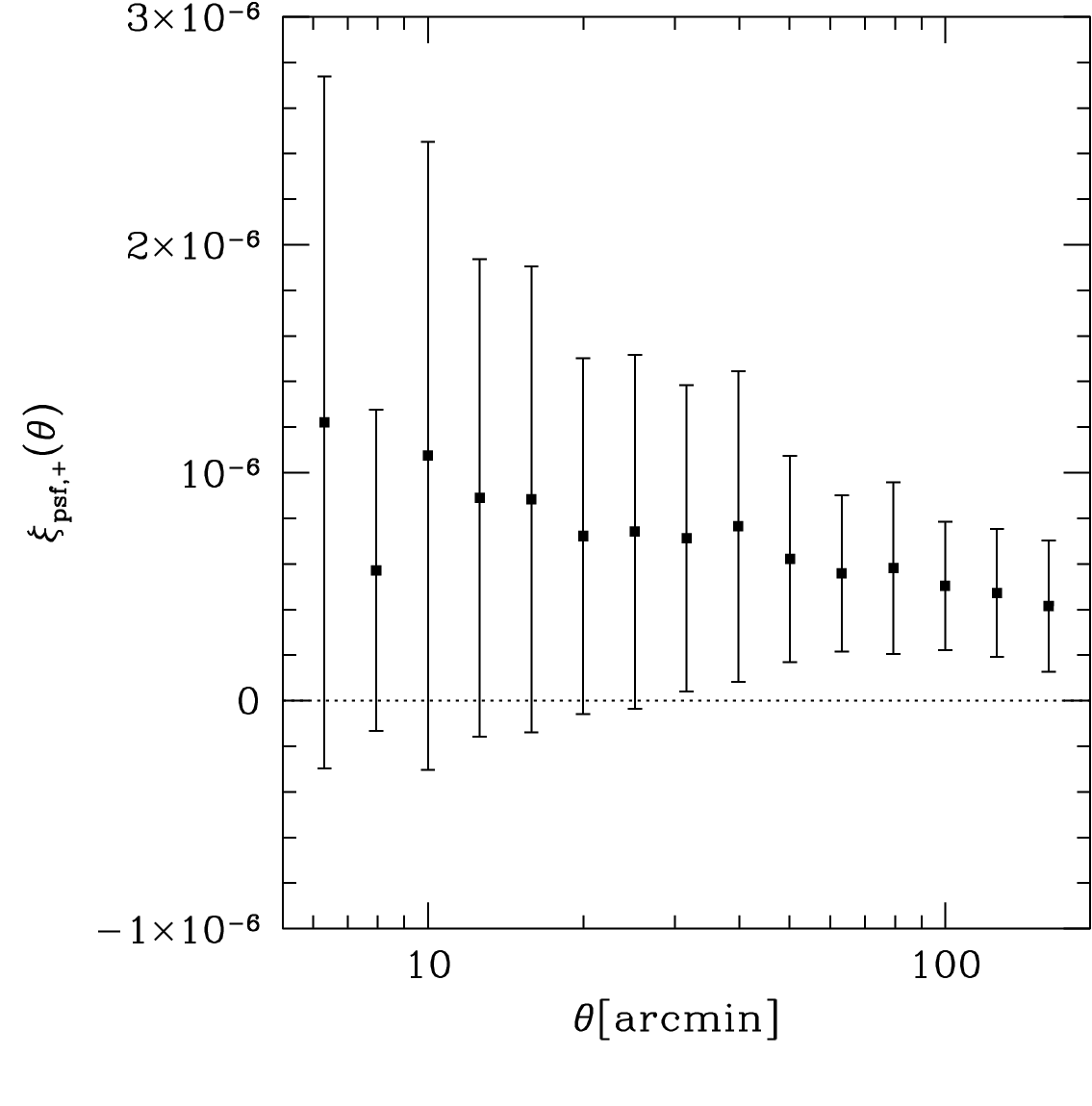}
  \end{center}
  \caption{$\xi_{\mbox{psf},+}$ defined in equation~(\ref{eq:xi_psf}) is
    shown. Here we adopt
    $\alpha_{\mbox{psf}}=0.029\pm 0.010$ and 
    $\beta_{\mbox{psf}}=-1.42\pm 1.11$. Errors are computed from those of
    $\alpha_{\mbox{psf}}$ and $\beta_{\mbox{psf}}$.
    \label{fig:xipsf}}
\end{figure}

Using the derived parameter values, we compute an estimate of the impact of
the PSF errors on $\xi_+$, namely $\xi_{\mbox{psf},+}$ defined in 
equation~(\ref{eq:xi_psf}).
The result is shown in Figure~\ref{fig:xipsf}, where 
error bars are computed from those of
$\alpha_{\mbox{psf}}$ and $\beta_{\mbox{psf}}$.
The derived $\xi_{\mbox{psf},+}$ should be considered as a rough 
estimate because it is based on the simple
linear model, equations~(\ref{eq:g_psf}) and (\ref{eq:xi_psf}).
Taking into account the large error bars, it is reasonable to
conclude that $\xi_{\mbox{psf},+}$ is about $10^{-6}$ on scales
$5'<\theta<60'$.

%
%
\begin{figure}
  \begin{center}
    \includegraphics[width=82mm]{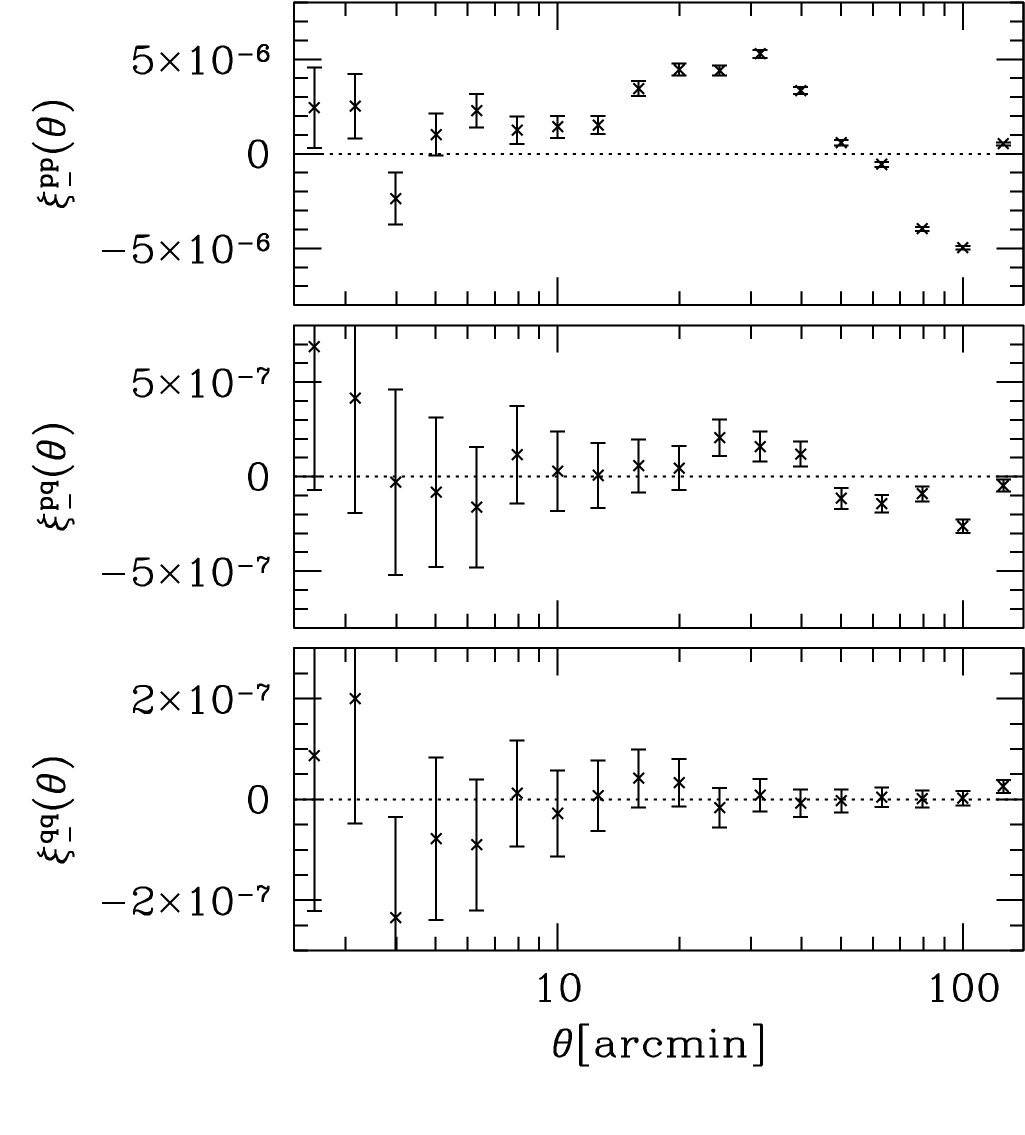}
  \end{center}
  \caption{From top to bottom panels, $\xi_-^{pp}$, $\xi_-^{pq}$, and
    $\xi_-^{qq}$ are shown. See Section~\ref{sec:cosmic-shear-TPCS} for
    their definitions and details of measurements. Error bars represent 
    the shape noise. 
    \label{fig:xim_pp_qq_pq}}
\end{figure}

Next we measure the $\xi_-$ component.
The measured $\xi_-^{pp,qp,qq}$ are shown in
Figure~\ref{fig:xim_pp_qq_pq} and  
$\xi_-^{gp,gq}$ are shown in Figure~\ref{fig:xipm_gp_gq}.
The $SN$s are lower compared with the corresponding $\xi_+$ components, 
$\xi_-^{qq}$ and $\xi_-^{gq}$ are especially noisy.
We thus cannot measure $\alpha_{\mbox{psf}}$ and $\beta_{\mbox{psf}}$
from $\xi_-$ alone. 
In order to examine the impact of PSF leakage and residual PSF
model errors on the cosmic shear $\xi_-$, we employ the estimates 
from $\xi_+$ instead.
Taking $\alpha_{\mbox{psf}}\sim0.03$ and $\beta_{\mbox{psf}}\sim-1.4$,
we find the additional PSF term in equation~(\ref{eq:xi_psf})
is about $-1\times 10^{-8}$ at $\theta \sim 1$ degree, which is more
than two orders of magnitude smaller than the cosmic shear signals.
Thus for $\xi_-$, we do not apply any correction for systematics 
caused by the residual PSF and PSF model.

%
%
\section{E/B-mode cosmic shear TPCFs}
\label{appendix:B-mode}

%
%
\begin{figure}
  \begin{center}
    \includegraphics[width=82mm]{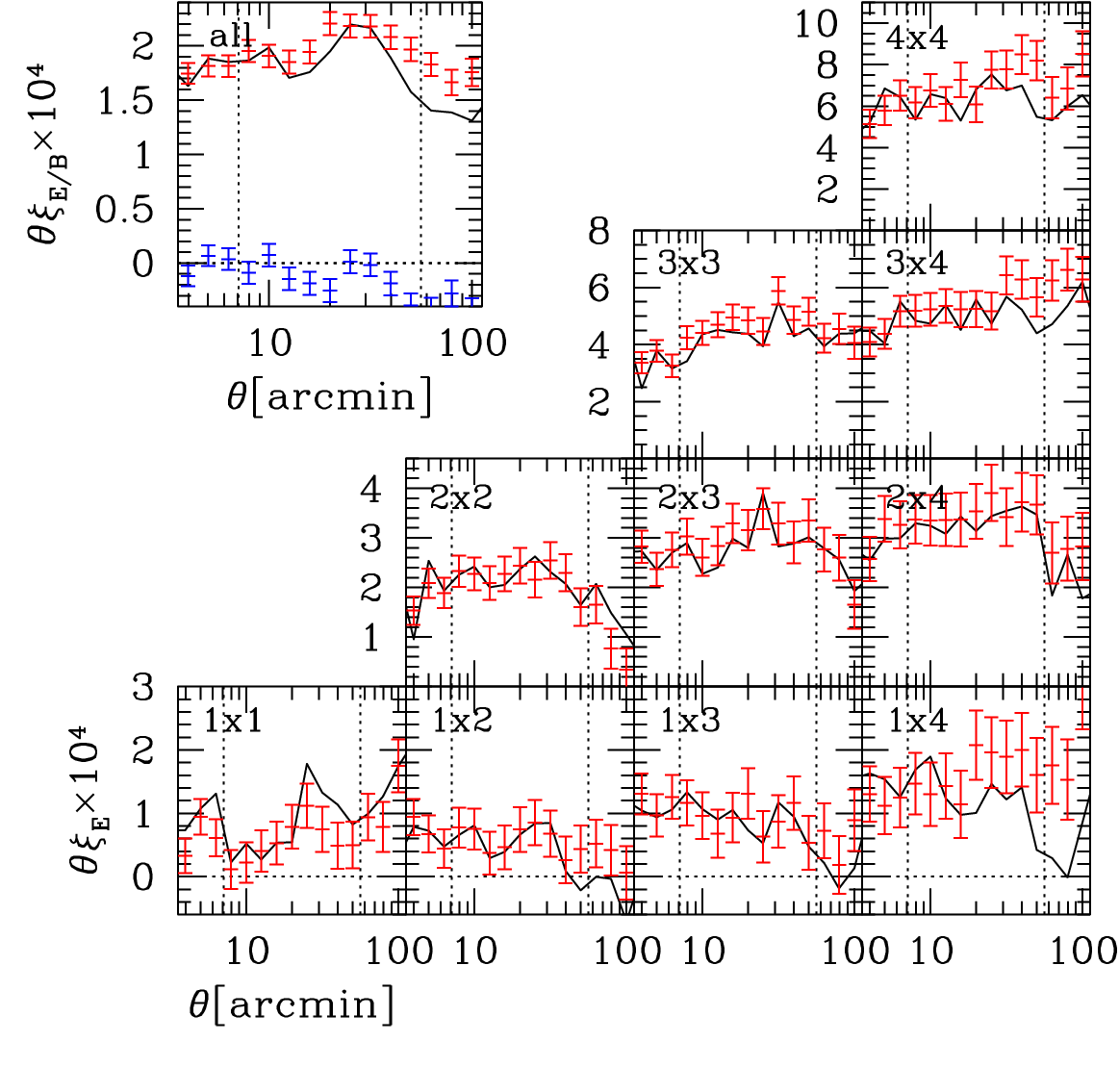}
  \end{center}
  \caption{Bottom right triangular tiled plots: The
    measured E-mode tomographic shear correlation function $\xi_E$ (red
    bars with error bars) compared with $\xi_+$ (black solid line).    
    Combinations of tomographic redshift bins are labeled in each plot.
    Top left panel: E-mode (red symbols) and B-mode (blue symbols)
    non-tomographic (the galaxies in all four tomographic bins
    $0.3<z<1.5$ are combined) shear correlation functions.
    Error bars represent the shape noise for $\xi_{E/B}$.
    Vertical dotted lines show the angular range (for $\xi_+$) used for
    the cosmological analysis.
    \label{fig:xielin_4x4}}
\end{figure}

%
%
\begin{figure}
  \begin{center}
    \includegraphics[width=82mm]{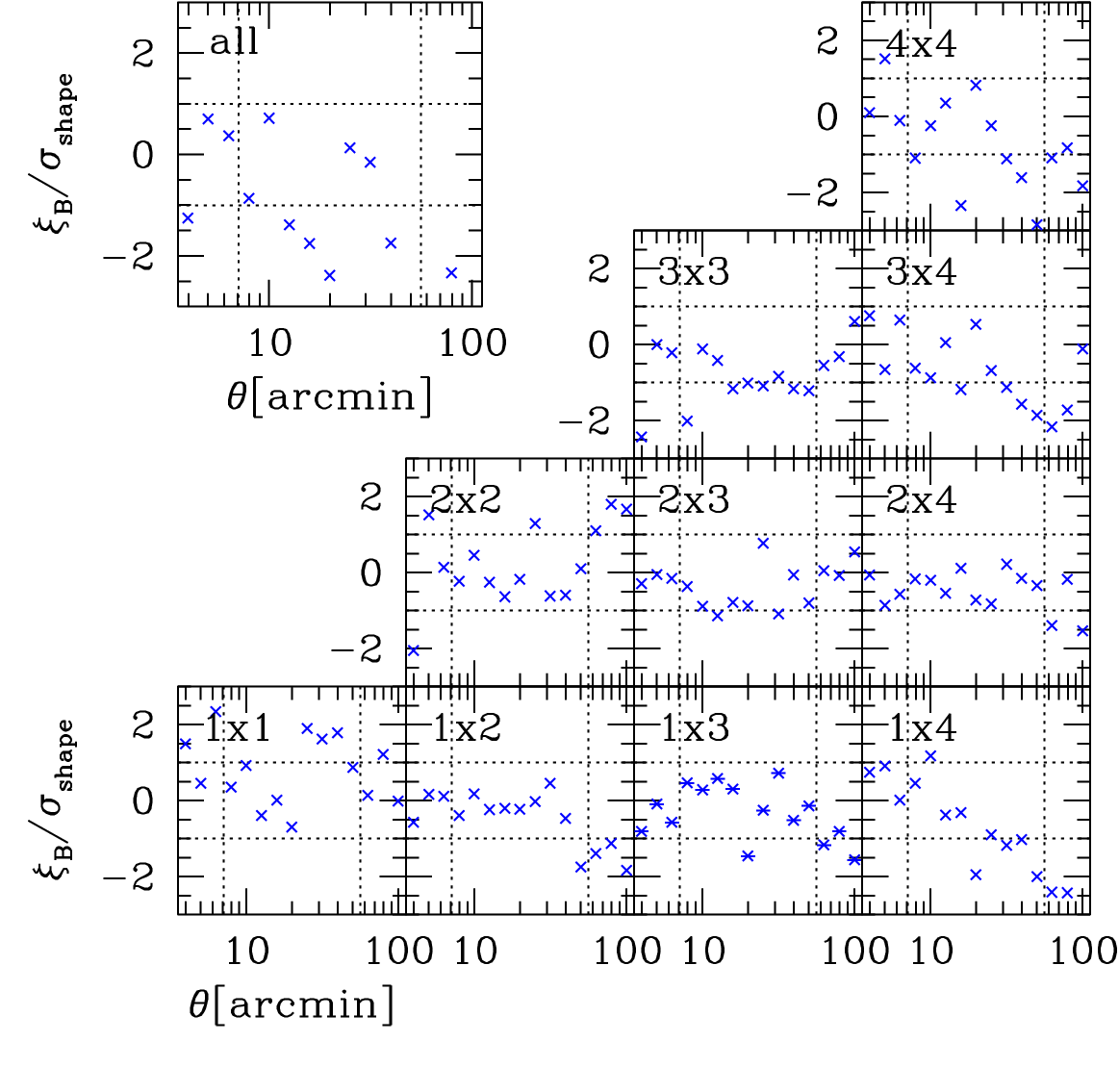}
  \end{center}
  \caption{Bottom right triangular tiled plots: The
    measured B-mode tomographic shear correlation function $\xi_B$
    normalized by the shape noise $\sigma$ for $\xi_B$.
    Top left panel: B-mode non-tomographic shear correlation functions
    normalized by the shape noise.
    Vertical dotted lines show the angular range (for $\xi_+$) used for
    the cosmological analysis, whereas the horizontal dotted lines
    represent $\pm 1$. 
    \label{fig:xiB_4x4}}
\end{figure}

In this Appendix, we present E/B-mode (gradient/curl-mode) decomposition
of the cosmic shear TPCFs \citep{2002ApJ...568...20C}.
The purpose here is to test our assumption that the cosmic shear field
we used for the cosmological analysis is consistent with being B-mode free
as expected from gravitational lensing by a scalar gravitational field
\citep{1992ApJ...388..272K}.
Note that the B-mode shear component in the HSC first year shear
catalog was examined by \citet{2018PASJ...70S..26O} and
\citet{2019PASJ...71...43H}:
The former looked into the B-mode aperture mass map, whereas the latter
used the cosmic shear power spectra in the multipole range of
$300<\ell<1900$, and both concluded that the B-mode 
component is consistent with zero.
Here we examine the E/B-mode tomographic shear TPCFs, allowing us to
closely examine B-mode signals both for individual tomographic bins and
for individual $\theta$-bins of $\xi_+(\theta)$.

The E/B-mode shear TPCFs are given via $\xi_\pm$ as 
\begin{eqnarray}
  \label{eq:xi_E}
  \xi_E(\theta) &=&
     {
       {\xi_+(\theta)+\xi'(\theta)}
       \over 2
     },\\
  \label{eq:xi_B}
  \xi_B(\theta) &=&
     {
       {\xi_+(\theta)-\xi'(\theta)}
       \over 2
     },
\end{eqnarray}
where
\begin{equation}
  \label{eq:xi_dash}
  \xi'(\theta) = \xi_-(\theta)
  + 4 \int_\theta^\infty {d\phi \over \phi} \xi_-(\phi)
  - 12\theta^2 \int_\theta^\infty {d\phi \over {\phi^3}} \xi_-(\phi).
\end{equation}
In the computation of the two integrals of
equation~(\ref{eq:xi_dash}), we measure $\xi_-(\theta)$ in the $\theta$-range
$0\farcm 16\le\theta\le 416'$ in equal log-intervals of
$\Delta \log \theta = 0.02$.
In order to complete the integrals in
equation~(\ref{eq:xi_dash}) beyond $\theta=416'$, we use the
theoretical model with the WMAP9 $\Lambda$CDM cosmology
\citep{2013ApJS..208...19H}. 
The result is not sensitive to the choice of the cosmological
model for the angular range we adopt for $\xi_+$ 
($7\farcm 1\leq \theta \leq 56'$). 

The measured E/B-mode TPCFs are shown in Figures~\ref{fig:xielin_4x4} and
\ref{fig:xiB_4x4}, where the error bars represent the shape noise for
$\xi_{E/B}$.  
In order to evaluate the significance of the B-mode, we compute the standard
$\chi^2$ value for the null signal, for tomographic B-mode TPCFs with 
the shape noise covariance estimated from the data.
We adopt the angular range of our fiducial choice for $\xi_+$, which is  
shown with dotted vertical lines in Figure~\ref{fig:xiB_4x4}, and we 
combine all 10 tomographic combinations.
We find $\chi^2=84.7$ for $N_d=90$, leading to a $p$-value of 0.64.
Therefore we safely conclude that no evidence for a significant 
B-mode shear is found. 

%
%
\section{Mock simulation data}
\label{appendix:mock}

Here we describe the HSC mock shape catalogs, focusing on aspects which are
directly relevant to this study.
See \citet{2019MNRAS.486...52S} for a full description of how the mock data
were constructed, and a comprehensive study of the covariance matrix.

Mock catalogs are constructed based on 108 realizations of the full-sky
gravitational lensing ray-tracing simulation through a large set of
cosmological $N$-body simulations
\citep{2017ApJ...850...24T}\footnote{The full-sky light-cone simulation
  data are freely available for download at {\tt
    http://cosmo.phys.hirosaki-u.ac.jp/takahasi/allsky\_raytracing/}.}. 
The simulations adopt a flat $\Lambda$CDM cosmology which is consistent
with the WMAP9 cosmology \citep{2013ApJS..208...19H} with $\Omega_c =
0.233$, $\Omega_b = 0.046$, the total matter density $\Omega_m
= \Omega_c + \Omega_ b= 0.279$, $\Omega_\Lambda = 1-\Omega_m = 0.721$,
$h = 0.7$, $\sigma_8 = 0.82$, and $n_s = 0.97$.
The lensing data (convergence and shear) are computed on {\tt HEALPix}
\citep{2005ApJ...622..759G} format grids with a grid spacing of 
$0\farcm42$, and on 38 source 
planes with a regular radial interval of comoving 150$h^{-1}$Mpc.
The most distant source plane is located at $z=5.3$.
The degree of independence in 108 full-sky realizations has been studied in 
\citet{2017MNRAS.470.3476S}, who show that the 108 full-sky maps can be 
safely regarded as independent realizations.

From each full-sky lensing data, 21 non-overlapping HSC footprints
are taken, yielding a total of $21\times 108 =2268$ independent mock samples.
Here we briefly describe the procedure for constructing HSC mock shape 
catalogs, referring interested readers to \citet{2019MNRAS.486...52S},
\citet{2014ApJ...786...43S}, \citet{2017MNRAS.470.3476S}, and 
\citet{2018PASJ...70S..26O} for more details. 
For each mock realization, galaxy positions are taken from the real HSC
shape catalog to keep exactly the same survey geometry including
masked regions.
The same tomographic redshift sampling as the real sample is made based
on the same point estimator of photo-$z$'s. 
The redshift of each galaxy is drawn randomly according to the photo-$z$ PDF
$P(z)$ for each mock realization. 
The intrinsic galaxy shape and shape measurement noise are taken from the two
component distortion $(e_1, e_2)$ of the real HSC shape catalog (an estimate of
measurement noise is also given in the catalog) but a random rotation is
applied to erase the cosmic shear signal in the real catalog.
This allows us to preserve both the intrinsic shape noise and
measurement noise in the statistical sense.
Finally, the lensing shear and convergence are taken from full-sky
simulation data for each galaxy, and mock distortion data, $(e_1, e_2)$,
were computed using the relationship between the observed (i.e., lensed)
and intrinsic galaxy shapes under the action of gravitational lensing
\citep[e.g.,][]{1991ApJ...380....1M,2002AJ....123..583B}.

%
%
\subsection{Covariance of mean shears over fields}
\label{appendix:cov_mean_shear}

%
%
\begin{figure}[t]
  \begin{center}
    \includegraphics[height=82mm,angle=-90]{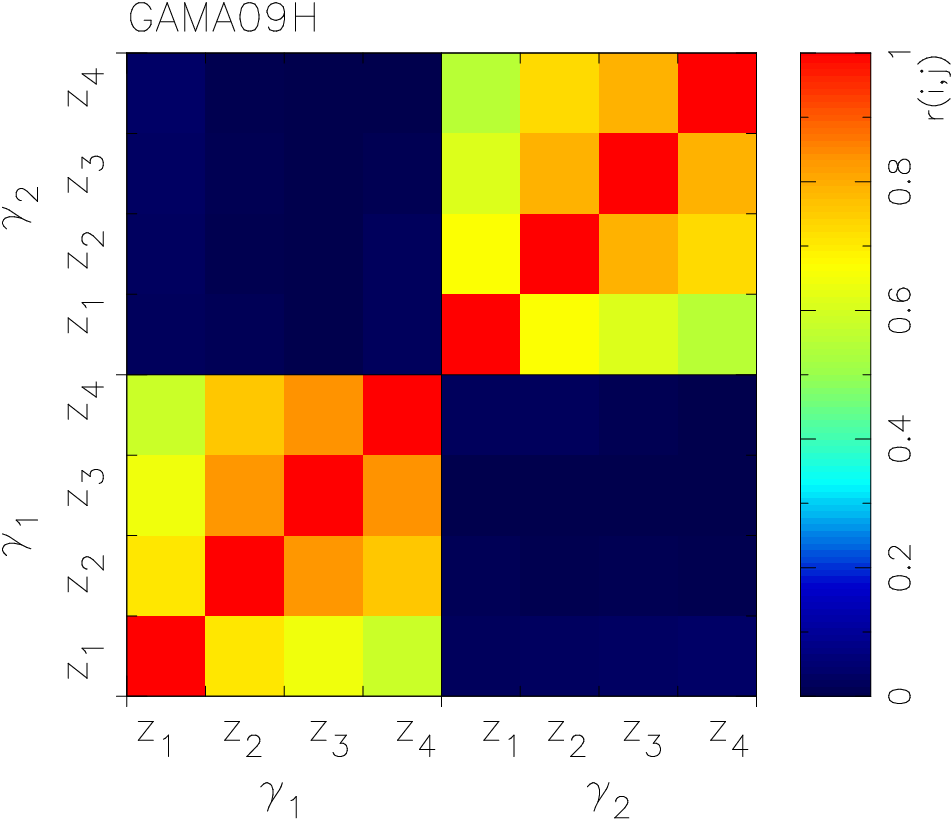}
  \end{center}
  \caption{Two-dimensional matrix plot showing the correlation coefficient 
    of the mean shear covariance matrix,
    $r(d_i,d_j)=\mbox{Cov}(d_i,d_j)/\sqrt{\mbox{Cov}(d_i,d_i),\mbox{Cov}(d_j,d_j)}$.
    Here we show the result for the GAMA09H field, but results in the
    other fields are almost identical to this.
    \label{fig:rcov_GAMA09H}}
\end{figure}

%
%
\begin{figure}
  \begin{center}
    \includegraphics[width=82mm]{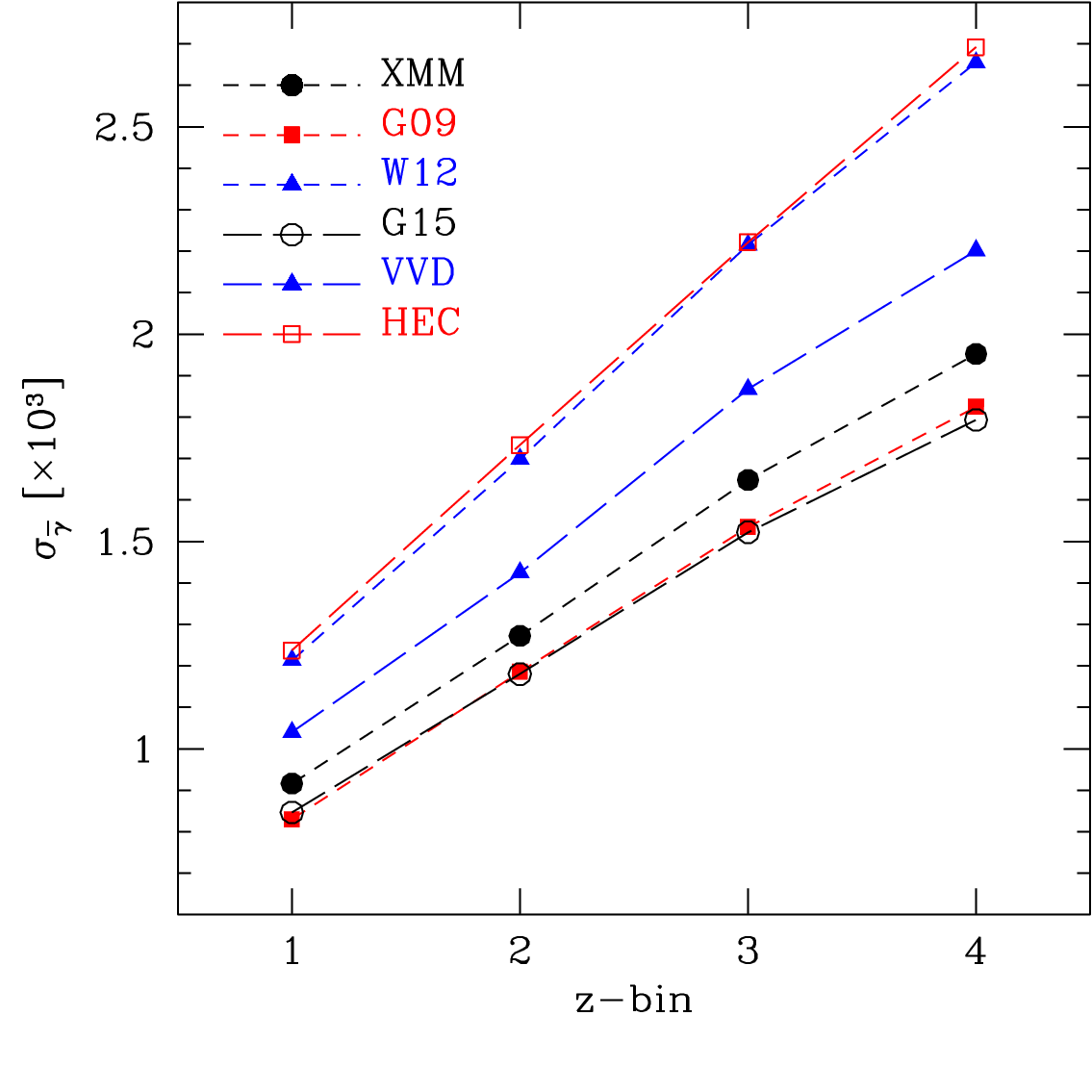}
  \end{center}
  \caption{The root-mean-square values of two component mean shears derived
    from the diagonal components of the covariance matrix, i.e.,
    $\sigma_{\bar{\gamma}}^a=[\mbox{Cov}(\bar{\gamma}_1^a,\bar{\gamma}_1^a)+
      \mbox{Cov}(\bar{\gamma}_1^a,\bar{\gamma}_1^a)]^{1/2}$.
    \label{fig:conste_sigma}}
\end{figure}

A mean shear over a field can naturally arise from the gravitational
lensing shear effect on scales larger than the field, and also can arise
from residual systematics in shear estimation and/or image processing.
The latter, if exists, can have an influence on the cosmological
inference.
In Appendix~\ref{sec:mean_shear}, we utilize the mock catalogs to check 
if the measured mean shears over each field in the real HSC shape 
catalog are consistent with the cosmic shear origin.
Here we describe the covariance matrix of mean shears which is used in 
this test.

We compute the mean shear of mock catalogs for each field and for each
tomographic sample.
It is computed by a simple mean with the shear weight ($w$),
$\bar{\gamma}_i^a=\sum w \gamma_i/\sum w$, where the subscript
$i$ denotes the two shear components, the superscript $a$ denotes
the tomographic bins, and the summation runs over all galaxies in
each tomographic sample and field.
We then define the data vector consisting of eight mean shear
components, 
\begin{equation}
  \label{eq:d_mean_shear}
d_i=(\bar{\gamma}_1^1,\bar{\gamma}_1^2,\bar{\gamma}_1^3,\bar{\gamma}_1^4,
\bar{\gamma}_2^1,\bar{\gamma}_2^2,\bar{\gamma}_2^3,\bar{\gamma}_2^4).
\end{equation}
Finally, for each field we compute the covariance matrix of the data 
vector using 2268 mock realizations, denoted by $\mbox{Cov}(d_i,d_j)$.
Figure~\ref{fig:rcov_GAMA09H} shows the correlation coefficients of the
covariance matrix, 
$r(d_i,d_j)=\mbox{Cov}(d_i,d_j)/\sqrt{\mbox{Cov}(d_i,d_i),\mbox{Cov}(d_j,d_j)}$
for the GAMA09H field as an example.
We find that the mean shears in different tomographic bins are
strongly correlated. This is the natural consequence of galaxies at
different redshifts being affected by the same large-scale structure 
along the line-of-sight. We also find that the correlation is tighter 
for  closer  tomographic redshift bins.
Figure~\ref{fig:conste_sigma} shows the root-mean-square values of two
component mean shears derived from the diagonal components of the
covariance matrix, i.e.,
$\sigma_{\bar{\gamma}}^a=[\mbox{Cov}(\bar{\gamma}_1^a,\bar{\gamma}_1^a)
  +  \mbox{Cov}(\bar{\gamma}_1^a,\bar{\gamma}_1^a)]^{1/2}$.
As expected, the root-mean-square value is higher for the higher 
redshift tomographic bins, as the gravitational lensing effect is stronger 
for sources at higher redshifts.
The difference in the root-mean-square values among different fields 
is due to the different field areas.
It is important to note that the expected value of the mean cosmic
shear over a field depends on the cosmological model, and thus
the covariance also does.
The root-mean-square values presented here are for the WMAP9 cosmology
adopted in the mock simulations.

%
%
\section{Connection with the power spectrum analysis}
\label{appendix:powerspectrum}

%
%
\begin{figure}
  \begin{center}
    \includegraphics[width=82mm]{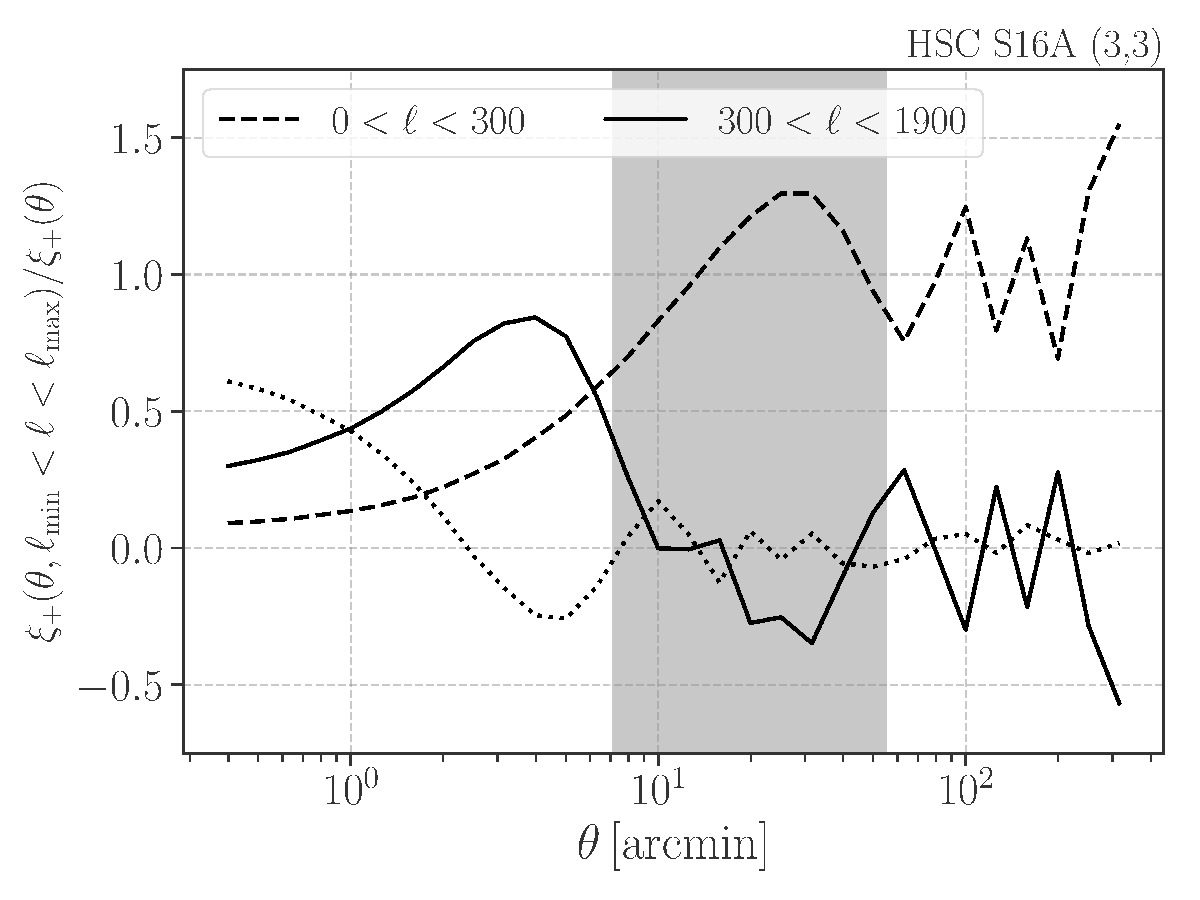}
    \includegraphics[width=82mm]{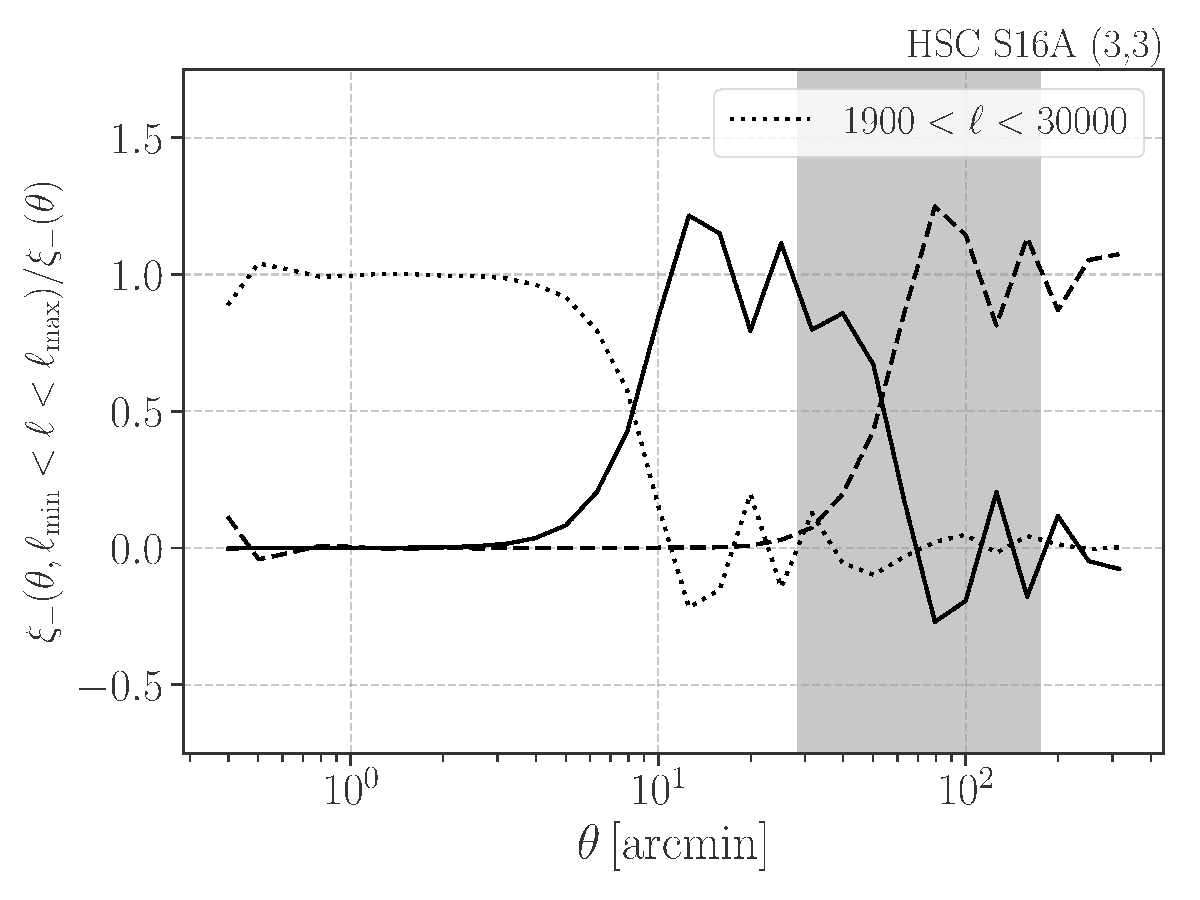}
  \end{center}
  \caption{The fractional contributions of $\xi_{\pm}$ coming from three
    disjoint $\ell$-ranges are shown; dashed lines for $\ell<300$, solid
    lines for $300<\ell<1900$, and dotted lines for $1900<\ell<30000$.
    Only the auto correlation in
    the third tomographic bin, $\xi_\pm^{33}$, are plotted, but the results
    are similar for a different combination of tomographic bins.
    Gray regions show the angular ranges used in this study.
    \label{fig:xi_lrange}}
\end{figure}

In this study, we used exactly the same tomographic galaxy samples as
those used in \citet{2019PASJ...71...43H}, but that study adopted
Fourier-space power spectra as rather than the real space TPCFs used in
this paper.
Here we compare the information content in the measured cosmic shear 
statistics between two studies.

To do so, we divide the $\ell$-integration range of the TPCFs into three
parts (see equation~\ref{eq:xipm}); 
$\ell<300$, $300<\ell<1900$, and $\ell>1900$.
The second $\ell$ range corresponds to the range adopted in
\citet{2019PASJ...71...43H} for their cosmological analysis.
We evaluate these partial contributions assuming the {\it WMAP9} 
cosmology and compute the fractions to the total TPCFs defined by
$\xi_{\pm}(\theta,\ell_{\mbox{min}}<\ell <
\ell_{\mbox{max}})/\xi_{\pm}(\theta)$.
The results are shown in Figure~\ref{fig:xi_lrange}, in which 
we show the result only for one combination of tomographic
bins, as we find that the results are quite similar for different 
combinations of tomographic bins. On the angular 
range used in this study, the dominant contribution to $\xi_+$ 
comes from $\ell<300$, especially on larger $\theta$ scales.
For $\xi_+$, on scales $\theta<60'$, the major contribution comes
from $300<\ell<1900$, whereas on larger scales the majority of 
the contribution comes from $\ell<300$. 
To summarize, a large part of the contribution to $\xi_{\pm}$ on 
scales adopted in this study comes from $\ell<300$, which was not
used in the cosmic shear power spectrum analysis in 
\citet{2019PASJ...71...43H}. 

We also evaluate the fractional contribution to the total signal-to-noise 
ratio from the above three $\ell$ ranges. 
We define the partial signal-to-noise ratio as
\begin{equation}
  \label{eq:partialSN}
  S/N^{\ell\mbox{-}part} = \sum_{i,j} d_i(\xi_\pm^{\ell\mbox{-}part})
  \mbox{Cov}_{ij}^{-1} d_j(\xi_\pm),
\end{equation}
where $\xi_\pm^{\ell\mbox{-}part}$ is the TPCFs computed from a limited
$\ell$-range. Again we assume the {\it WMAP9} cosmology and adopt 
the same angular bins for the data vector $d_i$ and covariance 
matrix as those used in the actual cosmological analysis in this 
study.  We find that the fractional contributions to the total 
$S/N$ are 57\% ($\ell<300$), 37\% ($300<\ell<1900$), and 6\% 
($\ell>1900$). It follows that, 
although \citet{2019PASJ...71...43H} and this study share the same dataset, 
in deriving cosmological constraints two studies utilize fairly different 
and complementary information. This also explains the relatively weak 
correlations of cosmological constraints derived from power spectrum
and TPCF analyses when analyzing the same mock catalogs 
(see Section~\ref{sec:comparison_Hikage}).
Note that the mock analysis presented in
Section~\ref{sec:comparison_Hikage} uses the realistic mock catalogs
in which the realistic shape noise and redshift distributions of
galaxies are included. We performed this test, instead of a
noise-less test, to experimentally examine the
correlations between the two analyses in the presence of such
realistic noises. Nevertheless, a noise-less test would be valuable to
examine a more theoretical aspect of the information content in the
real/Fourier-space cosmic shear measurements, which we leave for a
future study.

%
%
\section{On an error in a constraint on $S_8$ caused by 
  uncertainties in galaxy redshift distributions}
\label{appendix:photoz_uncertainties}

Here we derive a relationship between an
uncertainty in a galaxy redshift 
distribution and an error in a constraint on $S_8$ induced by it in an 
approximative but reasonably reliable manner.
Then we use the derived relationship to discuss a possible impact of an
error in the outlier fraction of galaxy redshift distributions on a
constraint on $S_8$. 

Since the constraint on $S_8$ primary comes from the amplitude of the
cosmic shear correlation function (or power spectrum) on linear to
quasi-nonlinear scales, we will focus on $\xi_+(\theta)$ at
$\theta=10$ arcmin.
For simplicity, we will not treat the full galaxy redshift
distribution but characterize the distribution with a single parameter,
the mean redshift denoted by $\bar{z}_s$.
We consider $\xi_+(\theta=10')$ for a single source
plane model (that is $p(z)=\delta_D(z-\bar{z}_s)$, where $\delta_D$ is
the Dirac's delta function).
The relation between $\xi_+(\theta=10')$ and $\bar{z}_s$ can be
approximated by the following power-law relation with good accuracy,
$\xi_+(\theta=10')\propto \bar{z}_s^u$ with $u\simeq 2.0$ (1.8) for
$0.1<\bar{z}_s<0.7$ ($0.7<\bar{z}_s<1.5$).
Also we find an accurate power-low relation with $S_8$ (for a range of
$0.4<S_8<1.2$), 
$\xi_+(\theta=10')\propto S_8^v$ with $v\simeq 2.8$, 2.3, 2.0, and 1.8
for $\bar{z}_s=0.44$, 0.77, 1.05, and 1.33, respectively.
From those two scaling relations, we have the following relationship,
\begin{equation}
\label{eq:dz_s8}
{{\delta S_8} \over S_8} = -{u\over v}  {{\delta \bar{z}_s} \over \bar{z}_s}. 
\end{equation}
Note that the scaling factor $u/v$ ranges
from 0.7 to 1 for our interested range of
$0.4 \lesssim \bar{z}_s \lesssim 1.4$.
It is also noted that the anti-relationship originates from the fact
that an over/under-estimate of the mean redshift leads to
an over/under-estimate of the theoretical prediction, resulting in an
under/over-estimate of $S_8$ to compensate.

In the re-analysis of DES-Y1 cosmic shear data presented in
\citet{2020A&A...638L...1J}, it is reported that the mean redshifts of
the galaxy redshift distributions derived bases on the COSMOS 30-band
photo-$z$ are systematically lower than those derived
based on spectroscopic samples. They found $\Delta \bar{z}_s$
(defined by $\bar{z}_s$[spec-$z$]$-\bar{z}_s$[COSMOS-30]) of
$+0.014$,  $+0.053$, $+0.020$,  and $+0.035$ for their four tomographic
redshift bins ($0.2<z<0.43$, $0.43<z<0.63$, $0.63<z<0.9$, and
$0.9<z<1.3$).
They found the best fit $S_8$ values of $0.763$ and $0.793$ for galaxy
redshift distributions bases on the COSMOS 30-band photo-$z$ and
spectroscopic samples, respectively, resulting in 
$\Delta S_8 = S_8$[spec-$z$]$-S_8$[COSMOS-30]$=-0.030$.
The relation between those values are in a good agreement with one
expected from the derived relationship, equation (\ref{eq:dz_s8}),
supporting its validity.

Finally, we discuss a possible impact of an error in the outlier
fraction of galaxy redshift distributions on a constraint on $S_8$ using
simple models. 
Suppose a galaxy redshift distribution has a bi-modal shape, such as
one shown in top-panel of Figure \ref{fig:pz}, consisting of a main
population with $\langle z_{\rm main} \rangle =  0.5$ and an outlier population
with $\langle z_{\rm out} \rangle =  3$.
Assuming the outlier fraction of 5\%, the mean redshift of this
distribution is $\bar{z}_s = 0.95\times 0.5 + 0.05 \times 3=0.625$.
If we suppose 10\% error in the outlier fraction, the error in the mean
redshift is $\delta \bar{z}_s = \pm 0.0125$, leading to  $\delta S_8 \simeq
\mp 0.016$ (here we used equation (\ref{eq:dz_s8}) with $u/v=0.8$).
Actual errors in the outlier fraction of our galaxy samples are not
understood well, but this rough estimate gives us a crude idea of its
possible impact on a constraint on $S_8$.

\end{document}